 \documentclass[final,5p,times,twocolumn, authoryear]{elsarticle}

\usepackage{amsmath, amssymb}
\usepackage{lipsum}
\usepackage{multirow, booktabs}
\usepackage{lineno}
\usepackage{hyperref}
\hypersetup{
   colorlinks = true,
   citecolor  = colorLink,
   urlcolor   = colorLink,
   linkcolor  = colorLink,
}

\journal{Journal of High Energy Astrophysics}

\begin{document}

\begin{frontmatter}


\author[1]{Swaraj Pratim Sarmah}
\ead{rs_swarajpratimsarmah@dibru.ac.in}

\author[1]{Pranjal Sarmah}
\ead{p.sarmah97@gmail.com}

\author[1]{Umananda Dev Goswami\corref{cor1}}
\ead{umananda@dibru.ac.in}
\cortext[cor1]{Corresponding author}

\affiliation[1]{organization={Department of Physics},%
            addressline={Dibrugarh University}, 
            city={Dibrugarh},
            postcode={786004}, 
            state={Assam},
            country={India}}

\title{Ultra-high energy cosmic rays with UFA-15 source model in Bumblebee 
gravity theory}


\begin{abstract}
We explore the effects of Bumblebee gravity on the propagation of ultra-high 
energy cosmic rays (UHECRs) using astrophysical sources modelled in the 
Unger-Farrar-Anchordoqui (UFA) framework (2015), which includes star formation 
rate (SFR), gamma-ray bursts (GRBs), and active galactic nuclei (AGN). 
We compute the density enhancement factor for various source separations 
distances ($d_\text{s}$s) up to 100 Mpc within the Bumblebee gravity 
scenario. Additionally, we calculate the CRs flux and their 
suppression, goodness-of-fit values obtained from comparisons with 
observational data from the Pierre Auger Observatory (PAO) and the Telescope 
Array experiment data for the flux and the Levenberg-Marquardt algorithm for 
suppression. The anisotropy in the CRs arrival directions is examined, 
with corresponding goodness-of-fit values obtained from the PAO 
surface detector data (SD 750 and SD 1500). Finally, we present skymaps of 
flux and anisotropy under different model assumptions, providing insights into 
the observational signatures of UHECRs in Bumblebee gravity. We show 
that Bumblebee gravity stands as a viable cosmological model for explaining key 
observational features of UHECRs, including spectrum, composition and 
anisotropy. Our results show that increasing the Bumblebee gravity 
parameter $l$ enhances the density factor $\xi$, particularly at low energies, 
highlighting Lorentz violation's impact on CRs' propagation. Larger 
$d_\text{s}$ values amplify deviations from the $\Lambda$CDM 
model, with AGN sources dominating at high energies and GRB/SFR sources at 
lower energies. The skymaps indicate the structured flux patterns at large 
$d_\text{s}$ and structured anisotropies at higher energies.
\end{abstract}




\end{frontmatter}




\section{Introduction}

\par
Cosmic rays (CRs), which are highly energetic charged particles (from protons
to iron nuclei) radiations emanating from outer space, span a wide range 
of energies, approximately from $10^9$ eV to $10^{20}$ eV, with three 
prominent features in their energy spectrum. The first feature, known as the 
``knee", occurs around $10^{15.6}$ eV, where the spectrum steepens
\citep{Antoni:2005wq}. The second feature, called the ``ankle", appears near 
$10^{\bf 18.7}$ eV, where the spectrum hardens~\citep{Bird:1993yi, hires2008, 
Abraham:2010mj}. And finally, there is a cutoff at roughly $10^{19.6}$ eV 
\citep{hires2008, Auger2008}. Additionally, between the knee and the ankle, 
there are more subtle features: a slight hardening of the spectrum around 
$2 \times 10^{16}$ eV~\citep{Apel:2012tda, Aartsen:2013wda, Knurenko:2013dia, 
Prosin:2014dxa}, followed by two softening points at approximately 
$10^{16.9}$ eV~\citep{Apel:2012tda, Aartsen:2013wda} and $10^{17.5}$ eV 
\citep{Bird:1993yi, AbuZayyad:2000ay, Knurenko:2013dia, Prosin:2014dxa}, the 
latter commonly referred to as the ``second knee". These variations in the 
energy spectrum are related to the processes of CRs' production, the 
distribution of their sources, and their propagation through space.

The first and second knees have relatively simple explanations. They represent 
the maximum energy limits of galactic magnetic confinement or the acceleration 
capacity of the sources, both of which scale linearly with the nuclear charge 
$Z$. The first knee marks the point where protons stop contributing to CRs 
flux, while the second knee corresponds to the point where the highest-$Z$ 
galactic CRs are no longer confined. As energy increases beyond the second 
knee and approaches the ankle, the composition transitions from heavy to 
light elements~\citep{Kampert:2012mx}, while the arrival directions of CRs 
remain nearly isotropic across this range \citep{Abreu:2011ve, Auger:2012an, 
ThePierreAuger:2014nja}. Beyond the ankle, the spectrum significantly hardens 
and the composition becomes gradually heavier as per the interpretations 
obtained by using the standard extrapolations of accelerator-constrained 
particle physics models~\citep{auger_prd9012, auger_prd90}. This evolution in 
the composition and spectral index of extragalactic CRs raises significant 
questions. A composition dominated by protons could match the observed 
extragalactic spectrum~\citep{berezinski_four_feat}, provided the experimental 
uncertainties in the energy scale are accounted for~\citep{Ahlers:2012az}. 
However, models that fit both the spectrum and composition at the highest 
energies often predict a noticeable gap between the end of galactic CRs and 
the beginning of extragalactic ones~\citep{Allard:2005cx, Allard:2007gx, 
Allard:2008gj, DeDonato:2008wq, Taylor:2013gga, Deligny:2014opa}. Although 
new models can be developed to bridge this gap, they require fine-tuning to 
align the new population precisely to fill the void~\citep{Gaisser:2013bla, 
Aloisio:2013hya, Giacinti:2015hva}.

The turbulent magnetic fields (TMFs) are permeated through the intergalactic 
medium (IGM) that play a pivotal role in the propagation of ultra-high-energy 
CRs (UHECRs) originating from extragalactic sources. When charged particles 
traverse a random magnetic field, their propagation is governed by the 
distance traveled by them relative to the scattering length, denoted as 
$l_{\text{D}} = 3D/c$, where $D$ represents the diffusion coefficient and $c$ 
denotes the speed of light. If the distance traveled by a particle is 
significantly shorter than its scattering length, its motion exhibits 
ballistic behavior. Conversely, if the travel distance is substantially 
greater, the motion becomes diffusive. Incorporating the effects of 
extragalactic TMFs and the finite density of sources in UHECRs propagation 
studies one can unveil a low-energy magnetic horizon effect, as mentioned in 
Ref.~\citep{horizon_jcap}. This phenomenon has the potential to reconcile 
observations with a higher spectral index, which aligns more closely with 
the predictions derived from diffusive shock acceleration. An alternative 
hypothesis posits that heavy nuclei are accelerated by extragalactic sources, 
leading to the photodisintegration and the generation of secondary nucleons. 
This process may account for the observed light composition below the ankle, 
as proposed in Ref.~\citep{Unger:2015laa} and further elaborated in 
Ref.~\citep{globus}. The propagation of UHECRs within the intergalactic magnetic 
fields can be investigated utilizing the Boltzmann transport equation or 
employing various simulation methodologies. In Ref.~\citep{Supanitsky}, a set 
of partial differential equations is introduced to describe UHECRs propagation 
in random magnetic fields, derived from the Boltzmann transport equation. This 
study underscores the diffusive nature of CRs propagation. An analytical 
solution to the diffusion equation for CRs in an expanding Universe is 
provided in Ref.~\citep{berezinskyGre}, while Ref.~\citep{harari} offers a 
numerical fitting of the diffusion coefficient $D(E)$ for both Kolmogorov 
and Kraichnan  turbulences. The effects of CRs diffusion within 
the magnetic field of the local supercluster on UHECRs originating from 
nearby extragalactic sources are the subject of further research and analysis. 
The energy spectra of UHECRs are extensively studied in Ref.~\citep{molerach}, 
where the authors propose a strong enhancement of the flux at specific energy 
ranges as a potential explanation for the observed features of the CR 
spectrum and composition. Ref.~\citep{aloisio} provides a comprehensive 
analytical study of UHE particle propagation in extragalactic magnetic fields 
by solving the diffusion equation while accounting for energy losses. 
Additionally, Ref.~\citep{berezinsky_modif} examines the ankle, instep, and 
GZK cutoffs in the UHECR spectrum by considering the modification factor that 
arises from the various energy losses experienced by CR particles as they 
traverse complex galactic or intergalactic medium. Similarly, 
Ref.~\citep{berezinski_four_feat} identifies four key features in the CR 
protons spectrum: the ankle, instep, second knee, and GZK cutoff, by 
considering extragalactic proton interactions with the CMB and assuming a 
power-law spectrum. Recent combined fits to UHECR spectrum and 
composition data suggest that astrophysical sources may exhibit hard 
injection spectra, low maximum energies, and heavy chemical 
compositions \citep{pao_jcap04}.

General relativity (GR), developed by Albert Einstein in 1915, is one of the 
most elegant, well-validated, and successful theories in physics formulated 
to describe gravitational interactions. The theory gained significant support 
with the detection of gravitational waves by the LIGO detectors in 2015 
\citep{ligo}, nearly a century after Einstein himself predicted them. Similarly,
the images of black holes at the centers of M87 and the Milky Way galaxies 
taken very recently by the Event Horizon Telescope collaboration 
\citep{m87a, m87b, m87c, m87d, m87e, m87f} are 
remarkable achievements in support of GR. These and other milestones have 
reinforced the importance of GR even after more than 100 years. However, GR 
faces major challenges, both theoretically and observationally. For instance, 
GR is not a quantum theory of gravity and is also not suitable to incorporate 
into a consistent quantum framework \citep{Hawking_1976, Oppenheim_2023}. On the observational side, GR struggles to explain the accelerated expansion of the 
Universe \citep{reiss1998, perlmutter1999, spergel2003, astier} without 
invoking dark energy \citep{Cognola:2007zu, sami, udg_prd, Odintsov:2020nwm, 
Odintsov:2019evb}. Additionally, the theory cannot fully account for the galaxy 
rotation curves, which suggest the presence of unseen mass, often attributed 
to dark matter \citep{Oort, Zwicky1, Zwicky2, Garrett, nashiba1}. In this 
context, the Bumblebee gravity model was introduced in 1989 
\citep{Kostelecky_1989} alongside the variety of other gravity theories 
\citep{Buchdahl_1970, Hehl_1976, Ferraro_2007, deRham_2011} developed 
over periods. This Bumblebee gravity model incorporates a 
vector field, known as the Bumblebee field, which modifies the Einstein field 
equations of GR. The model is a straightforward yet effective extension of the 
standard model, referred to as the standard model extension (SME) that 
operates on the principle of Lorentz symmetry breaking (LSB) through the 
introduction of a vector field \citep{Kostelecky_1989, Kostelecky_1989a}. By 
adding this field and its potential to the conventional Einstein-Hilbert (EH) 
action, the model alters the standard Einstein field equations, providing 
insights into various cosmological phenomena without the need to invoke 
exotic components like dark matter (DM) or dark energy (DE) 
\citep{Capelo_2015, Bertolami_2005}. The influence of SMEs in gravitational 
research is extensively discussed in Refs.~\citep{Bluhm_2008, Kostelecky_2009,  
Altschul_2005, Gogoi_2022, Karmakar_2023}.

In our previous works, we have explored the flux characteristics 
\citep{swaraj1, Sarmah:2025yoy} and anisotropic properties \citep{swaraj2} of UHECRs for a 
single source within the framework of $f(R)$ gravity. Moreover, for the 
multiple sources, we have examined their propagation \citep{swaraj4}, flux 
suppression \citep{swaraj3, swaraj7}, and anisotropy \citep{swaraj5} in different 
modified gravity theories (MTGs) along with the $\Lambda$CDM model. Building 
on these motivations, our current aim is to study different properties of CRs 
using the Unger-Farrar-Anchordoqui (UFA) source model, especially within the 
realm of Bumblebee gravity. To validate our results from the observational 
point of view, we utilize the data from Pierre Auger \citep{augerprd2020} and 
Telescope Array \citep{ta2019} experiments.

The paper is organized as follows: In Section \ref{secII}, we provide a 
discussion on Bumblebee gravity and derive the essential equations required 
for the analysis. Section \ref{secIII} 
focuses on the diffusion of CRs and turbulent magnetic fields, including the 
equations for the enhancement factor and flux within the context of Bumblebee 
gravity. In Section \ref{secIV}, we introduce the UFA-15 source model. Section 
\ref{secV} presents the numerical results, which are further subdivided into 
four subsections addressing density enhancement, flux, suppression, and 
anisotropy. Finally, Section \ref{secVI} offers a comprehensive summary and 
concluding remarks.

\section{Bumblebee Gravity Models}\label{secII}
In this section, we derive the Hubble parameter $H(z)$ in the isotropic 
Universe within the framework of the Bumblebee gravity model. We begin with 
the Friedmann-Lema\^itre-Robertson-Walker (FLRW) metric and then solve the 
modified Friedmann equations under the vacuum expectation value (VEV) 
condition and derive the evolution of the Hubble parameter as a function of 
redshift $z$.

In the presence of a Bumblebee field $B_\mu$, we consider the action of the 
model as \citep{Capelo_2015, Pranjal_Bumblebee2024}
\begin{align}
S = \int \sqrt{-g}& \left[ \frac{1}{2\kappa} \left( R + \zeta B^\mu B^\nu R_{\mu\nu} \right) 
    - \frac{1}{4} B^{\mu\nu} B_{\mu\nu} \right.  \nonumber \\
  & \left. - V \left( B^\mu B_\mu \pm b^2 \right) 
    + \mathcal{L}_M \right] d^4x,
\end{align}
where $ \kappa = 8\pi G $. The coupling constant $\zeta$ has the dimension 
$[\zeta] = M^{-2}$. The field-strength tensor is defined as 
$B_{\mu\nu} \equiv \partial_\mu B_\nu - \partial_\nu B_\mu$. The quantity 
$b^2 \equiv b_\mu b^\mu = \langle B_\mu B^\mu \rangle_0 \neq 0$ represents 
the expectation value of the contracted Bumblebee vector. The potential $V$ 
attains its minimum when $B_\mu B^\mu \pm b^2 = 0$ and the $\mathcal{L}_M$ 
denotes the Lagrangian density for the matter fields. 

In the isotropic Universe, the FLRW metric is given by
\begin{equation}
ds^2 = -\,dt^2 + a(t)^2 \left( dx^2 + dy^2 + dz^2 \right),
\end{equation}
where $a(t)$ is the scale factor. Under the vacuum expectation value (VEV) 
condition ($ V = V' = 0 $) and isotropy in space, the modified Friedmann 
equations in the Bumblebee gravity are \citep{Capelo_2015}
\begin{equation}
3H^2 = \frac{\kappa \rho}{1 - l}, \quad 3H^2 + 2\dot{H} = -\frac{\kappa P}{1 - l},
\end{equation}
where  $ \rho $ is the energy density, $ P $ is the pressure, and 
$ l = \zeta B_0^2 $ is the Lorentz violation parameter. The continuity equation 
for a perfect fluid in this gravity model with the equation of state 
$ P = \omega \rho $ can be written as \citep{Capelo_2015}
\begin{equation}
\dot{\rho} = -\,3H\rho\, (1 + \omega) + \frac{3\,l}{\kappa} \left( \frac{\ddot{a}}{a}H - \frac{\dddot{a}}{a}  \right),
\end{equation}
which can be further simplified to
\begin{equation}
\dot{\rho} = -\,3H\rho\, (1 + \omega) - \frac{3\,l}{2\, \kappa} \frac{d}{dt} \left( H^2 + \frac{\kappa\, \omega \, \rho}{1 - l}  \right). 
\end{equation}
The solution of this equation can be obtained as
\begin{equation}
\rho = \rho_0\, a^{-\frac{6(1-l)(1 + \omega)}{2-l(1-3\omega)}}.
\end{equation}
For matter ($ \omega = 0 $), radiation ($ \omega = \frac{1}{3} $), and dark 
energy ($ \omega = -1 $), the respective energy densities evolve as
\begin{equation}
\rho_m = \rho_{m0}\, a^{-\frac{6 (1-l)}{2-l}}, \quad \rho_r = \rho_{r0}\, a^{-4 (1-l)}, \quad \rho_\Lambda = \rho_{\Lambda0}. 
\end{equation} 

By substituting the total energy density $\rho(z) = \rho_m(z) + \rho_r(z) + 
\rho_\Lambda(z)$ into the first modified Friedmann equation, we obtain
\begin{align}
    3H^2(z)  &= \frac{\kappa}{1 - l}\\  \nonumber 
  &   \left[ \rho_{m0}\, (1 + z)^\frac{6 (1-1)}{2-l} + \rho_{r0}\, (1 + z)^{4 (1-l)} + \rho_{\Lambda0} \right].
\end{align}
To express this equation in terms of dimensionless density parameters, we use 
the critical density definition $\rho_{c0} = {3H_0^2}/{\kappa}$ and define 
the present-day density parameters as $\Omega_{m0} = {\rho_{m0}}/{\rho_{c0}}$, 
$\Omega_{r0} = {\rho_{r0}}/{\rho_{c0}}$, and $\Omega_{\Lambda0} = 
{\rho_{\Lambda0}}/{\rho_{c0}}$. Rewriting this Friedmann equation in terms of 
these parameters, we obtain
\begin{align}
    H^2(z) &= H_0^2 \\ \nonumber
 &   \frac{\Omega_{m0}\, (1 + z)^\frac{6 (1-1)}{2-l} + \Omega_{r0}\, (1 + z)^{4 (1-l)}  + \Omega_{\Lambda0}}{1 - l}.
\end{align}
Thus the final expression of the Hubble parameter in the isotropic Universe 
for the Bumblebee gravity model is
\begin{align}
    H(z) &= H_0 \\ \nonumber
&    \sqrt{ \frac{\Omega_{m0}\, (1 + z)^\frac{6 (1-1)}{2-l} + \Omega_{r0}\, (1 + z)^{4 (1-l)} + \Omega_{\Lambda0}}{1 - l}}.
\end{align}
Accordingly, the cosmological time evolution as a function of redshift can be 
expressed as
\begin{align}
&\bigg |\frac{dt}{dz}\bigg | = (H_0(1+z))^{-1} \nonumber \\
\times \, &\left[\frac{\Omega_{m0}\, (1 + z)^\frac{6 (1-1)}{2-l} + \Omega_{r0}\, (1 + z)^{4 (1-l)}  + \Omega_{\Lambda0}}{1 - l}\right]^{-1/2}.
\end{align}
\begin{figure}[!h]
\centering
\includegraphics[width=0.9\linewidth]{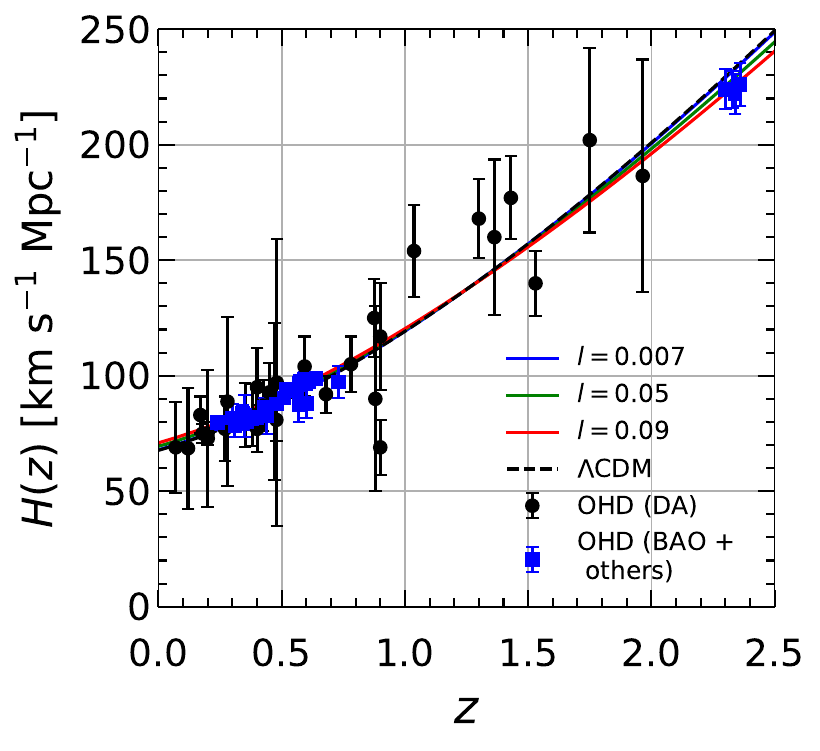}
\vspace{-0.3cm}
\caption{The Hubble parameter $H(z)$ as a function of $z$ for the Bumblebee 
gravity model with different values of the Lorentz violation parameter. The 
observational data are taken from the Baryon Acoustic Oscillations (BAO) and 
Differential Age (DA) method \citep{swaraj1, gogoi, solanki}. The $\Lambda$CDM 
model is taken as a reference for the comparison.}
\label{fig2a}
\end{figure}

In Fig.~\ref{fig2a}, the Hubble parameter $H(z)$ is shown as a function of 
redshift $z$ for the Bumblebee gravity model with different values of the 
Lorentz violation parameter ($l = 0.007, 0.05, 0.09$). The observational 
data are obtained from Baryon Acoustic Oscillations (BAO) and the Differential 
Age (DA) method \citep{swaraj1, gogoi, solanki}. The $\Lambda$CDM model is 
included as a reference for the comparison. The deviations among the predicted 
results increase at higher redshifts, although it is observed that all 
predicted results align well with the observational data.

\section{Diffusion of Cosmic Rays in Turbulent Magnetic Fields}\label{secIII}
Modelling of extragalactic magnetic fields presents significant challenges due 
to certain limitations \citep{han}. The precise values of these fields remain 
uncertain and vary depending on the specific region of extragalactic space 
\citep{hu_apj, urmilla}. To simplify the investigation, attention is directed 
toward the propagation of CRs in a turbulent and uniform extragalactic 
magnetic field. This field is characterized by its root mean square (RMS) 
strength $B$ and the coherence length $l_\text{c}$. The RMS strength 
$B$, defined as $\sqrt{\langle B^2(x)\rangle}$, can range from as low as 
$10^{-16}$~G in voids \citep{Neronov:2010gir} to values $\gtrsim 10$~nG in 
galaxy clusters, with typical strengths of $\sim 1$~nG in filaments
~\citep{feretti, Valle, Vazza}. The coherence length $l_\text{c}$ is 
usually taken between $0.01$~Mpc and $1$~Mpc~\citep{sigl}. The effective 
Larmor radius for a charged particle with charge $Ze$ and energy $E$ 
propagating through a TMF of strength $B$ can be expressed as
\begin{equation}\label{larmor} r_\text{L} = \frac{E}{ZeB} \simeq 1.1\, \frac{E/\text{EeV}}{ZB/\text{nG}}\;\text{Mpc}. \end{equation}

The concept of critical energy is fundamental for understanding the diffusion 
of charged particles in magnetic fields. It is defined as the energy at which 
the coherence length of a particle with charge $Ze$ equals its Larmor radius, 
i.e., $r_\text{L}(E_\text{c}) = l_\text{c}$. Thus the critical energy can be 
expressed as  
\begin{equation}\label{cri_energy}
 E_\text{c} = ZeBl_\text{c} \simeq 0.9 Z\, \frac{B}{\text{nG}}\, \frac{l_\text{c}}{\text{Mpc}}\;\text{EeV}. 
\end{equation}  
This energy delineates two distinct diffusion regimes: resonant diffusion at 
energies below $E_\text{c}$ and non-resonant diffusion at energies above 
$E_\text{c}$.

The diffusion coefficient $D$ as a function of energy is given by \citep{harari}
\begin{equation}\label{diff_coeff} 
D(E) \simeq \frac{c\,l_\text{c}}{3}\left[4 \left(\frac{E}{E_\text{c}} \right)^2 + a_\text{I} \left(\frac{E}{E_\text{c}} \right) + a_\text{L} \left(\frac{E}{E_\text{c}} \right)^{2-\gamma} \right],
\end{equation}  
where $\gamma$ is the spectral index, and $a_\text{I}$ and $a_\text{L}$ are 
coefficients. For the Kolmogorov spectrum in a TMF, $\gamma = 5/3$ with 
$a_\text{L} \approx 0.23$ and $a_\text{I} \approx 0.9$. As mentioned earlier,
the diffusion length $l_\text{D}$, representing the distance at which a 
particle’s overall deflection reaches about one radian, is defined as 
$l_\text{D} = 3D/c$. In the diffusive regime, the transport equation for UHE 
particles propagating through an expanding Universe from a source located 
at $x_\text{s}$ can be written as \citep{berezinskyGre}
\begin{align}\label{diff_eqn}
\frac{\partial n}{\partial t} + 3 H(t)\, n - b(E,t)\,\frac{\partial n}{\partial E}& -n\, \frac{\partial n}{\partial E} - \frac{D(E,t)}{a^2(t)}\,\nabla^2 n  \nonumber \\
&= \frac{\mathcal{N}(E,t)}{a^3(t)}\,\delta^3({x}-{x}_\text{s}),
\end{align}  
where the Hubble parameter $H(t)$ is expressed as $H(t) = \dot{a}(t)/a(t)$. 
Here $\dot{a}(t)$ represents the rate of change of the scale factor $a(t)$ 
with respect to cosmic time $t$. The coordinates ${x}$ refer to comoving 
positions, $n$ indicates the density of particles, and $\mathcal{N}(E)$
is the emissivity of sources. At a specific time $t$, associated with redshift 
$z$, the separation between the source and the particle is given by 
$r_\text{s} = {x} - {x_\text{s}}$. Energy losses experienced by the 
particles, caused by the expansion of the Universe and their interactions with 
the CMB, are accounted for by
\begin{equation}
\frac{dE}{dt} = -\, b(E,t),\;\; b(E,t) = H(t)E + b_\text{int}(E).
\end{equation}
In this context, $H(t)E$ represents the adiabatic energy losses caused by the 
expansion of the Universe, while $b_{\text{int}}(E)$ accounts for energy 
losses due to interactions. UHECRs can interact with both CMB and 
extragalactic background light (EBL). However, in this work, we consider 
energy losses due to interactions with the CMB only. These interaction losses, 
involve processes such as pair production and photopion production (for 
further details see \citep{harari}). The general solution to 
Eq.~\eqref{diff_eqn} was derived in Ref.~\citep{berezinskyGre} and is given by
\begin{equation}\label{density}
n(E,r_\text{s})= \int_{0}^{z_{i}} dz\, \bigg | \frac{dt}{dz} \bigg |\, \mathcal{N}(E_\text{g},z)\, \frac{\textrm{exp}\left[-r_\text{s}^2/4 \lambda^2\right]}{(4\pi \lambda^2)^{3/2}}\, \frac{dE_\text{g}}{dE},
\end{equation}
where $\lambda$ is the Syrovatskii variable and is formulated as \citep{syrovatsky_1959}
\begin{equation}\label{syro}  
\lambda^2(E, z) = \int_{0}^{z} dz \, \left| \frac{dt}{dz} \right| (1 + z)^2 D(E_\text{g}, z).  
\end{equation}
and $E_\text{g}(E, z)$ denotes the generation energy at redshift $z$ 
corresponding to an observed energy $E$ at $z = 0$. In the diffusive regime, 
the particle density is influenced by factors such as energy, the distance 
from the source, and the properties of the TMF. Due to diffusion, the 
observed density of CRs at a given distance from the source can be 
significantly higher than what would be expected under rectilinear 
propagation, which follows a $1/r^2$ behaviour. This enhancement arises from 
the delayed escape and multiple scattering of particles, and it can be 
quantified by the ratio of the diffusive particle density to the rectilinear 
desity \citep{molerach}. This is expressed by the enhancement factor, which 
is given by
\begin{equation}\label{enhancement}
\xi(E,r_\text{s})=\frac{4\pi r_\text{s}^2c\, n(E, r_\text{s})}{\mathcal{N}(E)}.
\end{equation}

The diffusion of CRs in TMFs has been extensively studied by various 
researchers \citep{berezinskyGre, blasi, globus, sigl, stanev, kotera, 
Yoshiguchi, lemoine1, hooper, hooper2, sigl2007, aloisio_ptep}. Berezinsky 
and Gazizov \citep{berezinsky_gazizov, berezinskyGre} extended the Syrovatskii 
solution \citep{syrovatsky_1959} to investigate the diffusion of protons in 
an expanding Universe. The flux from a CR source located at a distance 
$r_\text{s}$, significantly greater than the diffusion length $l_\text{D}$, 
can be determined by solving the diffusion equation in the framework of an 
expanding Universe \citep{berezinskyGre}. The resulting expression is given 
by \citep{Manuel}
\begin{align}\label{fluxeq}
J(E) = \frac{c}{4\pi} \int_{0}^{z_{\text{max}}} \!\! dz \, \left| \frac{dt}{dz} \right| \, \mathcal{N}&\left[E_\text{g}(E, z), z\right]\\ \nonumber  
&\frac{\exp\left[-r_\text{s}^2 / (4 \lambda^2)\right]}{(4 \pi \lambda^2)^{3/2}} \frac{dE_\text{g}}{dE},
\end{align}
where $z_{\text{max}}$ represents the highest redshift at which the source 
begins emitting CRs. $E_g$ is the generation energy of a particle at 
the redshift $z$ that corresponds to energy $E$ at $z = 0$.
The total source emissivity $\mathcal{N}$ is determined by summing the 
charge-specific emissivities $\mathcal{N}_\text{Z}$s for different nuclei. 
The charge-specific emissivity follows a power-law form with a rigidity cutoff 
$ZE_\text{max}$ and is expressed as $\mathcal{N}_\text{Z}(E, z) = \varepsilon_\text{Z} 
f(z) E^{-\gamma} / \cosh(E / ZE_\text{max})$ \citep{mollerachjcap}. Here, 
$\varepsilon_\text{Z}$ indicates the relative contribution of nuclei with 
charge $Z$ to the CRs flux, while $f(z)$ encapsulates the evolution of source 
emissivity as a function of redshift $z$.  
%
%
Eq.~\eqref{fluxeq} can be extended to nuclei when interpreted in terms of 
their rigidity. During photo-disintegration processes, the rigidity and 
Lorentz factor of the primary fragment is generally conserved, which 
minimally alters the diffusion characteristics of the particle. Nevertheless, 
these processes introduce challenges, as the source term $\mathcal{N}$ 
describes the primary nucleus responsible for producing the observed fragment 
and determining this relationship is difficult due to the stochastic nature of 
the disintegration. This issue was addressed in Ref.~\citep{mollerachjcap}, and 
we build on this discussion within the frameworks of modified and 
alternative gravity theories \citep{swaraj2, swaraj3, swaraj4}. As our focus 
shifts to multiple sources rather than a single one, we apply the propagation 
theorem \citep{aloisio} to aggregate contributions from all sources, which can 
be represented as
\begin{equation}\label{lim1}
\int_{0}^{\infty} dr \, 4\pi r^2 \frac{\exp\left[-r^2 / (4 \lambda^2)\right]}{(4 \pi \lambda^2)^{3/2}} = 1.
\end{equation}
                                                      
To analyse how the finite distance to sources affects suppression, we compute 
the sum based on a specific set of distance distributions. These distributions 
assume a uniform density of sources, with the distances from the observer 
given by \citep{mollerachjcap, Manuel}
\begin{equation}  
r_\text{i} = \left(\frac{3}{4\pi}\right)^{\!1/3}\!\!\! d_\text{s}\, \frac{\Gamma(i + 1/3)}{(i - 1)!},  
\end{equation}
where $d_\text{s}$ represents the separation distance between the sources.
For a discrete distribution of sources, summing over all sources introduces 
a specific suppression factor \citep{mollerachjcap, Manuel}
\begin{equation} \label{F_supp}  
F \equiv \frac{1}{n_\text{s}} \sum_i \frac{\exp\left[-r_\text{i}^2/4 \lambda^2\right]}{(4\pi \lambda^2)^{3/2}},  
\end{equation}
instead of obtaining Eq.~\eqref{lim1}, where $n_\text{s}$ denotes the 
source density. In Eq.~\eqref{fluxeq}, after summing contributions from all 
sources, the modified flux for an ensemble of sources in the Bumblebee gravity 
can be expressed as
\begin{align}\label{flux}
J_\text{mod}(E)  \simeq  \frac{R_\text{H} n_\text{s}}{4\pi}\!& \int_{0}^{z_\text{max}}\!\!\!\!\! dz\, (1+z)^{-1}  \Big | \frac{dt}{dz} \Big | \nonumber \\
&\times \mathcal{N}\left[E_\text{g}(E, z), z\right]\, \frac{dE_\text{g}}{dE}\, F,
\end{align}
where $R_\text{H} = c/H_0=4.3$ Gpc is the Hubble radius. We can rewrite 
Eq.~\eqref{syro} in terms of $R_\text{H}$ and from 
Eq.~\eqref{diff_coeff} as
\begin{align}\label{ad}
\lambda^2(E,z) = &\,\frac{H_0R_\text{H} l_\text{c}}{3} \int_{0}^{z} \!\! dz\, \bigg| \frac{dt}{dz} \bigg| (1+z)^2 
\bigg[4 \left( \frac{(1+z)\,E}{E_\text{c}} \right)^2  \nonumber \\
    & + a_\text{I} \left( \frac{(1+z)\,E}{E_\text{c}} \right) 
    + a_\text{L} \left( \frac{(1+z)\,E}{E_\text{c}} \right)^{2-\gamma}
\bigg].
\end{align}

\section{Source Model}\label{secIV}
\par
We adopt the  UFA-15 CR source model~\citep{Unger:2015laa}, whose 
further details can be found in Refs.~\citep{Muzio:2019leu, Muzio:2021zud}. 
This model explains the observed UHECR spectrum and composition through 
interactions with photons and gas in the source environment. It uses general 
parameters, such as the number of interactions before escape and the photon 
field temperature. Following Ref. \citep{Muzio:2023skc}, we consider two 
UFA-like populations: (i) a baseline 
population driving most of the observed UHECR spectrum and (ii) a population 
accelerating a pure-proton spectrum to $\gtrsim10$ EeV.

In this model, the source evolution function $f(z)$, describing the comoving 
CR power density at redshift $z$ relative to the present time is modelled as  
\begin{align}
    f_{m,z_0}(z) = 
    \begin{cases} 
      (1+z)^m & z \leq z_0, \\
      (1+z_0)^m e^{-(z-z_0)} & z > z_0,
   \end{cases}
\end{align}  
with $-7 \leq m \leq 7$ and $1 \leq z_0 \leq 5$. This simple parametrization 
captures key features of many observed evolutions, including the star 
formation rate (SFR) evolution~\citep{Robertson:2015uda},
\begin{align}
    f_\mathrm{SFR}(z) \propto \frac{(1+z)^{3.26}}{1+[(1+z)/2.59]^{5.68}},
\end{align} 
an active galactic nuclei (AGN) evolution \citep{Stanev:2008un}, 
\begin{align}
    f_\mathrm{AGN}(z) \propto 
    \begin{cases} 
      (1+z)^5 & z \leq 1.7 \\
      (1+1.7)^5 & 1.7 < z \leq 2.7 \\
      (1+1.7)^5 e^{-(z-2.7)} & z > 2.7
   \end{cases},
\end{align}
and a gamma-ray burst (GRB) evolution \citep{Kistler:2007ud,Yuksel:2008cu}, 

\begin{align}
    f_\mathrm{GRB}(z) \propto \frac{(1+z)^{1.5}}{\left[(1+z)^{-34} + \left(\frac{1+z}{5160}\right)^{3} + \left(\frac{1+z}{9}\right)^{35}\right]^{0.1}}.
\end{align}
In this study, we used these parameterized evolution functions as discussed in
the following section.

\section{Numerical Analysis}\label{secV}
In this section, we perform a detailed numerical analysis of the density 
enhancement factor $\xi$, flux computation and its fitting with chi-square 
($\chi^2$) analysis, flux suppression effects and its fitting with 
Levenberg-Marquardt algorithm \citep{Levenberg_1944, Marquardt_1963}, anisotropy studies, and the visualization of flux and anisotropy through HEALPix-based 
skymaps \citep{Gorski_2005}. In all calculations, we adopt a uniform 
turbulent extragalactic magnetic field with strength $B = 1$~nG.

\subsection{Density Enhancement}
\begin{figure}[!h]
\centering
\includegraphics[width=0.9\linewidth]{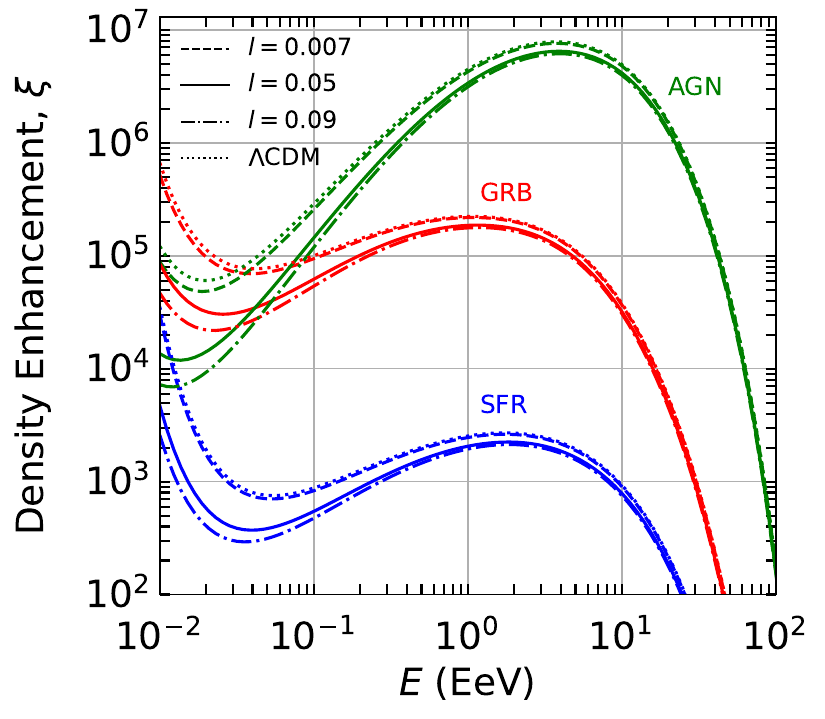}
\vspace{-0.3cm}
\caption{Density enhancement factor $ \xi $ as a function of CRs' energy for 
different source distributions in Bumblebee gravity. The parameter $ l $ is 
varied as $ l = 0.007 $ (dashed lines), $ l = 0.05 $ (solid lines), and 
$ l = 0.09 $ (dash-dotted lines), while the $ \Lambda $CDM model (dotted lines)
serves as a reference.}
\label{fig5a}
\end{figure}

Fig.~\ref{fig5a} illustrates the effect of the Bumblebee gravity parameter 
$ l $ on the density enhancement factor $ \xi $ for different distributions of 
CRs' sources. Here, three values of $ l $ are considered, viz., $ l = 0.007 $ 
(dashed lines), $ l = 0.05 $ (solid lines), and $ l = 0.09 $ (dash-dotted 
lines), while the $\Lambda$CDM model case (dotted lines) serves as a 
reference. As $ l $ increases, the density enhancement systematically rises 
across all source types, indicating that stronger modifications to gravity 
amplify CRs' densities. The effect is more pronounced at lower energies, where 
the enhancement increases significantly with $ l $. The deviation from the
$\Lambda$CDM model becomes more significant as $ l $ increases, reinforcing 
the idea that the modified gravity enhances CRs' densities across a broad 
energy range. The enhancement factor $\xi$ from Bumblebee gravity is 
lower than that from the $\Lambda$CDM model, which implies that the effective 
density enhancement due to diffusive propagation is somewhat suppressed in 
Bumblebee gravity. The primary reason for this lies in the modified 
cosmological background, particularly the $l$-dependent Hubble expansion 
rate $H(z)$, which alters the energy losses and co-moving distances. Since 
the pattern is similar across different values of $l$, we have adopt the 
$l=0.05$ throughout the rest of this study.
\begin{figure*}
     \centerline{
    \includegraphics[scale=0.44]{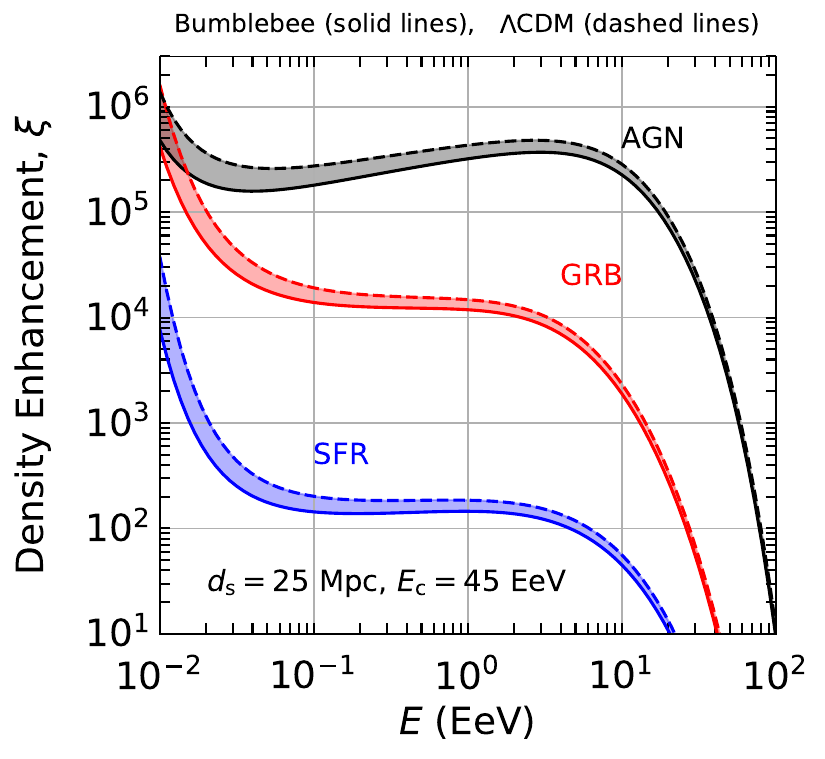}
    \includegraphics[scale=0.44]{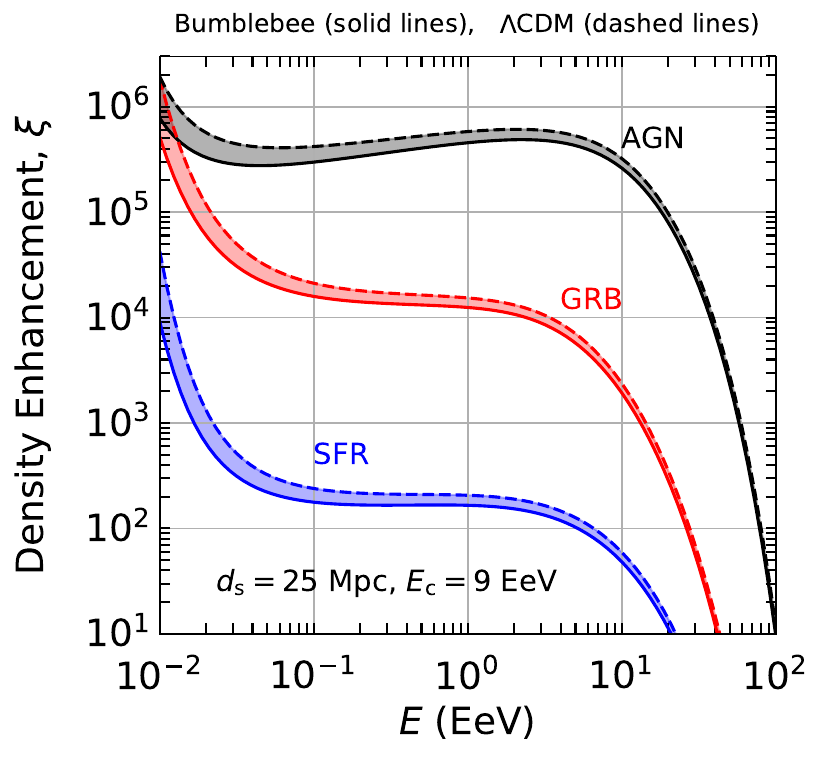}
    \includegraphics[scale=0.44]{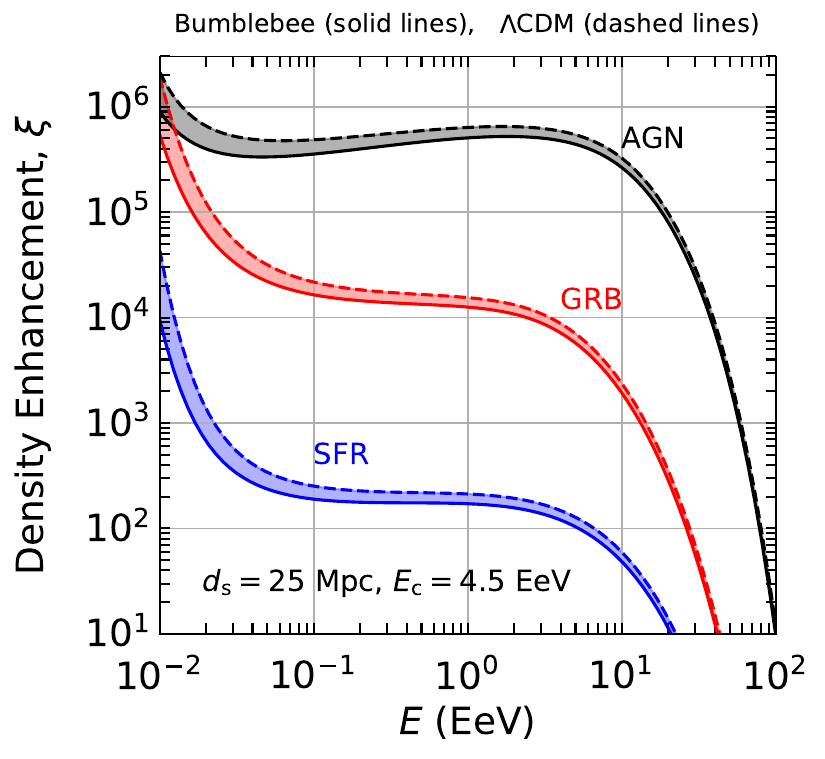}}
   
    \centerline{
   \includegraphics[scale=0.44]{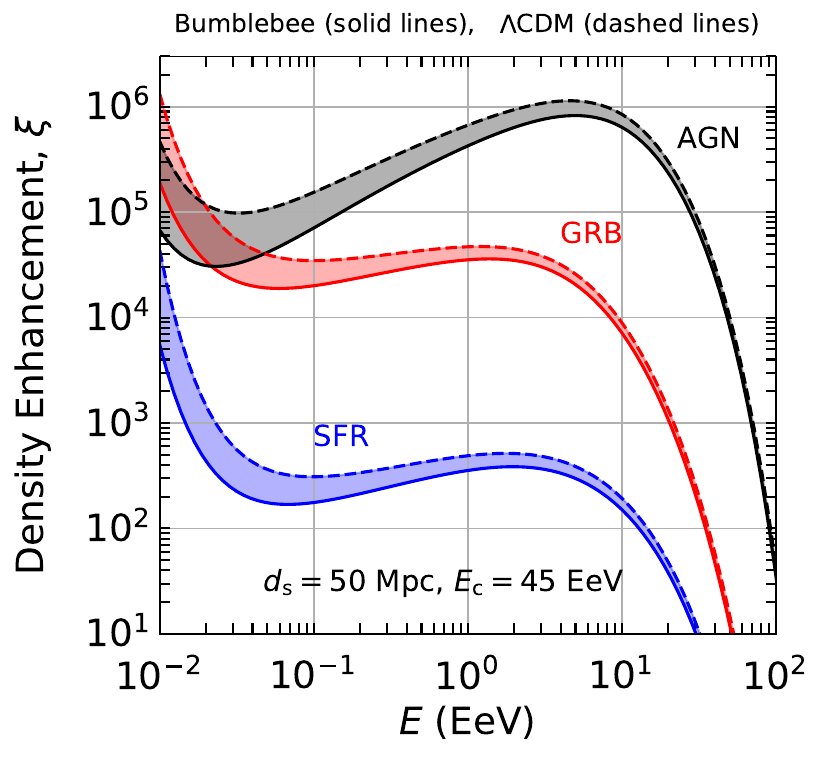}
   \includegraphics[scale=0.44]{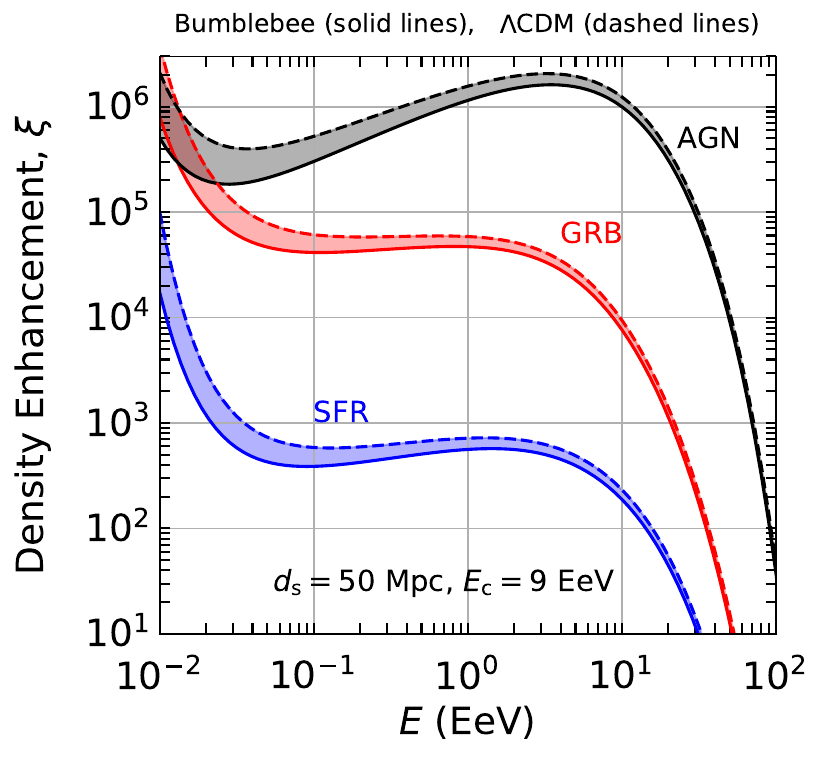}
   \includegraphics[scale=0.44]{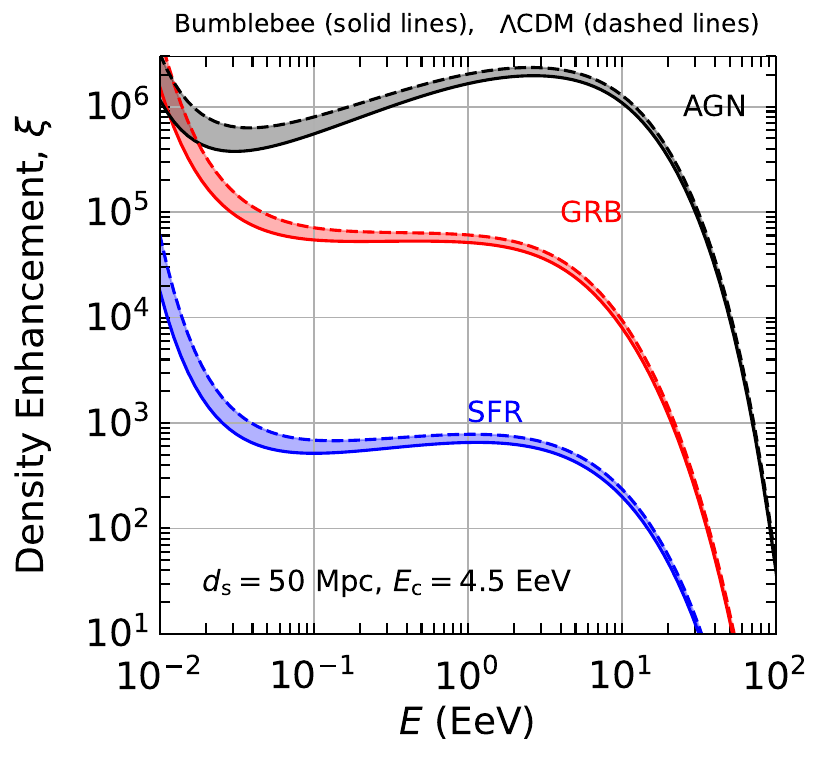}}
  
  \centerline{
  \includegraphics[scale=0.44]{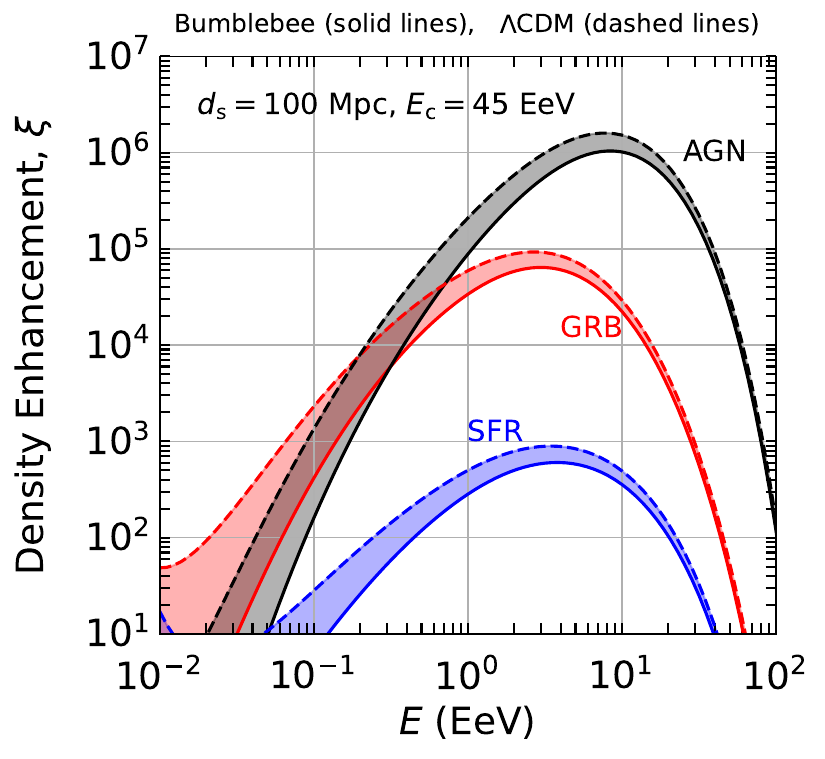}
  \includegraphics[scale=0.44]{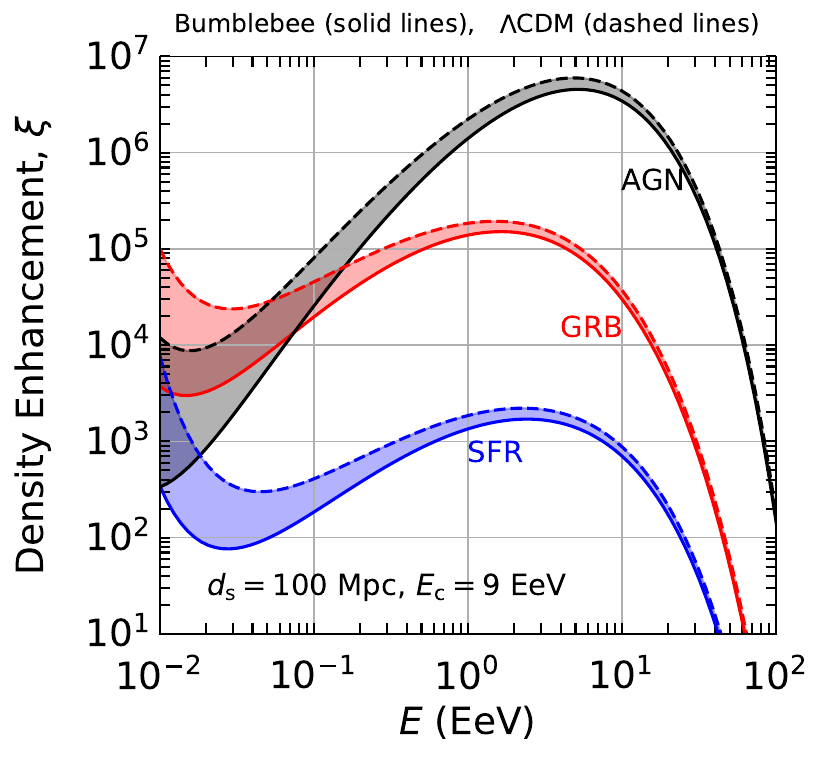}
  \includegraphics[scale=0.44]{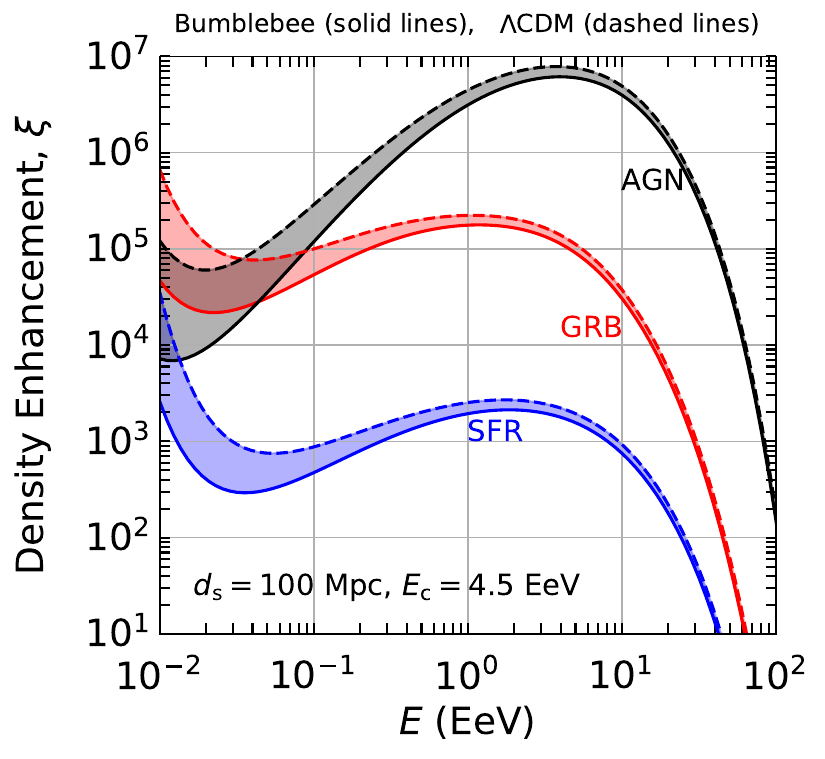}}
\caption{Density enhancement factor $ \xi $ for CRs as a function of energy, 
varying the critical energy $ E_\text{c} $ and source distance $d_\text{s}$. 
The three panels correspond to $ E_\text{c} = 45 $ EeV (left), 
$ E_\text{c} = 9 $ EeV (middle), and $ E_\text{c} = 4.5 $ EeV (right), each 
analyzed for source distances $ d_\text{s} = 25 $ Mpc (top), 
$ d_\text{s} = 50 $ Mpc (middle), and $ d_\text{s} = 100 $ Mpc (bottom). Solid 
lines represent Bumblebee gravity, while dashed lines correspond to the 
$\Lambda$CDM model case.}
\label{fig5b}
\end{figure*}

In Fig.~\ref{fig5b}, we analyze the density enhancement $\xi$ of CRs with
respect to their primary energy $E$ while varying their critical energy 
$E_\text{c}$ with $E_\text{c} = 45$ EeV (left), $E_\text{c} = 9$ EeV (middle), 
and $E_\text{c} = 4.5$ EeV (right) across three source distances: 
$d_\text{s} = 25$ Mpc, $d_\text{s} = 50$ Mpc, and 
$d_\text{s} = 100$ Mpc. In all of the plots in this figure, the AGN (black) 
sources exhibit the highest enhancement, followed by GRB (red) and SFR 
(blue). As the source separation distance $d_\text{s}$ increases, differences 
in density enhancement for different $E_\text{c}$s become more pronounced due 
to extended propagation effects, including energy losses, magnetic diffusion, 
and cosmic expansion. For example, in the case of $d_\text{s} = 25$ Mpc, the 
overall variation of enhancement among different $E_\text{c}$ values is 
relatively very small due to the shorter propagation distance. At 
$d_\text{s} = 50$ Mpc, the effects of CR propagation across different 
$E_\text{c}$ become more significant. This effect is highly visible in 
the case of $d_\text{s}= 100$ Mpc distance i.e.~the highest value of distance 
we have considered in this work. Again, for higher $E_\text{c}$ values, the 
enhancement is low if we compare that with a lower value as we can see the 
transitions from $E_\text{c}=45$ to $4.5$ EeV throughout the plots.
Also, by increasing the $d_\text{s}$ values, we get lower $\xi$ in the 
low-energy regime while a comparatively higher $\xi$ in the high-energy 
regime before the suppression region. 
%
Again, the difference between the Bumblebee gravity model (solid lines) and 
the $\Lambda$CDM model (dashed lines) completely depends on the $d_\text{s}$ 
parameter. It is seen that as the value of $d_\text{s}$ increases,
the cosmological effect becomes more pronounced as compared to the lower 
separation distance.
\subsection{Flux}
\begin{figure}
\centering
\includegraphics[width=\linewidth]{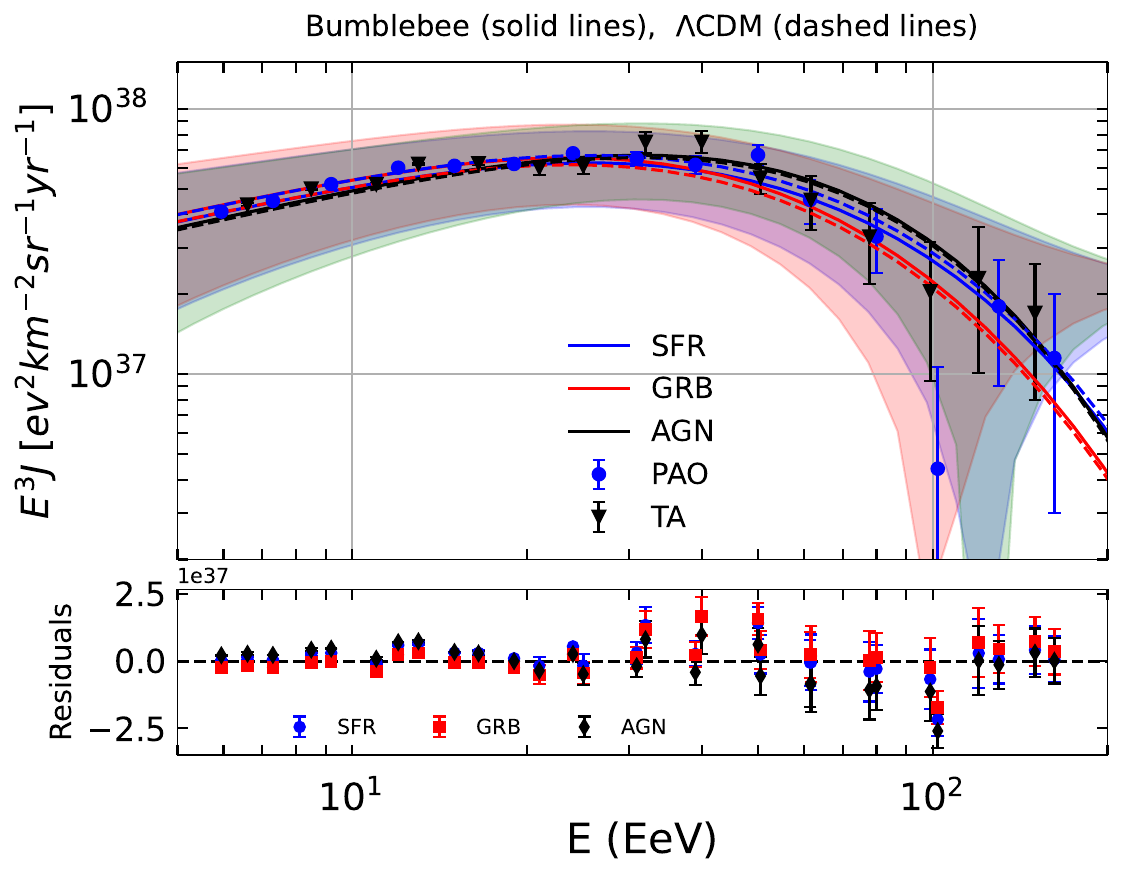}
\vspace{-0.3cm}
\caption{ UHECRs flux as a function of energy $E$ for different astrophysical 
source models: SFR (blue), GRB (red), and AGN (black) as predicted by the 
Bumblebee gravity (solid lines) and the $\Lambda$CDM model (dashed lines) 
for the pure proton composition. The observational data are taken from 
the PAO and the TA experiment \citep{augerprd2020, ta2019}.}
\label{fig5c}
\end{figure}

In Fig.~\ref{fig5c}, we present the UHECRs flux as a function of 
energy $E$ for different astrophysical source models: SFR (blue), GRB (red), 
and AGN (black) for the pure proton composition. The analysis is performed for both the Bumblebee gravity (solid lines) and the $\Lambda$CDM model (dashed 
lines).  Unless otherwise stated, we assume a spectral index 
$\gamma = 2$ and a maximum rigidity cutoff $R_{\max} = 50$~EV for all source 
models. The predicted fluxes are compared 
with observational data from the PAO (blue circles) and the TA experiment 
(black triangles), which shows a good agreement with the fitted models. Since 
the PAO and the TA experiment have different spectra, the energy scaling of 
data is adopted from Ref.~\citep{bergmanepj}. For the fitting, we have 
used different source separation distances ($d_\text{s}$) for each model. In 
Bumblebee gravity, we set $d_\text{s} = 74$ Mpc for SFR, $d_\text{s} = 8.2$ 
Mpc for GRB, and $d_\text{s} = 1.4$ Mpc for AGN. In the $\Lambda$CDM model, the 
corresponding values are $d_\text{s} = 69$ Mpc, $d_\text{s} = 7.2$ Mpc, 
and $d_\text{s} = 1.25$ Mpc, respectively. The different calculated flux 
spectra follow the expected trend, with a gradual rise at lower 
energies, a flattening in the intermediate range, and a sharp suppression at 
the highest energies due to the GZK effect. The Bumblebee gravity model shows 
slight deviations from the $\Lambda$CDM model, particularly at higher 
energies, where modified gravity effects may play a role in altering 
propagation characteristics. The residual plot in the lower panel shows the 
difference between the predicted flux and observed data. The residuals remain 
centered around zero, indicating that both models provide a reasonable fit 
to the data. The small fluctuations observed at higher energies suggest 
potential statistical uncertainties or minor deviations due to source modeling 
assumptions. Overall, the results indicate that both the Bumblebee gravity 
model and the $\Lambda$CDM model provide a consistent description of UHECRs 
flux, with source separation distances playing a crucial role in shaping the 
observed spectrum. We provide the $\chi^2$ test for both cosmological models 
with PAO and TA data, and it is defined as
\begin{equation}
    \chi^2 = \sum_{i} \frac{(\text{J}^i_\text{th}-\text{J}^i_\text{obs})^2}{\sigma_i^2},
\end{equation}
where $\text{J}^i_\text{th}$ are the theoretical values of flux obtained 
from our calculations and $\text{J}^i_\text{obs}$ are the observed flux 
values, which are obtained from the PAO and the TA experiment.
$\sigma$ denotes the total uncertainty in each energy bin, 
incorporating both statistical and systematic errors.
The corresponding $\chi^2$ and $\chi^2_\text{red}$ values are given in 
Table \ref{tab:chi2_values}.
\begin{table*}
    \centering
        \caption{Chi-square and reduced chi-square values of the fittings of 
the predicted UHECRs fluxes to observational data of the Pierre Auger 
Observatory (PAO) and the Telescope Array (TA) experiment for different sets
of considered models.}
\vspace{3pt}
    \begin{tabular}{l @{\hspace{0.7cm}} c @{\hspace{0.7cm}} c @{\hspace{0.7cm}} c @{\hspace{0.7cm}} c @{\hspace{0.7cm}} c @{\hspace{0.7cm}} c}
        \toprule
        \textbf{Dataset} & \multicolumn{3}{c}{\textbf{Bumblebee Gravity Model}} & \multicolumn{3}{c}{\textbf{$\Lambda$CDM Model}} \\
        \cmidrule(lr){2-4} \cmidrule(lr){5-7}
        & \textbf{SFR} & \textbf{GRB} & \textbf{AGN} & \textbf{SFR} & \textbf{GRB} & \textbf{AGN} \\
        \midrule
        \multicolumn{7}{c}{\textbf{Chi-square ($\chi^2$)}} \\
        \midrule
        AUGER & 18.98 & 15.61 & 21.67 & 18.09 & 20.46 & 20.08 \\
        TA & 21.76 & 13.39 & 18.54 & 11.87 & 27.70 & 26.76 \\
        AUGER + TA & 40.75 & 28.99 & 40.21 & 29.96 & 48.17 & 46.84 \\
        \midrule
        \multicolumn{7}{c}{\textbf{Reduced chi-square ($\chi^2_{\text{red}}$)}} \\
        \midrule
        AUGER & 3.16 & 2.60 & 3.61 & 2.58 & 2.92 & 2.86 \\
        TA & 2.42 & 1.49 & 2.06 & 1.19 & 2.77 & 2.68 \\
        AUGER + TA & 2.40 & 1.71 & 2.37 & 1.66 & 2.67 & 2.60 \\
        \bottomrule
    \end{tabular}
    \label{tab:chi2_values}
\end{table*}

To model CRs flux distributions, we employ the HEALPix \citep{Gorski_2005} 
framework with a resolution of $N_{\text{side}} = 256$, corresponding to 
$N_{\text{pix}} = 12 \times N_{\text{side}}^2$ pixels covering the full sky. 
The source positions are distributed in right ascension (RA) and declination 
(Dec) using a uniform distribution consideration, where RA ranges from 
$0^\circ$ to $360^\circ$ and Dec from $-90^\circ$ to $90^\circ$. These 
coordinates are converted into HEALPix angular coordinates $(\theta, \phi)$ 
using the transformations: $\theta = 90^\circ - \text{Dec}$ and 
$\phi = \text{RA}$. The corresponding HEALPix pixel indices are then 
determined using the \texttt{healpy.ang2pix} function. To construct the flux 
maps, we assign predefined flux values to the HEALPix pixels and sum the 
contributions from all sources falling into each pixel. To ensure spatial 
smoothness, we apply a Gaussian smoothing filter with an angular scale of 
$2^\circ$ using the \texttt{healpy.smoothing} function. The resulting maps 
are normalized using a logarithmic stretch to enhance contrast, replacing 
zero-valued pixels with a minimum nonzero value to avoid numerical artifacts. 
For visualization, we employ Mollweide projections \citep{Gorski_2005} with a 
logarithmic color scale, allowing for a clear representation of flux 
variations across 
the sky. The longitude and latitude markers are added to aid interpretation, 
ensuring that the maps provide an accurate depiction of the modeled CR flux 
distributions.
\begin{figure*}
    \centerline{
    \includegraphics[scale=0.26]{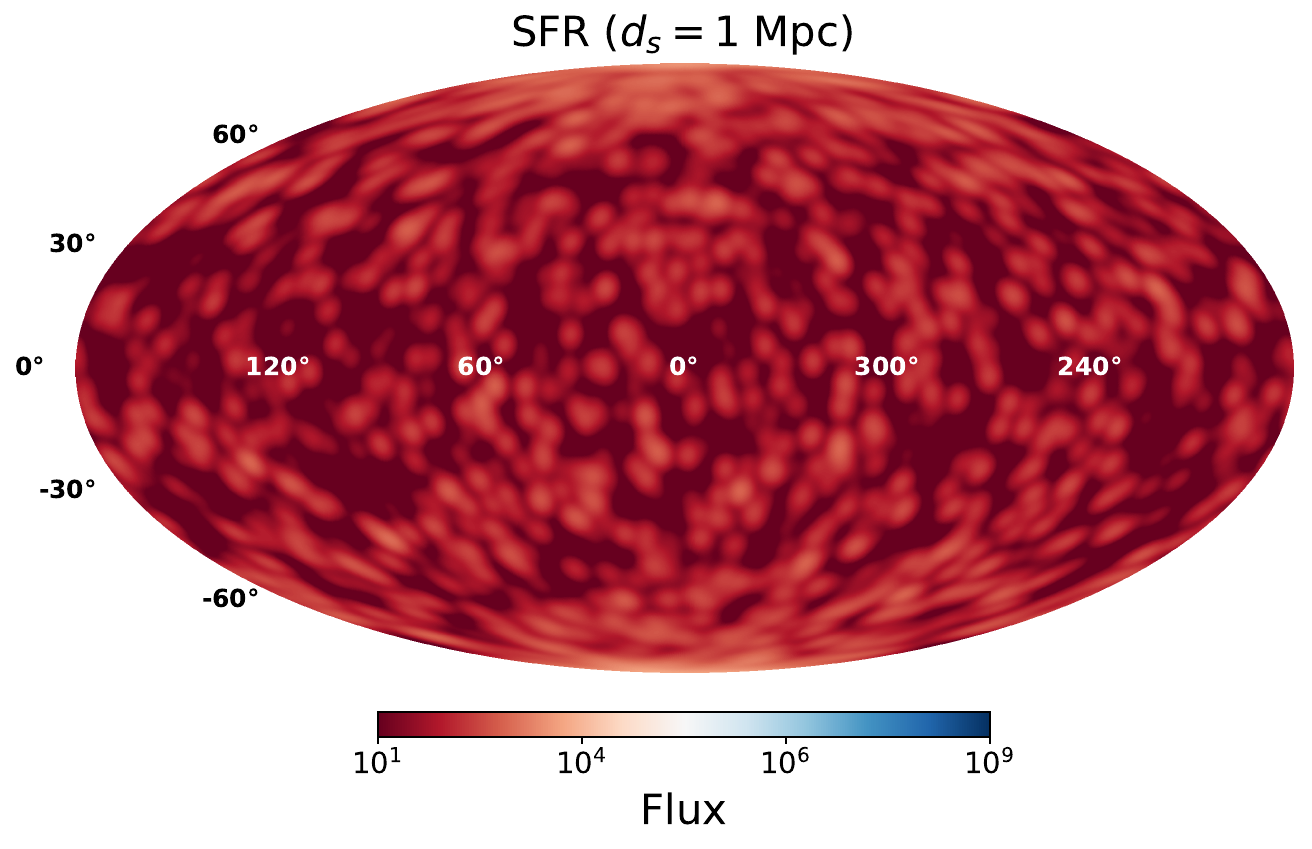}
   \includegraphics[scale=0.26]{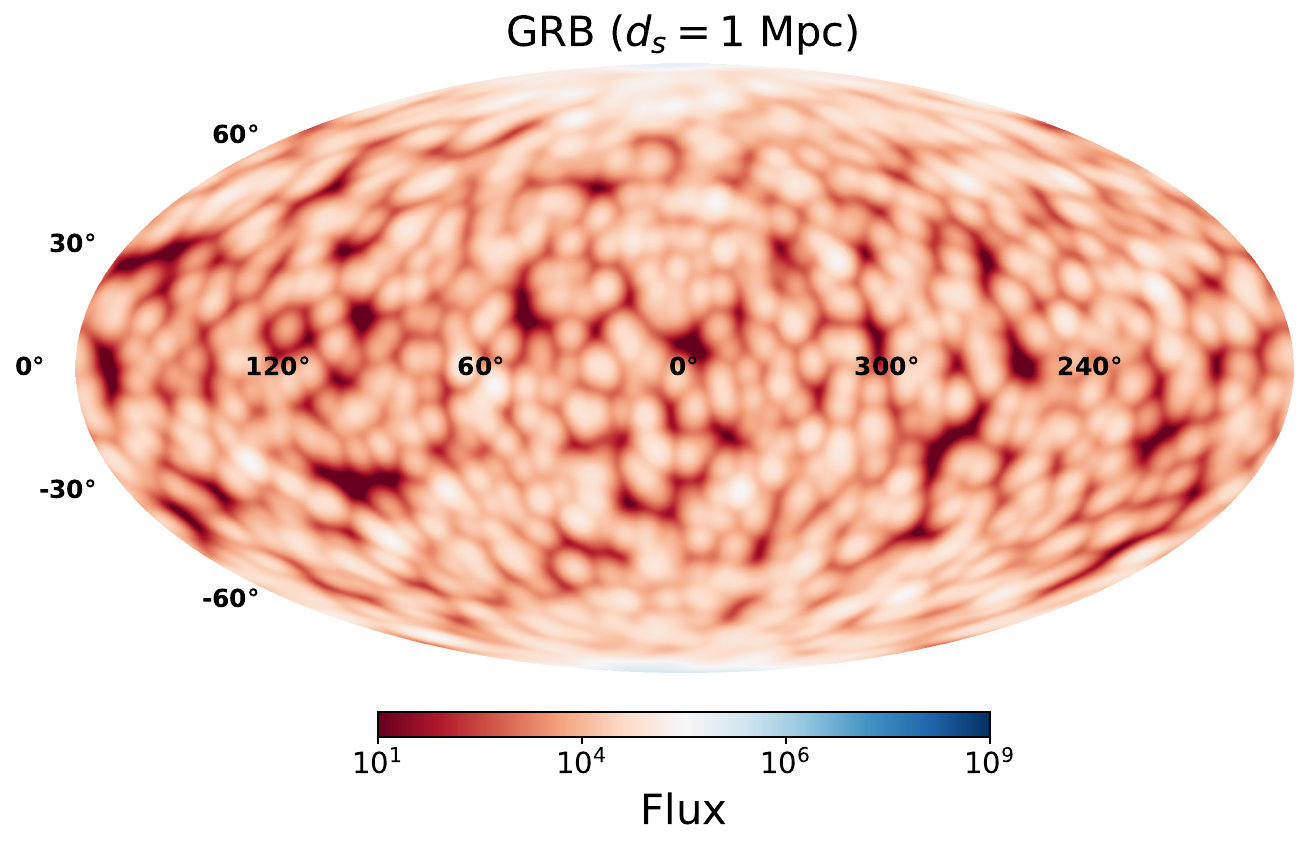}
   \includegraphics[scale=0.26]{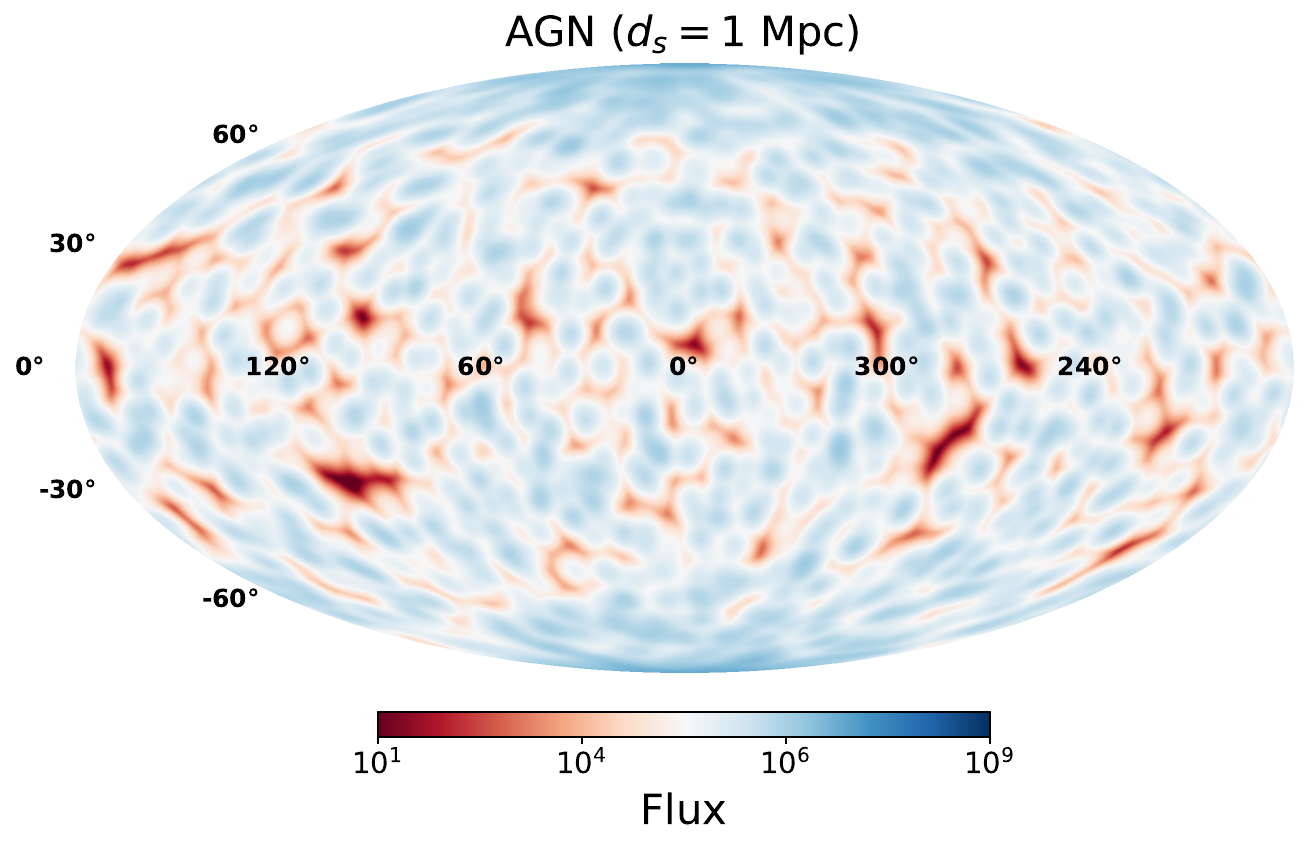}}
   
    \centerline{
   \includegraphics[scale=0.26]{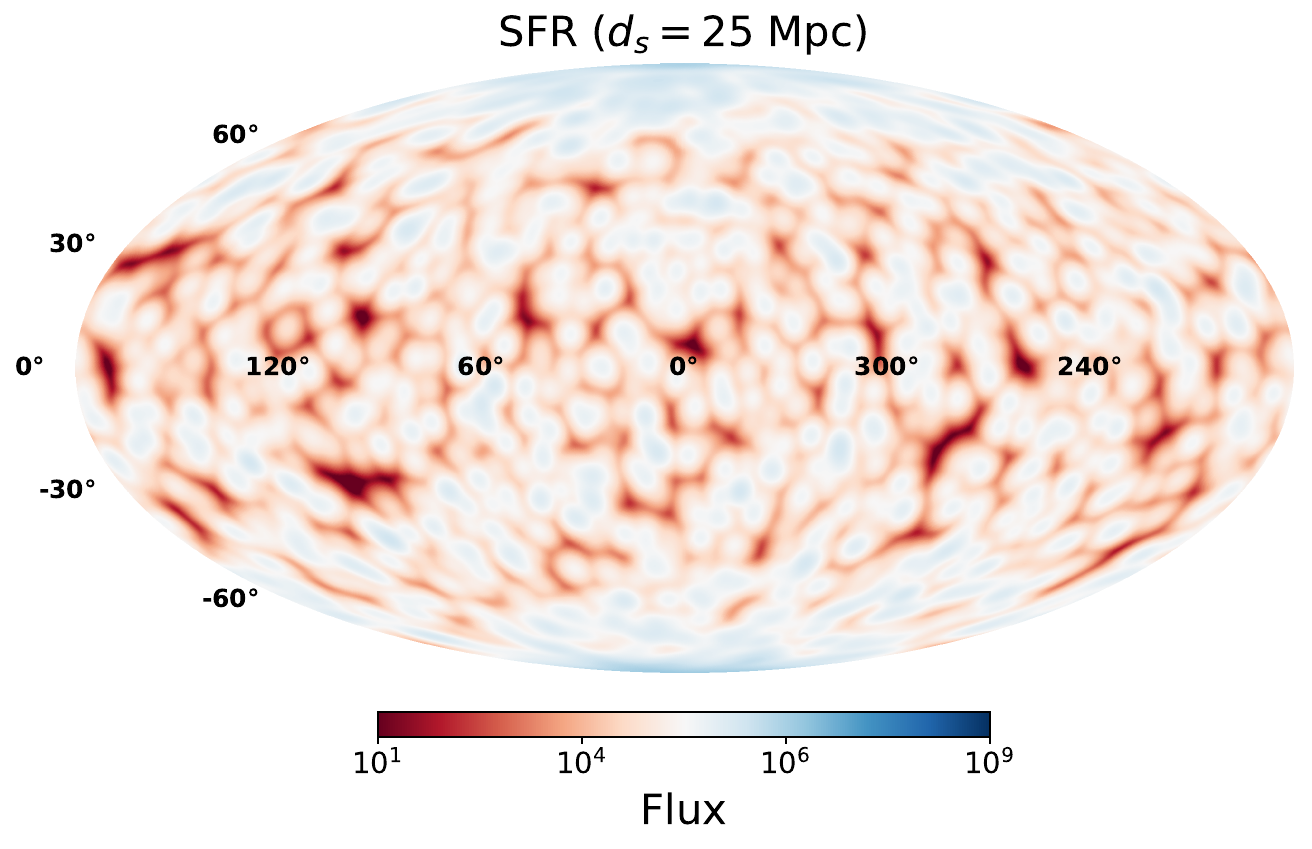}
  \includegraphics[scale=0.26]{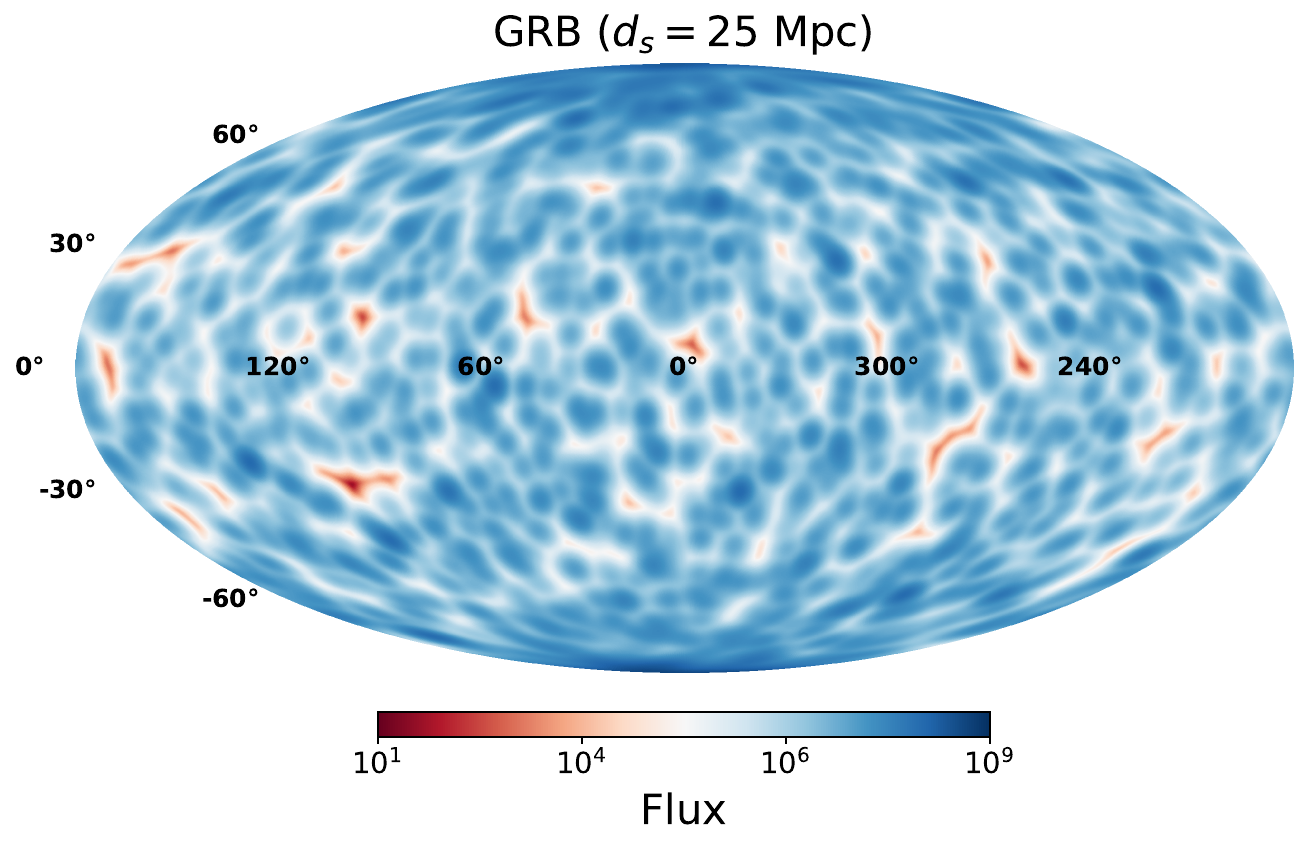}
  \includegraphics[scale=0.26]{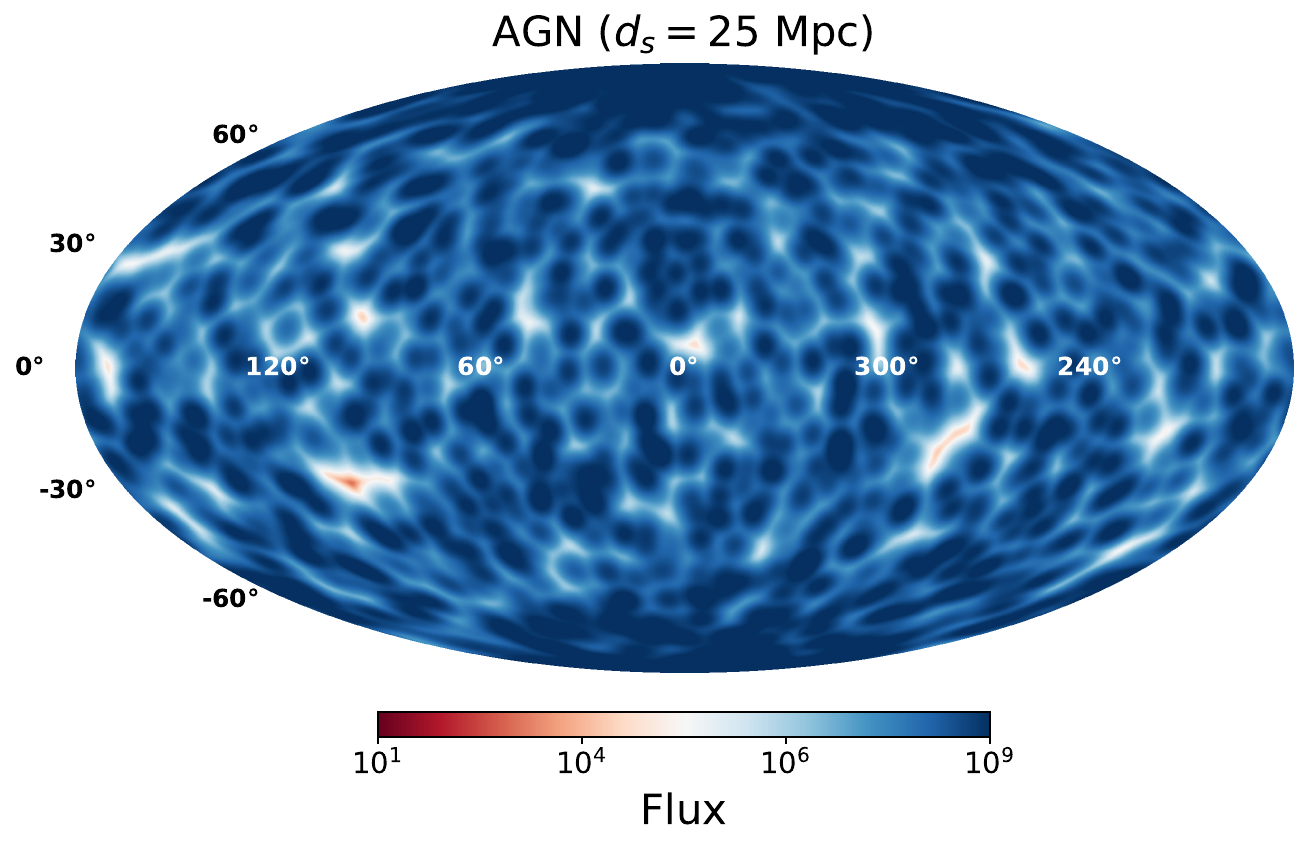}}
  
  \centerline{
  \includegraphics[scale=0.26]{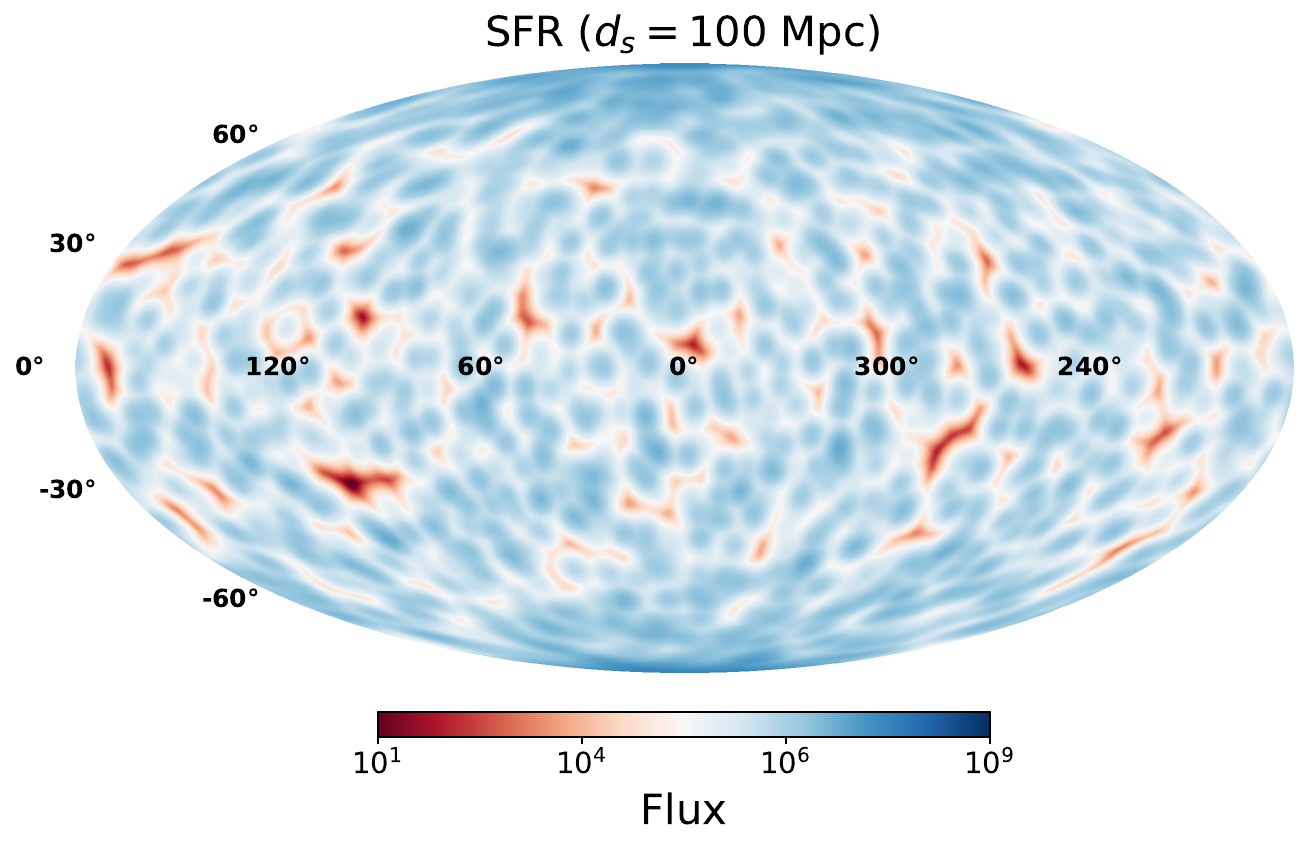}
   \includegraphics[scale=0.26]{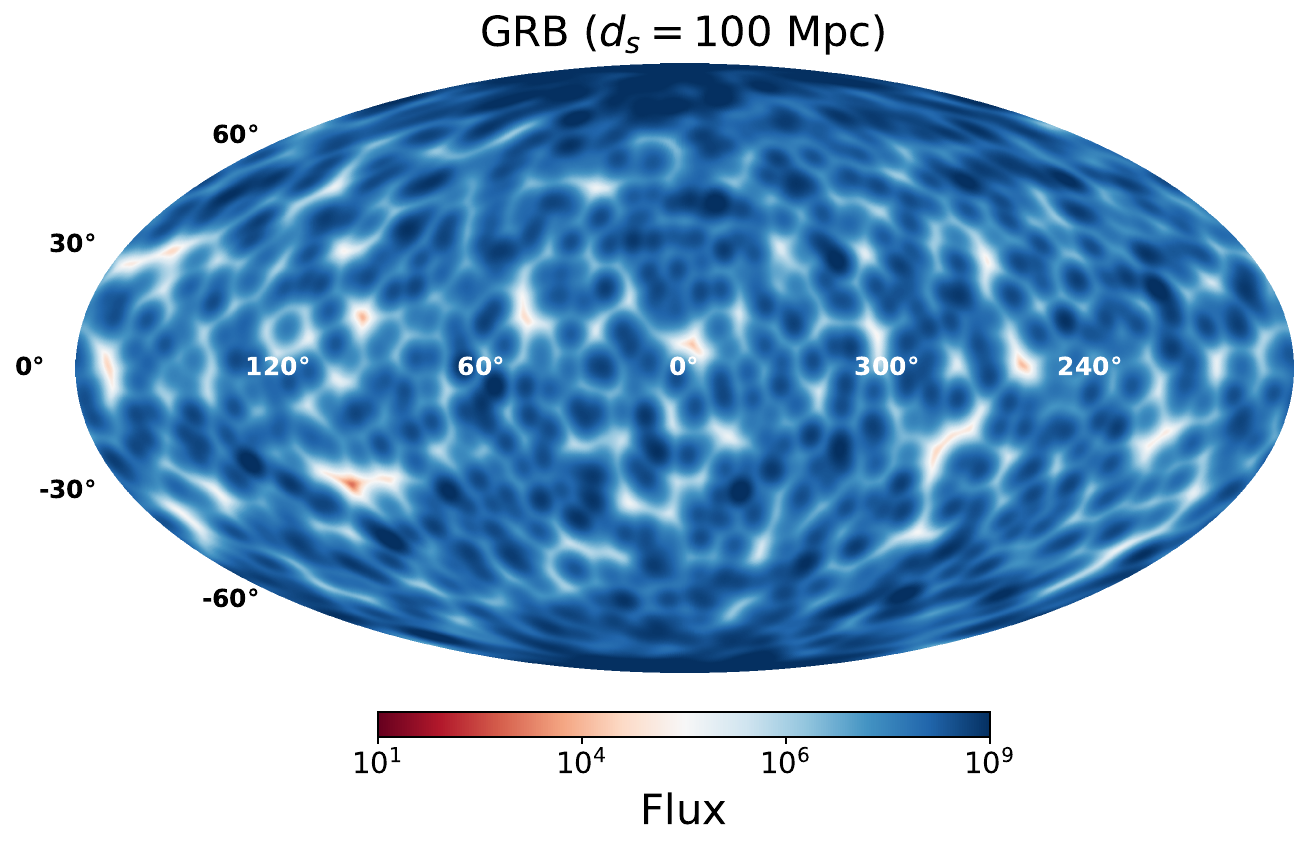}
  \includegraphics[scale=0.26]{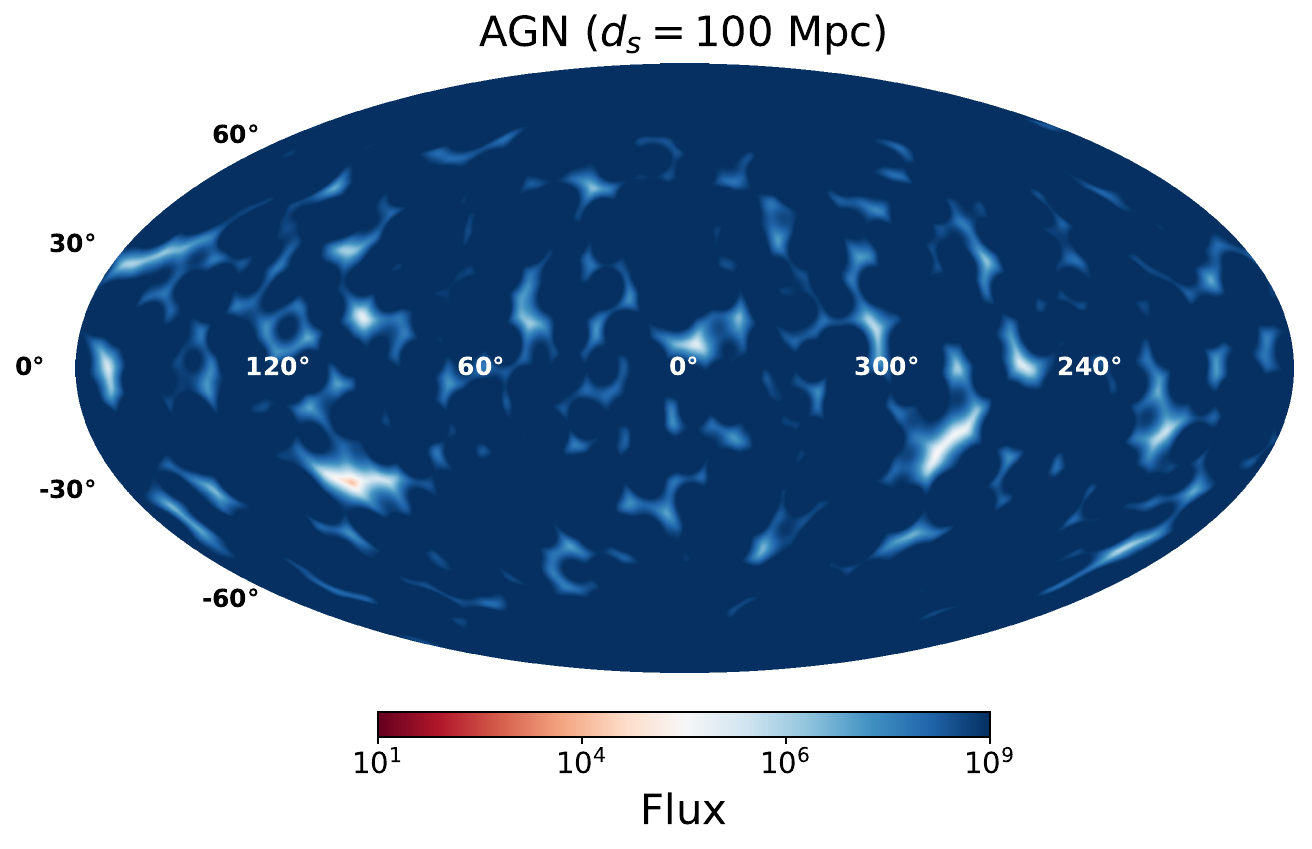}}
	\caption{Skymaps of CRs flux distributions for different astrophysical 
source models at varying source separation distances $ d_\text{s} = 1 $, $25$, 
and $ 100 $ Mpc. Each set of panels corresponds to sources following the SFR, 
GRB, and AGN distributions. The flux is shown on a logarithmic color scale, 
with blue indicating higher flux values and red indicating lower flux values.}
	\label{fig5d}
\end{figure*}

The skymaps in Fig.~\ref{fig5d} illustrate the CRs flux distributions
for uniform source distribution models of different astrophysical 
sources across HEALPix pixels, with random sampling of (RA, Dec) coordinates 
converted into HEALPix $(\theta, \phi)$ angles at varying source separation 
distances, 
specifically $d_\text{s} = 1$, $25$, and $100$ Mpc for the Bumblebee gravity 
scenario. Each set of panels corresponds to sources following the SFR, GRB, 
and AGN, respectively. The flux is represented using a logarithmic color 
scale, where blue regions indicate higher flux values and red regions indicate 
lower flux values. The differences in flux distributions arise from the 
spatial distribution of the sources, with AGNs being more localized and 
strongly clustered compared to the more evenly distributed SFR and GRB 
sources. For $d_\text{s} = 1$ Mpc, the SFR and GRB models exhibit relatively 
smooth flux distributions due to the more uniform nature of their respective 
sources. In contrast, the AGN model displays a more structured and 
inhomogeneous flux pattern. As the separation distance between sources 
increases to $d_\text{s} = 25$ Mpc and then $100$ Mpc, the flux distributions 
become more discrete. The SFR model begins to show enhanced contrast due to 
the increased source separation. The GRB model exhibits a more distinct 
structure, while the AGN model displays an even more pronounced clustering 
effect, with strong flux variations across the sky. Thus the increased source 
separation leads to a more discrete flux distribution, making individual 
source contributions more apparent. This behavior is expected, as increasing 
the source separation reduces the overall number density of sources within the 
observable volume, leading to a more structured and discrete flux 
distribution. For some individual energy ranges, the flux maps are shown in 
Appendix~\ref{appnA}.

\begin{figure*}
\centering
\includegraphics[scale=0.7]{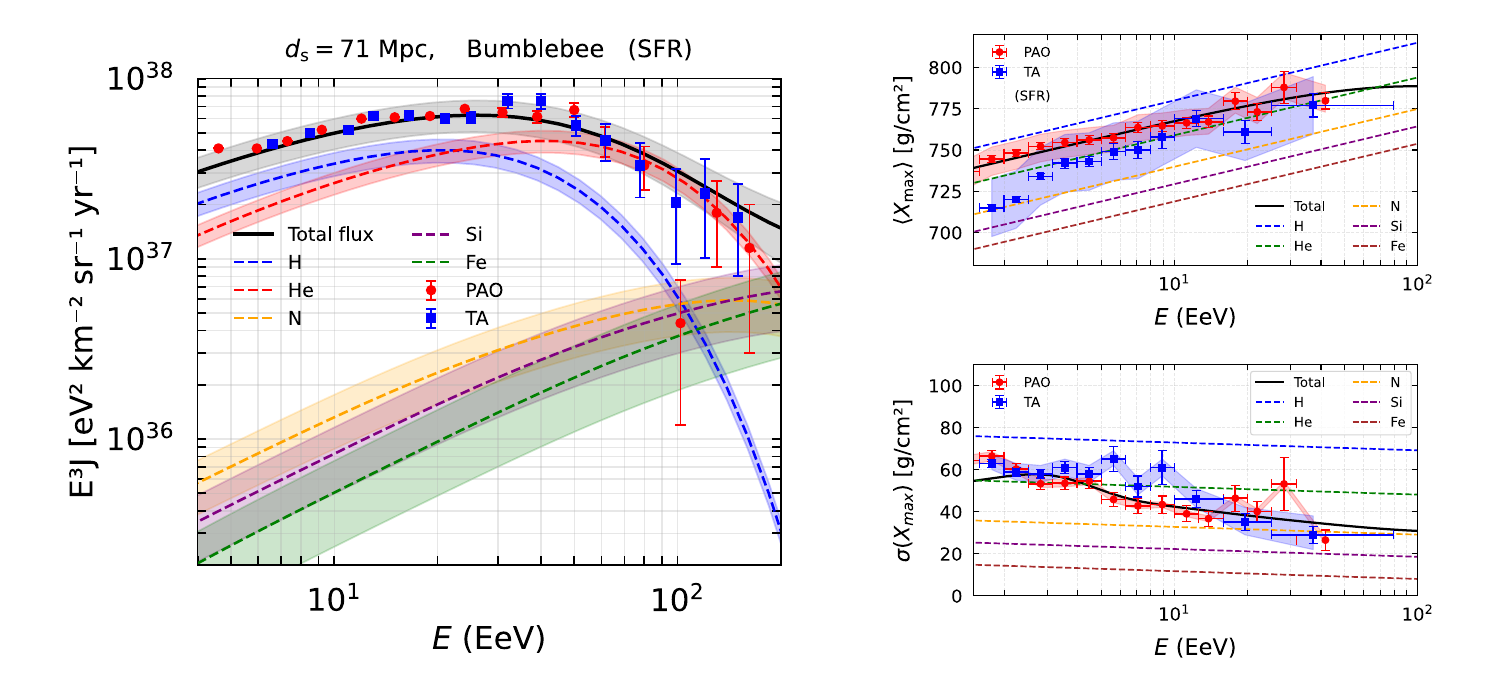}\\
\includegraphics[scale=0.7]{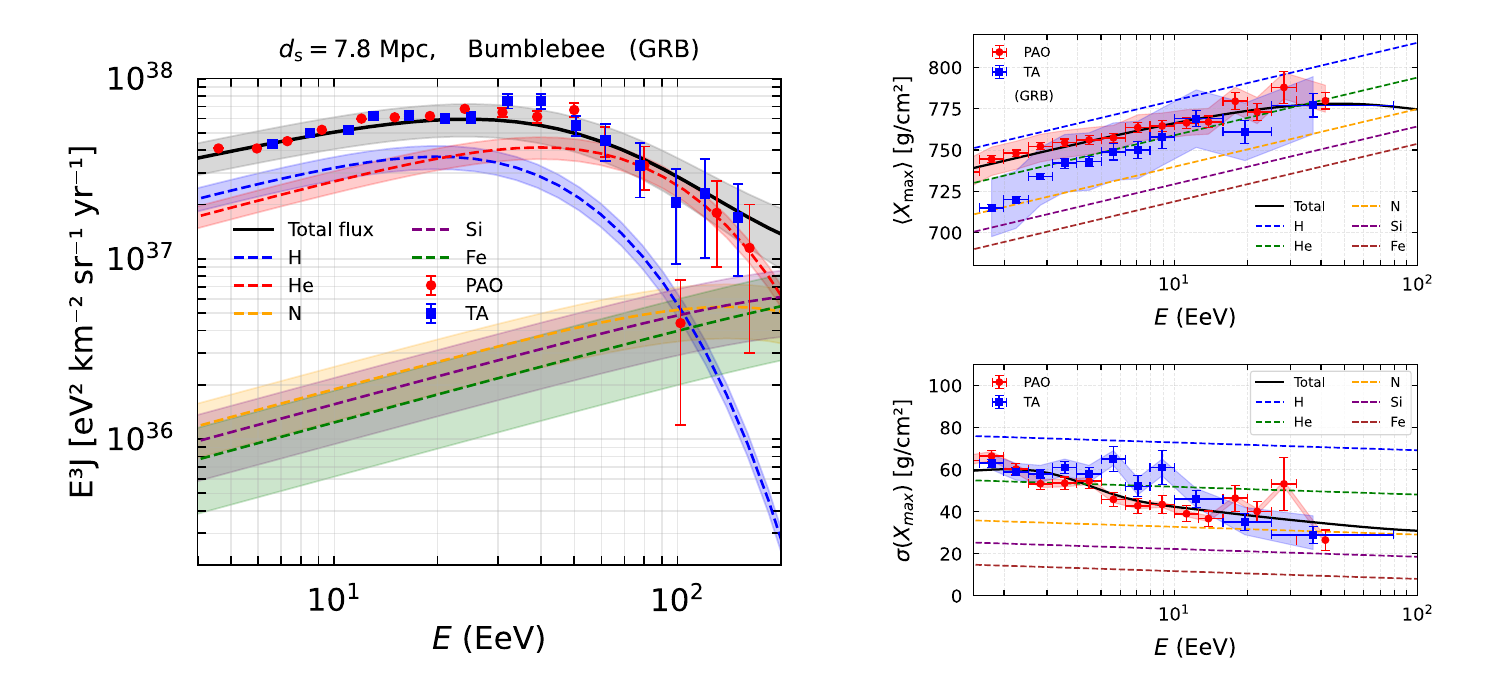}\\
\includegraphics[scale=0.7]{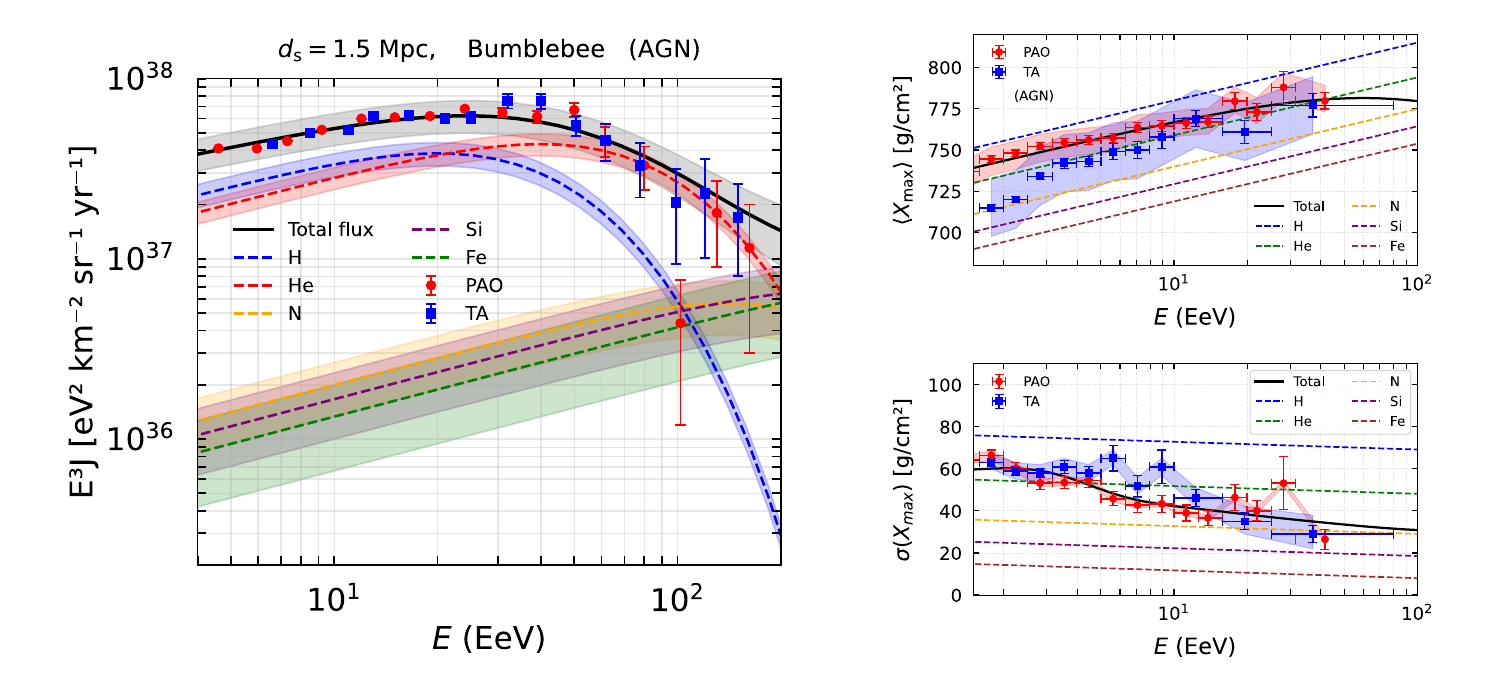}
\vspace{-0.4cm}
\caption{UHECRs fluxes for mixed compositions as predicted by
the Bumblebee gravity model for different source separation distances 
$d_\text{s}$ in comparison with the PAO \citep{augerprd2020} and 
TA \citep{ta2019} data.with corresponding 
$\langle X_{\text{max}} \rangle$ and $\sigma \left( X_\text{max} \right )$ results that are compared with the PAO \citep{xmax_pao} and TA \citep{xmax_ta} $\langle X_{\text{max}}\rangle $ data. }
\label{mixed1}
\end{figure*}

\begin{figure*}
\centering
\includegraphics[scale=0.7]{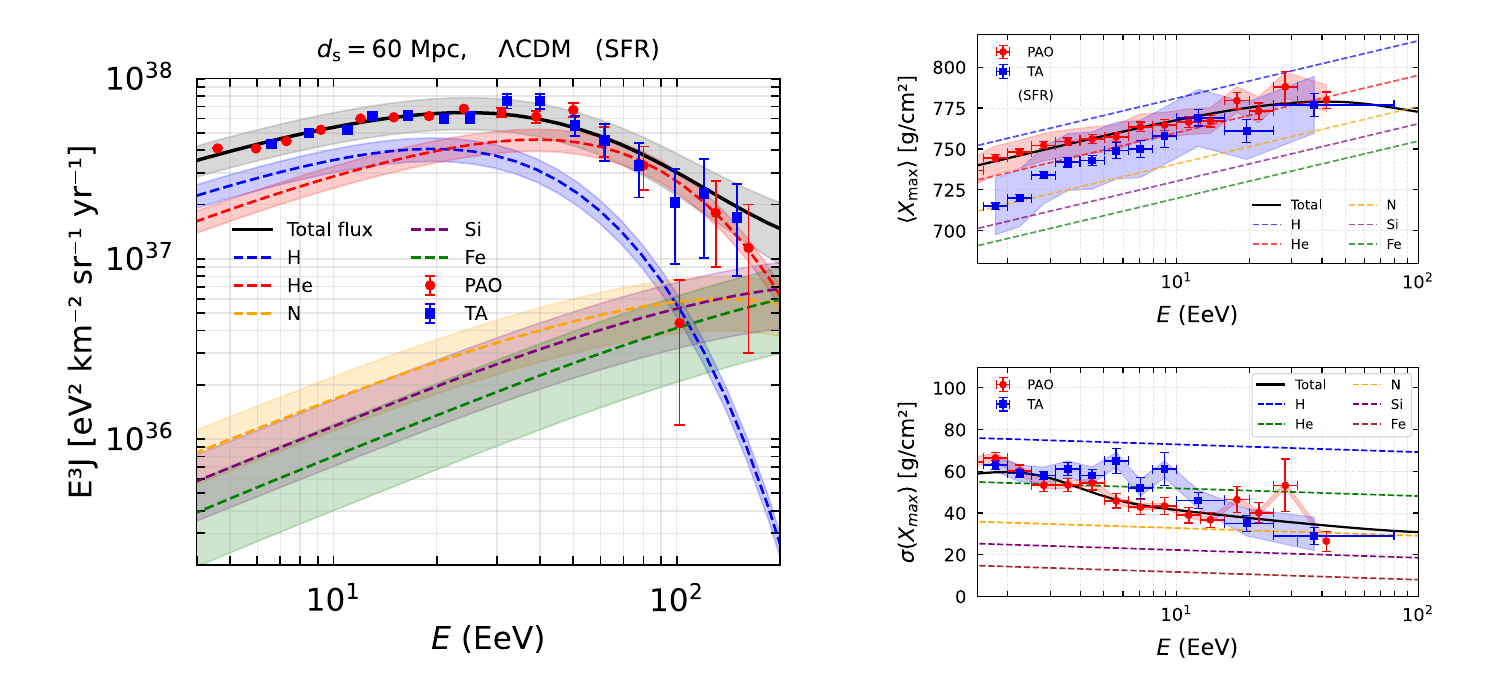}\\
\includegraphics[scale=0.7]{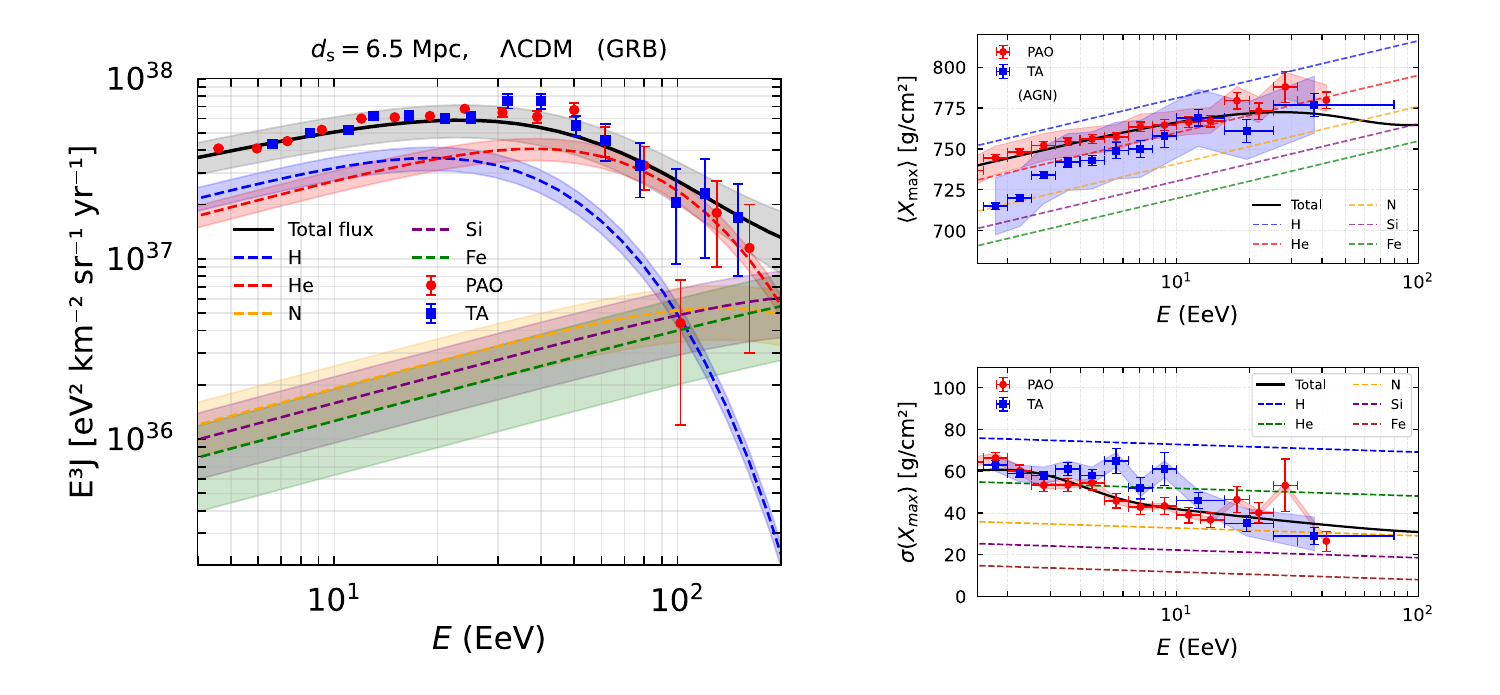}\\
\includegraphics[scale=0.7]{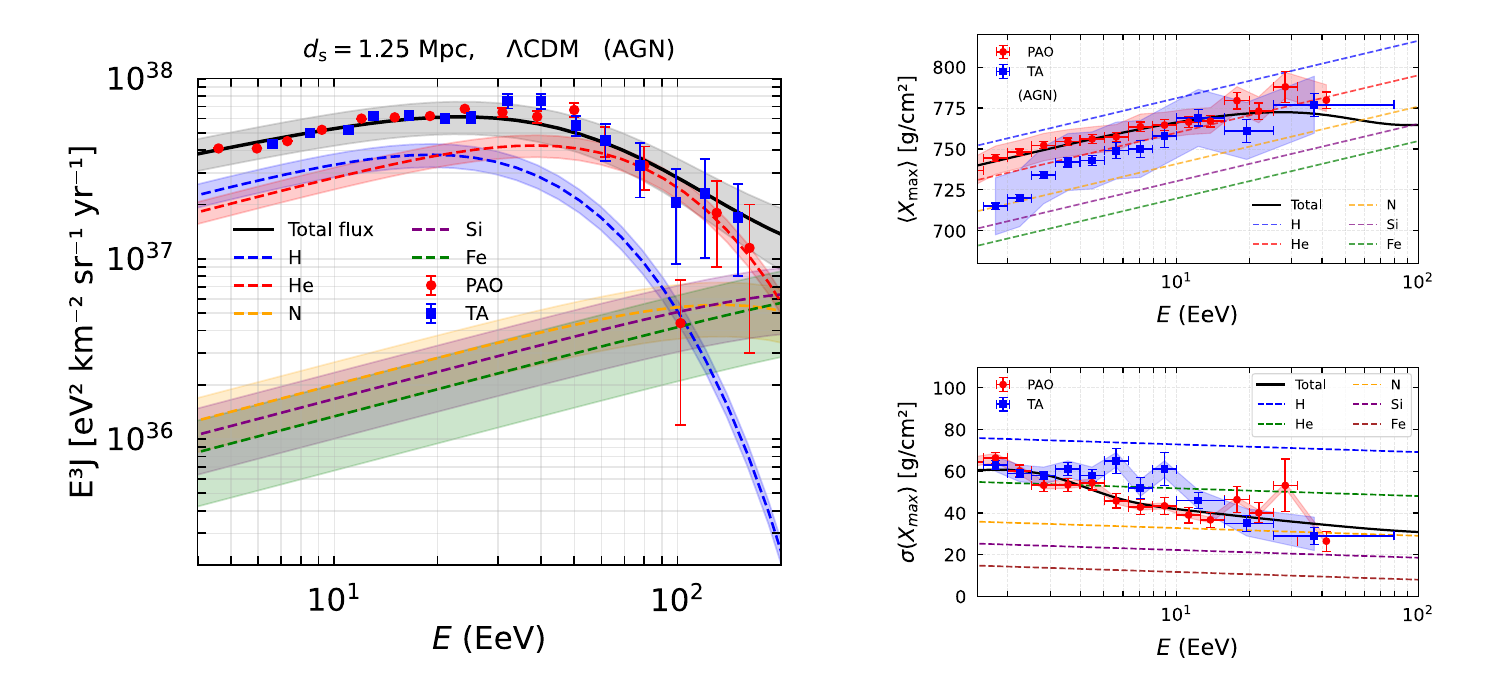}
\vspace{-0.4cm}
\caption{UHECRs fluxes for mixed compositions as predicted by 
the $\Lambda$CDM model for different source separation distances
$d_\text{s}$ in comparison with the PAO \citep{augerprd2020} and 
TA \citep{ta2019} data with corresponding 
$\langle X_{\text{max}} \rangle$ and $\sigma \left( X_\text{max} \right )$ results that are compared with the PAO 
\citep{xmax_pao} and TA \citep{xmax_ta} $\langle X_{\text{max}}\rangle $ data. }
\label{mixed2}
\end{figure*}

For the mixed composition case, we plot the Fig.~\ref{mixed1} for the fluxes 
along with the mean depths of shower maximum $\langle X_\text{max}\rangle$ as 
predicted by the Bumblebee gravity at different source distances and the 
Fig.~\ref{mixed2} for the same as predicted by the $\Lambda$CDM model for the 
reference. We take the contribution of nuclear abundance of 
p, He, N, Si and Fe as 0.4, 0.3, 0.15, 0.1 and 0.05 respectively \citep{mollerachjcap}.
The depth of the shower maximum, $X_\text{max}$, is determined 
using a parametrization based on air shower physics 
\citep{Auger2010, gaisser1990}. For a nucleus with mass number $A$ and energy 
$E$, it is expressed as \citep{Auger2010, gaisser1990}
\begin{equation}
X_\text{max}(E,A) = X_0 + \nu \ln\left(\frac{E}{A}\right),
\end{equation}
where $X_0$ and $\nu$ are parameters influenced by hadronic interactions 
\citep{Auger2010}. In our analysis, we have used the values 
$X_0 = 700\, \text{g cm}^{-2}$ and $\nu = 50\, 
\text{g cm}^{-2}$ \citep{gaisser1990, gaisser2016}.
For scenarios involving a mixed composition of CRs, the flux-weighted mean 
depth of the shower maximum is computed as
\begin{equation}
\langle X_\text{max} \rangle = \frac{\sum\limits_i J_i(E) \cdot X_\text{max, i} (E,A_i)}{\sum\limits_i J_i(E)},
\end{equation}
where $J_i(E)$ represents the flux of each nuclear species.
These two Figs.~\ref{mixed1} and \ref{mixed2} illustrate the fluxes for a 
mixed composition of nuclei along with the corresponding 
$\langle X_\text{max} \rangle$s and $\sigma \left( X_\text{max} \right )$s for the Bumblebee gravity and the $\Lambda$CDM 
model respectively at redshift $z=1$. The parametrizations exhibit deviations 
from a purely protonic composition, and the impact of the cosmological models 
is also evident.
The observational data for $\langle X_\text{max} \rangle$s from PAO and TA are 
taken from Refs.~\citep{xmax_pao, xmax_ta}. These datasets provide 
$\langle X_\text{max} \rangle$ values along with statistical errors and 
uncertainties. In the $\langle X_\text{max} \rangle$ and $\sigma \left( X_\text{max} \right )$ plots, statistical 
uncertainties are depicted as shaded bands. The agreement between our results 
and observational data supports the viability of the Bumblebee gravity in 
explaining the CRs spectra. The results shown in these two figures 
correspond to different best-fit source separation distances $d_s$ for the 
Bumblebee and $\Lambda$CDM models. For example, the fitted values of $d_s$ 
(in Mpc) for (Bumblebee, $\Lambda$CDM) are: (71, 60) for SFR; (7.8, 6.5) for 
GRB; and (1.5, 1.25) for AGN. Although the visual differences are subtle, the 
Bumblebee model consistently prefers slightly larger $d_s$ values, reflecting 
the effect of its modified cosmological background on UHECRs propagation.

\subsection{Suppression of flux}
The suppression factor of CRs' flux can be expressed as
\begin{equation}\label{suppression}
G\left(\frac{E}{E_\text{c}}\right) \equiv \frac{J_\text{mod}(E)}{J_\text{mod}(E) \big |_{d_\text{s} \to 0}},
\end{equation}
which represents the ratio of the flux obtained from a discrete source 
distribution to that of a continuous source distribution in the limit 
$d_\text{s} \to 0$. The continuous source distribution corresponds to the 
case where $F = 1$ in Eq.~\eqref{flux}, implying that the flux remains 
unaffected by CRs' propagation effects. The suppression factor depends on 
both the coherence length $l_\text{c}$ and the inter-source separation
distance $d_\text{s}$, according to the relation given in 
Refs.~\citep{mollerachjcap, crpropa} as
\begin{equation}\label{xs}
X_\text{s} = \frac{d_\text{s}}{\sqrt{R_\text{H} l_\text{c}}}.
\end{equation} 
Here, $X_\text{s}$ represents the finite density factor and it will be used 
in the suppression calculations in  Eq.~\eqref{F_supp}-\eqref{ad}.
\begin{figure}
\centering
\includegraphics[width=0.9\linewidth]{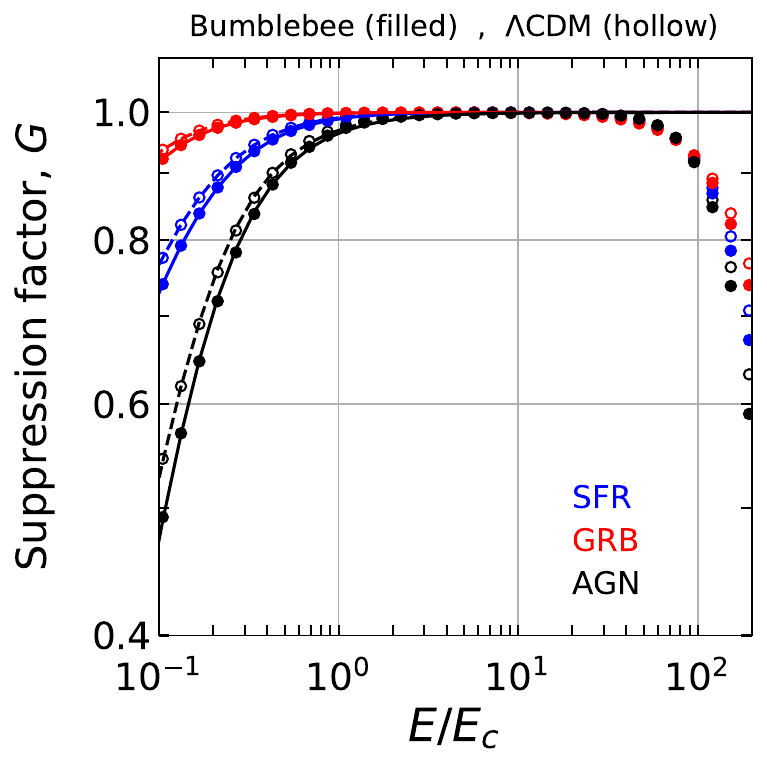}
\vspace{-0.3cm}
\caption{Suppression factor $ G $ as a function of normalized energy 
$ E / E_\text{c} $ with the finite density factor $X_\text{s}=1$ and with 
different astrophysical source models: AGN (black), SFR (blue), and GRB (red) 
as predicted by the Bumblebee gravity model in comparison with the 
$ \Lambda $CDM model. Solid lines represent the fits for the Bumblebee gravity 
model, while dashed lines correspond to the $ \Lambda $CDM model.}
\label{fig5e}
\end{figure}

In Fig.~\ref{fig5e}, the suppression factor $G$ as a function of the 
normalized energy $E/E_\text{c}$ is analyzed for different astrophysical 
source models: AGN (black), SFR (blue), and GRB (red). The solid and dashed 
lines represent the corresponding fits for the Bumblebee gravity and the 
$\Lambda$CDM model, respectively. Across all models, $G$ increases with 
energy, reaching a plateau near unity before slightly declining at very high 
energies. The GRB model (red) exhibits the highest suppression, attaining 
$G \approx 1$ more rapidly, while the AGN model (black) shows the weakest 
suppression at lower energies. The SFR model (blue) follows an intermediate 
trend, reflecting its broader source distribution. Comparing the two 
cosmological models, the Bumblebee gravity model induces slightly lower 
suppression at low energies relative to the $\Lambda$CDM model, particularly 
in the AGN and SFR cases. This suggests that Lorentz-violating effects in the 
Bumblebee gravity model enhance CR attenuation at low energies. However, the 
differences between the models diminish as $E/E_\text{c}$ increases.  

The fitting equation for the numerical results in Fig.~\ref{fig5e} is given 
by \citep{Manuel}          
\begin{equation}\label{fit}
R(x) = \text{exp}\left[-\left( \frac{a X_\text{s}}{x + b (x/a)^\beta}  \right)^{\!\alpha}\, \right],
\end{equation}
where $a$, $b$, $\alpha$ and $\beta$ are fitting parameters and their values 
corresponding to the fitting are given in Table \ref{tab:parameters}.
We fitted the computed suppression to the model functions using a nonlinear 
least squares method, minimizing the residual sum of squares 
\citep{Levenberg_1944, Marquardt_1963} as given by
\begin{equation}
\min_{\theta} \sum_{i=1}^{N} \left( y_i - f(E_{1,i}, \theta) \right)^2,
\end{equation}
where $y_i$ are the observed data points, $f(E_{1,i}, \theta)$ represents the 
fitting model with parameters $\theta = (a, b, \beta, \alpha)$, and $N$ is 
the total number of data points. The minimization was performed using the 
Levenberg-Marquardt algorithm, which interpolates between the Gauss-Newton 
and gradient descent methods. The fitting was implemented using the 
\texttt{curve\_fit} function from the \texttt{scipy.optimize} module in 
Python.
\begin{table}
    \centering
    \caption{Fitting parameters' values of Eq.~\eqref{fit} to the results of
suppression factor $G$ for the Bumblebee gravity and the $\Lambda$CDM model 
with different source types.}
    \renewcommand{\arraystretch}{1.2}
    \begin{tabular}{l @{\hspace{0.3cm}} c @{\hspace{0.3cm}}c @{\hspace{0.3cm}}c @{\hspace{0.3cm}}c @{\hspace{0.3cm}}c}
        \toprule
        \textbf{Model} & \textbf{Source} & \textbf{a} & \textbf{b} & \boldmath{$\beta$} & \boldmath{$\alpha$} \\
        \midrule
        \multirow{3}{*}{Bumblebee}  
        & SFR  & 0.1538 & 0.1760 & 0.1520 & 2.1630 \\
        & GRB  & 0.0551 & 0.0628 & 0.1602 & 2.1782 \\
        & AGN  & 0.0988 & 0.0729 & 2.9588 & 0.5050 \\
        \midrule
        \multirow{3}{*}{$\Lambda$CDM}  
        & SFR  & 0.1429 & 0.1709 & 0.1488 & 2.1765 \\
        & GRB  & 0.0502 & 0.0629 & 0.1648 & 2.1749 \\
        & AGN  & 0.0760 & 0.0386 & 2.9924 & 0.4928 \\
        \bottomrule
    \end{tabular}
    \label{tab:parameters}
\end{table}

\subsection{Anisotropy}

\begin{figure}
\centering
\includegraphics[width=0.9\linewidth]{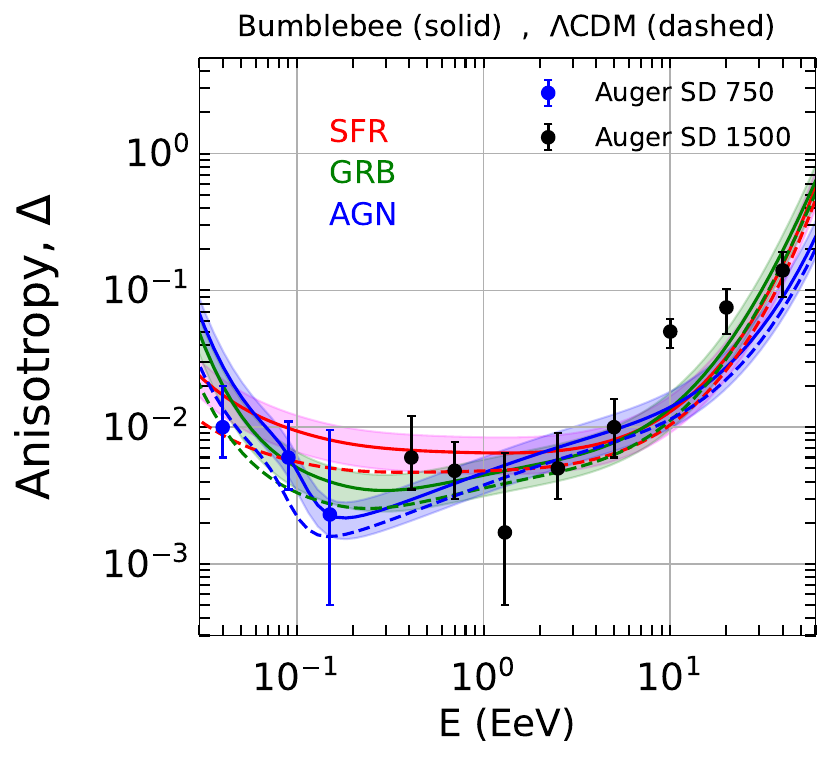}
\vspace{-0.3cm}
\caption{Anisotropy $ \Delta $ as a function of energy $ E $ for different 
astrophysical source models: SFR (red), GRB (green), and AGN (blue). The 
analysis is conducted within the Bumblebee gravity framework, with the 
$ \Lambda $CDM model included as a reference. Solid and dashed lines 
represent the predictions of Bumblebee gravity and the $ \Lambda $CDM model, 
respectively in comparison with the PAO surface detector data: SD 750 (blue) 
and SD 1500 (black) \citep{apj891}. The shaded regions indicate the 
uncertainties in the Bumblebee model predictions corresponding to different 
source models.}
\label{fig5f}
\end{figure}

We calculate the CRs' anisotropy using the methodology adopted in 
Ref.~\citep{Supanitsky} and it is given as
\begin{equation} \label{aniso}
\Delta = 3 ~ \frac{\eta}{\xi},
\end{equation}
where $\eta$ is the modification factor and is given by
\begin{equation}
\eta = \frac{J_\text{mod}(E)}{J_{0}(E)}.
\end{equation}
Here, $J_{0}(E)$ is the CRs' flux without any kind of energy losses, which 
is given as
\begin{equation}\label{flux0eq}
J_{0}(E) = \frac{c}{4\pi} \int_{0}^{z_{\text{max}}} \!\! dz \, \left| \frac{dt}{dz} \right| \, \mathcal{N}_{z\rightarrow 0}(E) \frac{\exp\left[-r_\text{s}^2 / (4 \lambda^2)\right]}{(4 \pi \lambda^2)^{3/2}}.
\end{equation} 
We present the anisotropy $\Delta$ as a function of energy $E$ in 
Fig.~\ref{fig5f} for different astrophysical source models: SFR (red), GRB 
(green), and AGN (blue). The analysis is performed within the Bumblebee 
gravity framework, with the $\Lambda$CDM model included as a reference for 
comparison. The solid and dashed lines correspond to the predictions of the 
Bumblebee gravity and the $ \Lambda$CDM model, respectively in comparison 
with the PAO surface detector data: SD 750 (blue) and SD 1500 (black) 
\citep{apj891}. The shaded regions indicate the uncertainties in the Bumblebee 
model predictions corresponding to different source models. For the fitting
of the predictions with the data, we have 
taken the strength of the magnetic field as $B = 1$ nG and chosen the source 
separation distances as $85$ Mpc for SFR, $10$ Mpc for GRB, and $1$ Mpc for 
AGN. These values influence the level of anisotropy observed for each source 
model. The $\chi^2$ and the $\chi_\text{red}^2$ values for the fitting are 
provided in Table \ref{tab:aniso}. The anisotropy generally decreases with 
energy at lower values, reaching a minimum around $E \approx 1$ EeV, and 
then increases at higher energies. The variation in anisotropy between 
different source models reflects the effect of source distribution and 
separation distances. The SFR model, with the largest source separation 
($85$ Mpc), shows more pronounced anisotropy at lower energies, while the 
AGN model, with the smallest source separation ($1$ Mpc), exhibits 
comparatively lower anisotropy. This trend arises because widely separated 
sources result in a more inhomogeneous CR sky, while closely spaced sources 
lead to a more isotropic distribution. The GRB model follows an intermediate 
trend, consistent with its moderate source separation of $10$ Mpc. At higher 
energies, the anisotropy increases for all source models due to reduced CR 
diffusion, allowing the intrinsic distribution of sources to become more 
apparent. The comparison between the Bumblebee and $\Lambda$CDM models shows 
that while the modified gravity effects in Bumblebee gravity slightly enhance 
anisotropy at lower energies, the differences between the two models diminish 
as energy increases. These results emphasize the role of energy-dependent 
propagation effects and source clustering in shaping CRs anisotropy.
\begin{table}[!htb]
    \centering
    \caption{Chi-square and reduced chi-square values for different models as
the CRs anisotropy predictions are fitted to the PAO surface detector data 
\citep{apj891}.}
    \begin{tabular}{l@{\hspace{0.5cm}} c@{\hspace{0.5cm}} c@{\hspace{0.5cm}}c @{\hspace{0.5cm}}}
        \toprule
        \textbf{ } & \textbf{SFR} & \textbf{GRB} & \textbf{AGN} \\
        \midrule
        \multicolumn{4}{c}{\textbf{Bumblebee}} \\
        \midrule
       Chi-square ($\chi^2$) & 35.70 & 26.87 & 34.04 \\
        Reduced chi-square ($\chi_{red}^2$)  & 3.96 & 2.99 & 3.78 \\
        \midrule
        \multicolumn{4}{c}{\textbf{$\Lambda$CDM}} \\
        \midrule
        Chi-square ($\chi^2$) & 23.90 & 17.71 & 22.86 \\
        Reduced chi-square ($\chi_{red}^2$) & 2.39 & 1.77 & 2.29 \\
        \bottomrule
    \end{tabular}
    \label{tab:aniso}
\end{table}
\begin{figure*}
      \centerline{
    \includegraphics[scale=0.28]{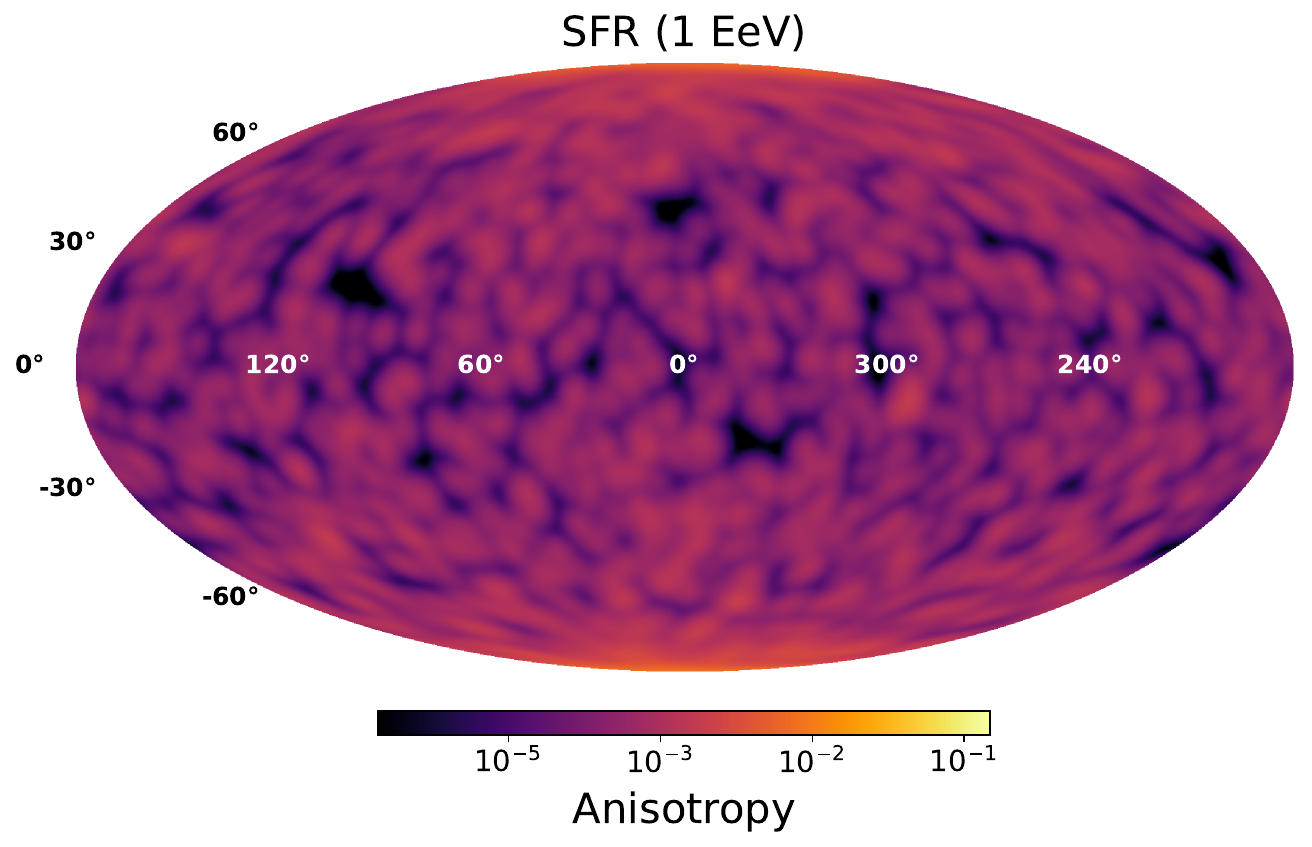}
   \includegraphics[scale=0.28]{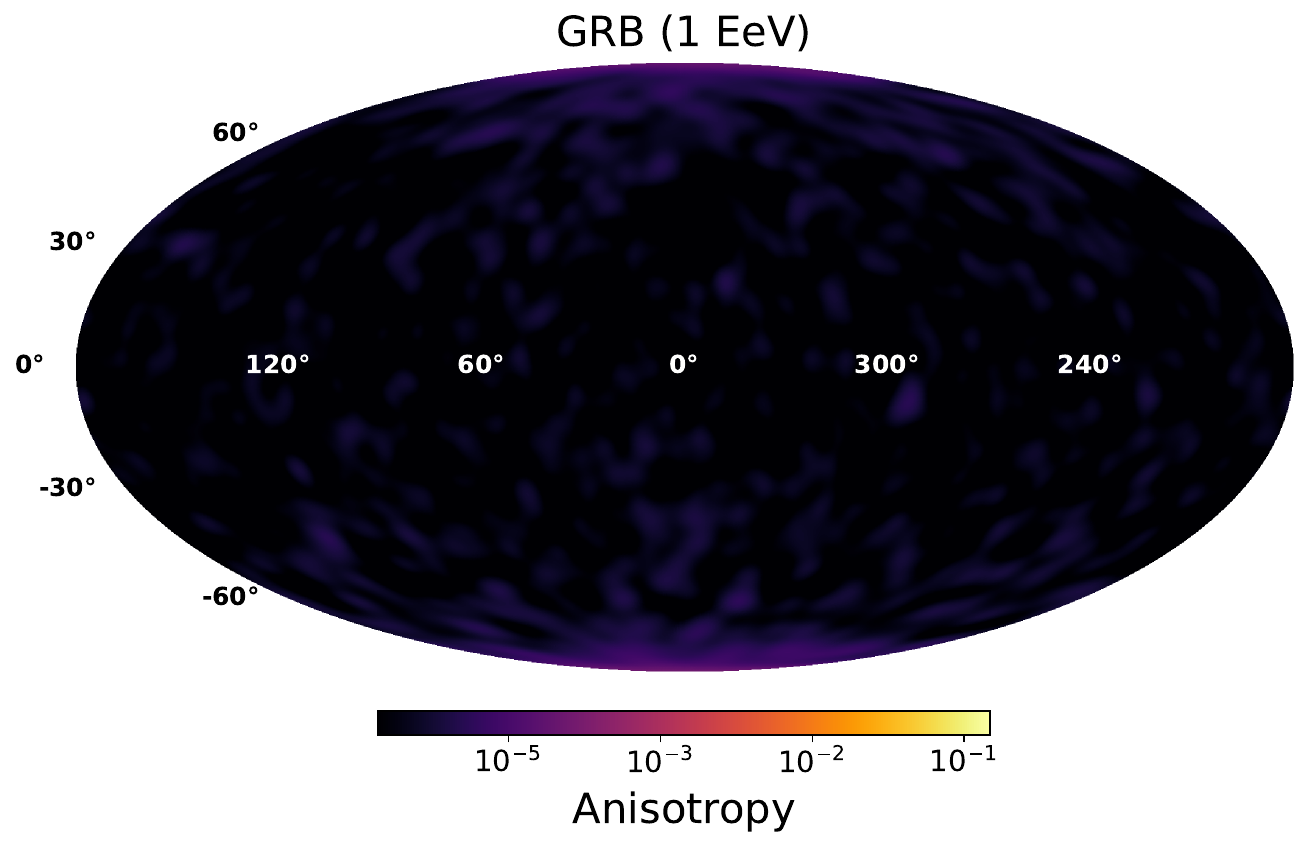}
   \includegraphics[scale=0.28]{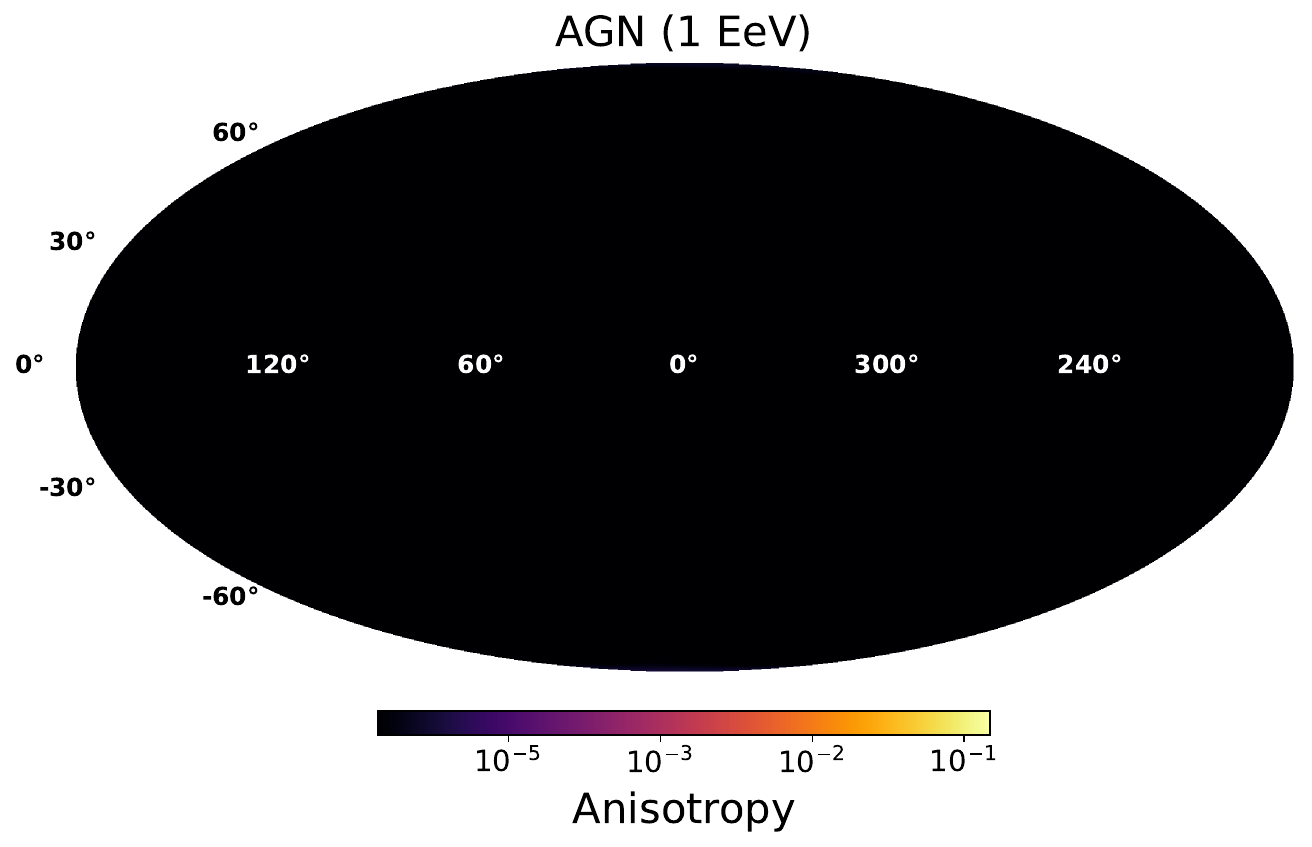}}
   
    \centerline{
   \includegraphics[scale=0.28]{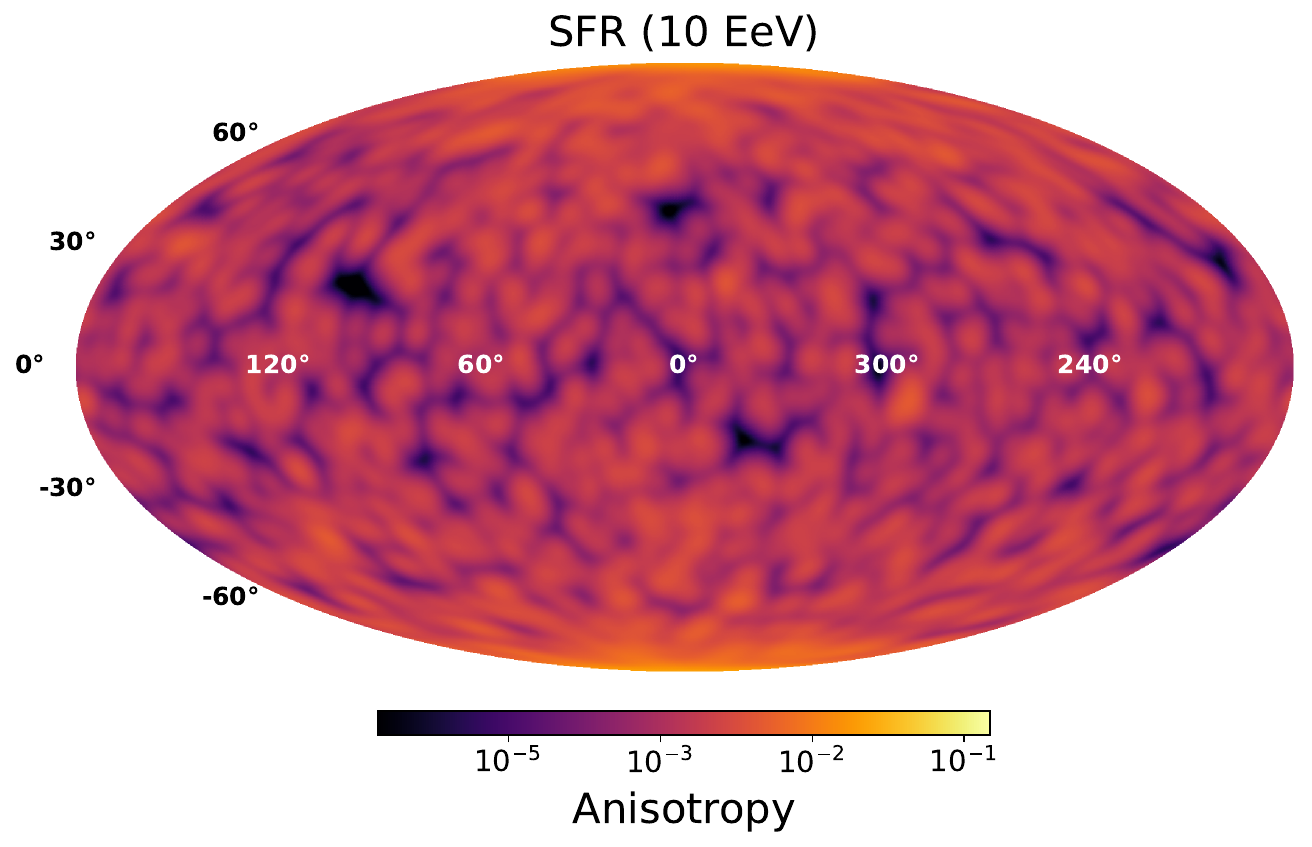}
  \includegraphics[scale=0.28]{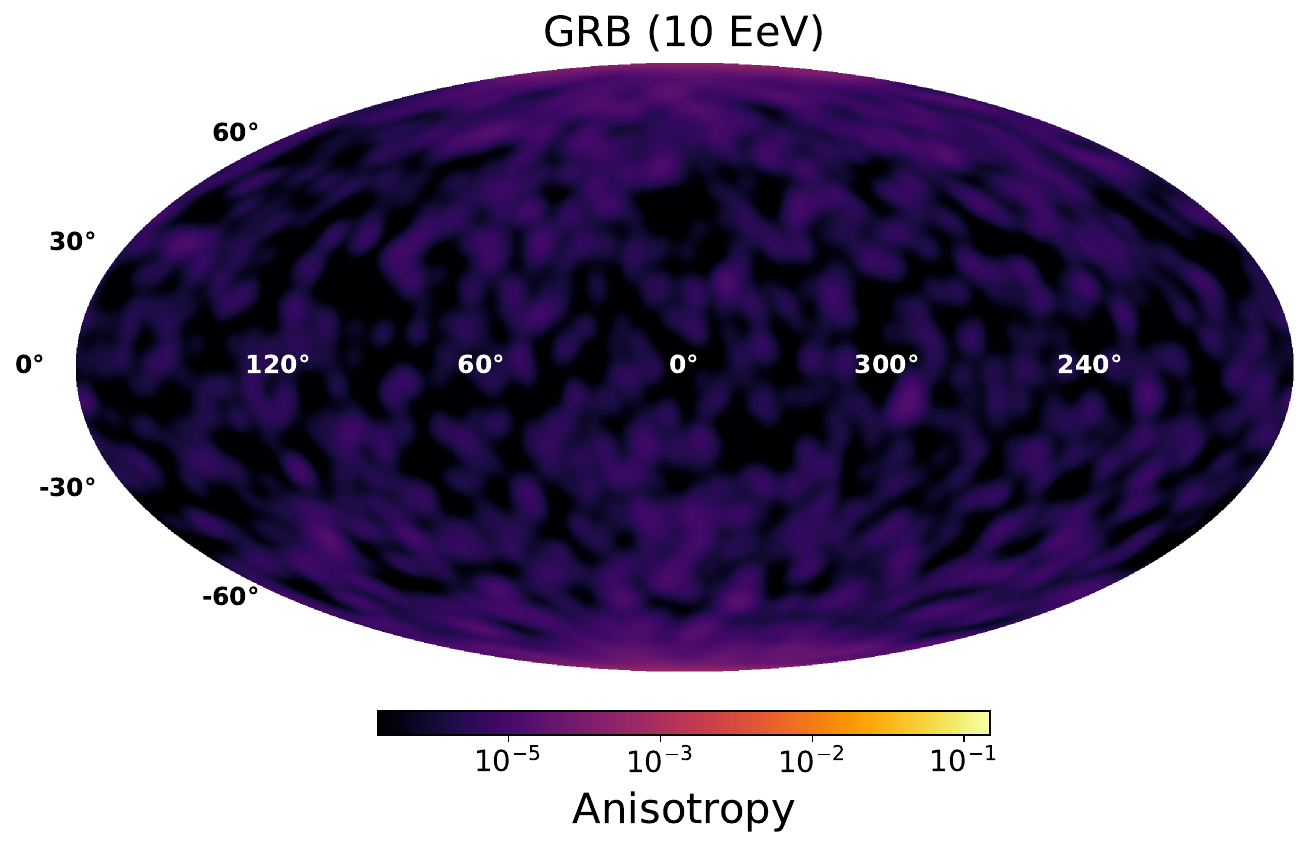}
  \includegraphics[scale=0.28]{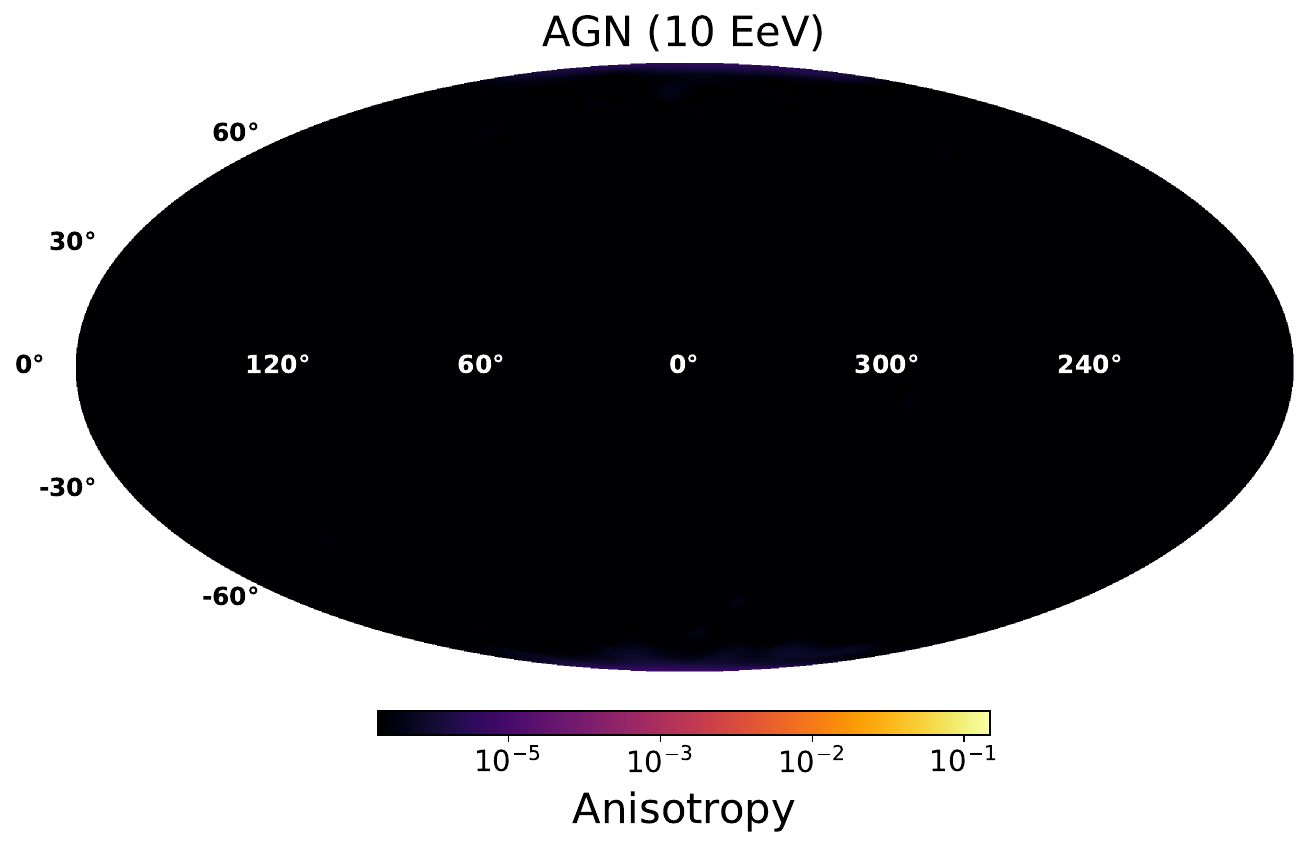}}

  \centerline{
  \includegraphics[scale=0.28]{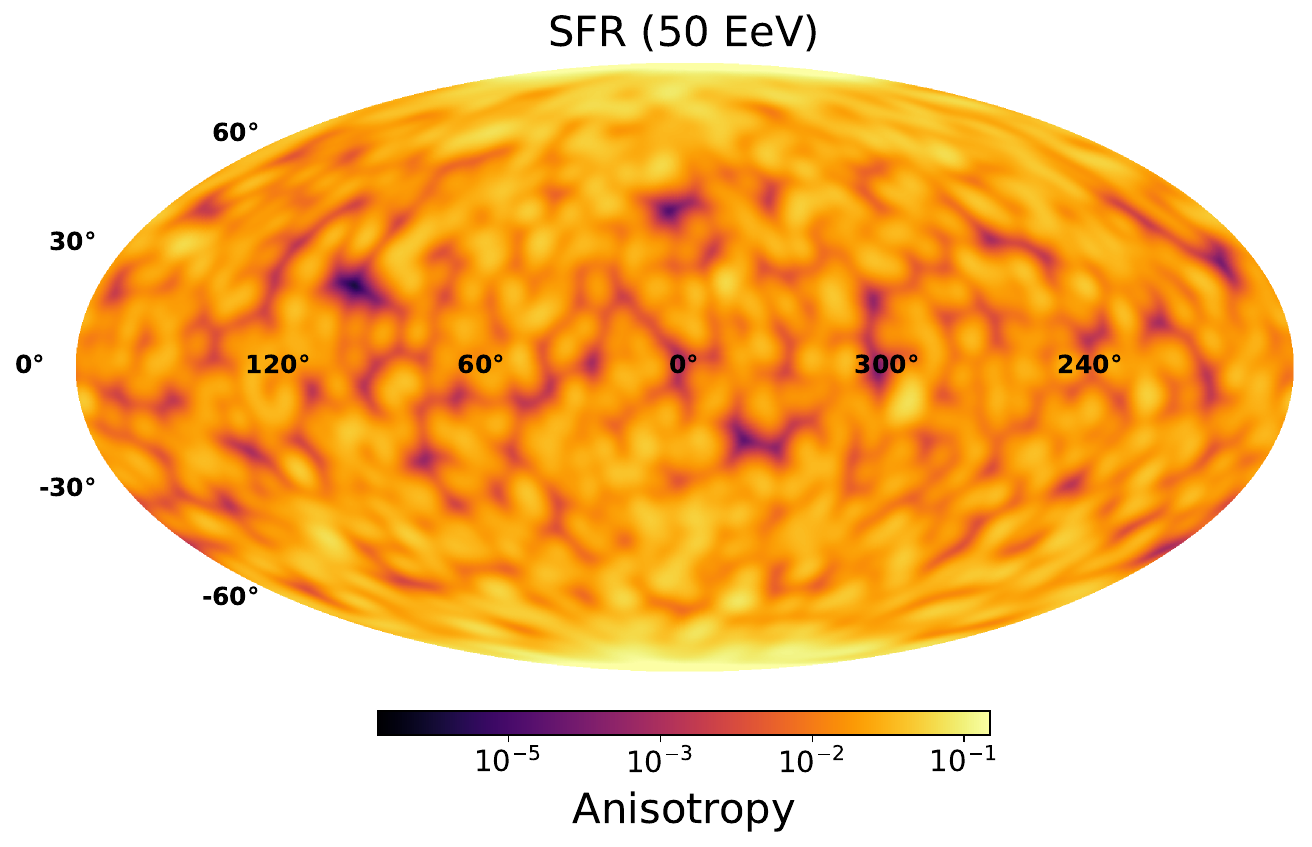}
   \includegraphics[scale=0.28]{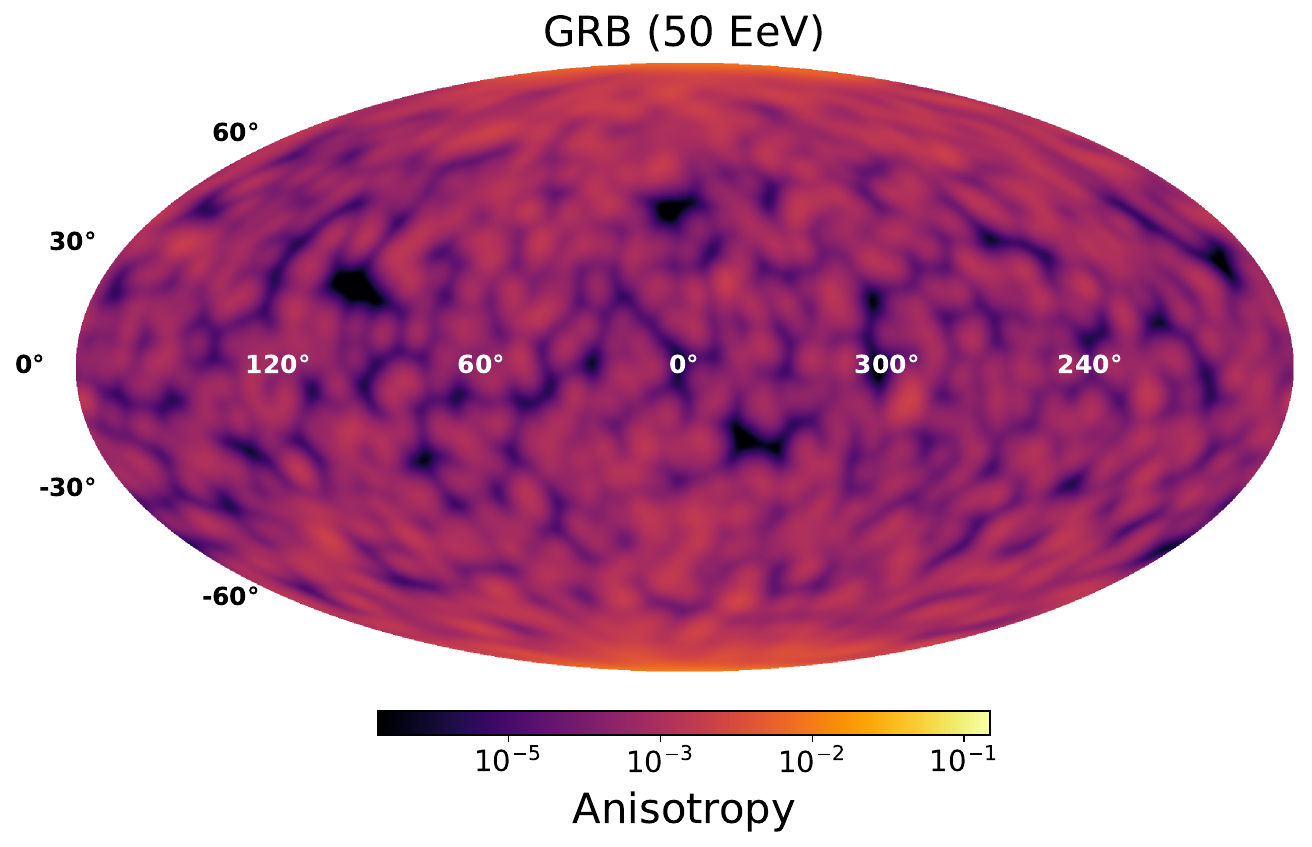}
  \includegraphics[scale=0.28]{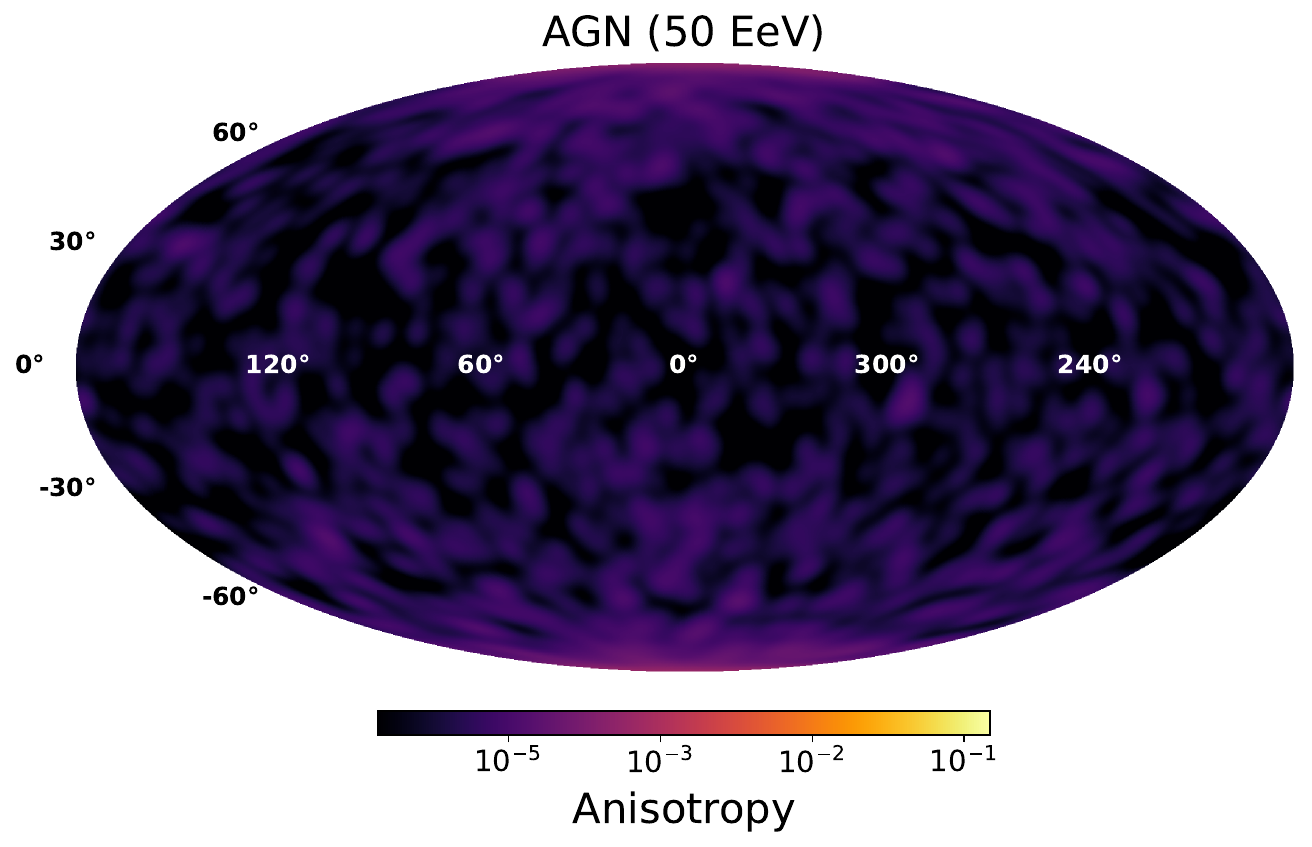}}
\caption{Anisotropy skymaps for SFR, GRB, and AGN sources within the Bumblebee 
gravity framework. To facilitate comparison, a fixed source separation 
distance of $ d_\text{s} = 30 $ Mpc is used. The three vertical panels 
correspond to source models, the SFR, GRB, and AGN from left to right, with CR 
arrival directions shown at $ E = 1 $ EeV, $ E = 10 $ EeV, and $ E = 50 $ EeV. 
The color scale represents the anisotropy amplitude, where yellow regions 
indicate stronger deviations from isotropy, while black regions correspond to 
nearly isotropic distributions.}
\label{fig5h}
\end{figure*}

In Fig.~\ref{fig5h}, we plot the skymaps for the SFR, GRB, and AGN sources 
within Bumblebee gravity results. To make the comparison more distinct, we 
employ the same source separation distance $d_\text{s}=30$ Mpc between all 
the sources. These anisotropy skymaps illustrate the variations in CRs arrival 
directions for different astrophysical source models at $E = 1$ EeV, $E = 10$ 
EeV, and $E = 50$ EeV energies. The three vertical panels correspond to 
sources following the SFR, GRB, and AGN from left to right. The color scale 
represents the anisotropy amplitude, where yellow regions indicate stronger 
deviations from isotropy, while black regions correspond to nearly isotropic 
distributions. These black regions (zero-valued pixels) in the anisotropy 
skymaps indicate directions with negligible anisotropy amplitude, 
particularly at low energies and for highly isotropic source distributions 
such as AGNs.

At $E = 1$ EeV energy, the anisotropy skymaps reveal distinct differences 
among the source models. The SFR model exhibits the highest anisotropy, 
characterized by large-scale structures and significant flux variations 
across the sky. This results from the relatively widespread but structured 
distribution of SFR sources, leading to enhanced directional flux variations. 
In contrast, the GRB model shows much weaker anisotropy, providing nearly 
isotropic CRs distribution. The AGN model demonstrates the lowest anisotropy, 
with the skymap appearing almost completely black, indicating that at $E = 1$ 
EeV energy, CRs from AGN sources undergo strong diffusion, leading to an 
effectively isotropic distribution.

At $E = 10$ EeV energy, the anisotropy patterns become more pronounced. The 
SFR model continues to show a strong anisotropic signal, with even greater 
contrast in flux variations. The GRB model also exhibits a slight increase in 
anisotropy compared to $E = 1$ EeV, suggesting that at higher energies, CRs
from GRB experience reduced diffusion, allowing the source distribution effect 
to emerge more clearly. However, the AGN model remains highly isotropic in the 
CRs distribution, implying that even at this energy, CRs originating from AGNs 
are still significantly affected by diffusion. These results highlight the 
interplay between source distribution and energy-dependent propagation 
effects. As energy increases, the CRs diffusion decreases, revealing stronger 
anisotropy in the SFR and GRB models, while AGN sources remain largely 
isotropic in their CRs distribution due to their sparse distribution.

At $E = 50$ EeV energy, the anisotropy becomes significantly more pronounced 
across all source models. The SFR model exhibits the strongest anisotropy. 
The GRB model now shows substantial anisotropy, with prominent flux 
fluctuations reflecting the discrete nature of GRB sources at high energies. 
Notably, the AGN model, which was nearly isotropic at lower energies, now 
displays significant anisotropy. This suggests that at $E = 50$ EeV, the CRs 
have undergone less diffusion, and their arrival directions more closely 
trace the intrinsic distribution of AGN sources also. 

These results underscore the impact of energy-dependent propagation effects 
on CRs anisotropy. At lower energies, strong diffusion leads to an effectively 
isotropic distribution, particularly for AGN sources. As energy increases, 
reduced diffusion allows the intrinsic anisotropy of the source distribution 
to emerge more clearly. The differences in anisotropy among the SFR, GRB, and 
AGN models further emphasize the role of source clustering in shaping the 
observed CR sky. For a complete energy range of $0.1-60$ EeV, the 
distributions of the anisotropy skymap are shown in Appendix~\ref{appnb}.

\section{Summary and conclusion}\label{secVI}
In this work, we have explored the propagation of UHECRs in the 
context of Bumblebee gravity, a Lorentz-violating modification of GR. Our 
motivation stems from the interest in testing alternate cosmological models 
beyond the standard $\Lambda$CDM scenario to investigate whether they can 
influence observable features of UHECRs such as energy spectrum, composition 
and anisotropy. While $\Lambda$CDM successfully explains a wide range of 
cosmological observations, alternative gravity models like Bumblebee gravity 
provide a unique opportunity to probe new physics in high-energy astrophysical 
processes.
We study various properties of CRs using the UFA-15 source 
models that include SFR, GRB, and AGN, within the realm of Bumblebee gravity 
and also we take the standard $\Lambda$CDM model as a reference for comparison.
For simplicity, the sources are assumed to be uniformly distributed 
in angular coordinates (RA, Dec) using the HEALPix grid in this work.
Fig.~\ref{fig5a} shows the effect of the Bumblebee gravity parameter $ l $ on 
the density enhancement factor $ \xi $ for different CR source distributions. 
Three values of $ l $ are considered: $ l = 0.005 $, $ l = 0.07 $, and 
$ l = 0.09 $, with the $\Lambda$CDM model as a reference. As $ l $ increases, 
$ \xi $ systematically rises, especially at lower energies, indicating that 
stronger gravity modifications amplify the CRs' density. 

We also analyzed the density enhancement factor $\xi$ of CRs considering 
different critical energies, $E_\text{c} = 45$ EeV, $9$ EeV, and $4.5$ EeV, 
for three representative source distances: $d_\text{s} = 25$ Mpc, $50$ Mpc, 
and $100$ Mpc, as shown in Fig.~\ref{fig5b}. Our findings reveal that AGN 
sources consistently exhibit the highest enhancement, followed by GRB and 
SFR sources. As the source distance $d_\text{s}$ increases, the influence of 
propagation effects becomes more significant, resulting in a more pronounced 
variation in $\xi$ across different $E_\text{c}$ values. For shorter distances 
like $d_\text{s} = 25$ Mpc, these differences remain modest. However, at 
$d_\text{s} = 50$ Mpc and especially at $100$ Mpc, the impact of propagation 
becomes clearly visible. Lower $E_\text{c}$ values consistently lead to 
higher enhancements, highlighting a trend of increasing $\xi$ as $E_\text{c}$ 
decreases from $45$ to $4.5$ EeV.
Furthermore, the contrast between the 
Bumblebee gravity model and the standard $\Lambda$CDM model becomes more 
evident with increasing source distance. This underscores the growing 
importance of cosmological effects at larger separations of sources, further 
influencing the observed CR density enhancement.

Fig.~\ref{fig5c} shows the UHECR flux as a function of energy $ E $ for 
different astrophysical sources. Flux predictions are given for both Bumblebee 
gravity and $ \Lambda $CDM model and are compared with PAO and TA data, 
showing good agreement with fitted models. For fitting, different source 
separation distances $ d_\text{s} $ are used. In the Bumblebee gravity, 
$ d_\text{s} = 74 $ Mpc for SFR, $ 8.2 $ Mpc for GRB, and $ 1.4 $ Mpc for AGN
are used. In the $ \Lambda $CDM model, the used $ d_\text{s} $ values 
are $ d_\text{s} = 69 $ Mpc, $ 7.2 $ Mpc and $ 1.25 $ Mpc, respectively. 
Larger $ d_\text{s} $ results in a smoother flux profile, while smaller values 
enhance clustering. The flux spectrum follows the expected trend: a gradual 
rise at low energies, flattening in the intermediate range, and sharp 
suppression at high energies due to the GZK effect. The Bumblebee gravity 
shows slight deviations from $ \Lambda $CDM, particularly at high energies, 
suggesting the modified gravity effect on CRs propagation. The residual plot 
in the lower panel shows the difference between the predicted and observed 
flux. Residuals remain centered around zero, indicating a good fit for both 
models. Small fluctuations at high energies suggest statistical uncertainties 
or minor deviations from source modeling.

The skymaps in Fig.~\ref{fig5d} depict the CRs flux distribution for different 
astrophysical source models at varying source separation distances, 
specifically at $ d_\text{s} = 1 $, $ 25 $, and $ 100 $ Mpc. At a short source 
distance of $d_\text{s} = 1$ Mpc, the SFR and GRB models produce relatively 
smooth flux distributions, reflecting the uniform nature of their source 
populations. In contrast, the AGN model displays a more structured and 
inhomogeneous flux pattern. As the source separation increases to 
$d_\text{s} = 25$ Mpc and further to $100$ Mpc, the flux distributions for 
all models become increasingly discrete. The SFR model starts exhibiting 
greater contrast, the GRB model develops clearer structural patterns, and 
the AGN model shows even stronger clustering effects, characterized by 
significant flux variations across the sky. This trend is a natural outcome 
of increasing source separation, which lowers the number density of sources 
within the observable volume and amplifies the visibility of individual 
source contributions, leading to a more structured and discrete CRs flux 
distribution. A mixed composition scenario along with the 
$\langle X_\text{max} \rangle$ for both the cosmological models is shown in 
Fig. \ref{mixed1} and Fig. \ref{mixed2}, which depicts the viability of 
Bumblebee gravity in CRs studies. Here, the best-fit values of source 
separation $d_s$ (in Mpc) for (Bumblebee, $\Lambda$CDM) are: (71, 60) 
for SFR; (7.8, 6.5) for GRB; and (1.5, 1.25) for AGN. These shifts arise due 
to changes in the energy-redshift relation and cosmic expansion history in the 
modified gravity framework.

Fig.~\ref{fig5e} shows the suppression factor $ G $ as a function of 
normalized energy $ E/E_\text{c} $ for different astrophysical sources. Across 
all models, $ G $ increases with energy, reaching a plateau near unity before 
slightly declining at very high energies. The GRB model exhibits the highest 
suppression, attaining $ G \approx 1 $ more quickly, followed by the AGN and 
the SFR model. An analytical fit is also shown for both of the cosmological 
models. The suppression factor in the $\Lambda$CDM model is slightly higher
than that of the Bumblebee gravity model. This effect is more pronounced in 
AGN and GRB distribution.

Fig.~\ref{fig5f} shows the anisotropy $ \Delta $ as a function of energy 
$ E $ for various astrophysical source models within the Bumblebee gravity 
framework, with the $ \Lambda$CDM model included for comparison. The analysis 
uses a magnetic field strength of $ B = 1 $ nG and source separation 
distances of $ 85 $ Mpc for SFR, $ 10 $ Mpc for GRB, and $ 1 $ Mpc for AGN. 
The anisotropy decreases with energy at lower values, reaching a minimum 
around $ E \approx 1 $ EeV, then increases at higher energies. The anisotropy 
varies among source models, with the SFR model showing more pronounced 
anisotropy at lower energies due to its larger source separation ($ 85 $ Mpc), 
while the AGN model, with a smaller separation ($ 1 $ Mpc), exhibits lower 
anisotropy. The GRB model shows intermediate behaviour. At higher energies, 
the reduced CR diffusion increases anisotropy for all models, revealing the 
intrinsic distribution of sources. In the lower energy region, the 
anisotropy amplitude in the Bumblebee gravity model is little higher than that 
of the standard $\Lambda$CDM model for the considered parameters, but it is 
almost same in the higher energy range.

Fig.~\ref{fig5h} shows skymaps for SFR, GRB, and AGN sources in Bumblebee 
gravity with $d_\text{s} = 30$ Mpc, illustrating CR anisotropy at $E = 1$, 
$10$, and $50$ EeV energies. At $E = 1$ EeV energy, SFR sources show the 
highest anisotropy due to their structured distribution, while GRB sources are 
nearly isotropic, and AGN sources show the lowest anisotropy due to strong 
diffusion.  At $E = 10$ EeV energy, anisotropy increases, with the SFR model 
retaining strong anisotropy, GRB sources showing more flux contrast, and AGN 
sources remaining nearly isotropic. At $E = 50$ EeV energy, anisotropy is 
prominent across all models: SFR sources show the strongest anisotropy, GRB 
sources display notable flux variations, and AGN sources, previously 
isotropic, exhibit significant anisotropy. These results highlight 
energy-dependent propagation effects — strong diffusion causes isotropy at 
low energies, especially for AGNs, while the reduced diffusion at high energies 
enhances intrinsic source anisotropy. The differences among models underscore 
the impact of source clustering on the CRs anisotropy.

In this work, we have adopted the UFA-15 composition models primarily 
for consistency with earlier studies and for illustrative purposes. We 
acknowledge that more recent models, such as those presented in 
Refs.~\citep{AlvesBatista:2018zui, Das:2020nvx, Heinze:2019jou}, offer improved 
parametrisations and incorporate additional astrophysical 
considerations. Our focus here is on the gravitational effect that is 
introduced by Bumblebee gravity model, and we leave a comparative study 
involving those advanced composition models for future works.
Our results indicate that increasing the Bumblebee gravity parameter $l$ 
enhances the density factor $\xi$, especially at low energies, highlighting 
the Lorentz violation role in CR propagation. Density enhancement $\xi$ depends 
on critical energy $E_\text{c}$ and source distance $d_\text{s}$, with larger 
$d_\text{s}$ increasing deviations from the $\Lambda$CDM model. AGN sources 
enhance flux at high energies, while GRB and SFR sources dominate at lower 
energies. UHECRs flux shows slight deviation of the Bumblebee gravity from the 
$\Lambda$CDM model, with stronger suppression $G$ at low energies and 
increasing anisotropy $\Delta$ at high energies. Skymaps reveal flux and 
anisotropy variations: low-energy CR diffusion leads to isotropy, while 
reduced diffusion at high energies enhances anisotropy. Overall, the Bumblebee 
gravity alters UHECRs propagation, with higher $l$ amplifying deviations from 
the $\Lambda$CDM model. Future observations can test these effects and refine 
modified gravity models.

\section*{Acknowledgement}
SPS and UDG gratefully acknowledge the anonymous reviewer
for his/her valuable comments and suggestions. UDG is thankful to the Inter-University Centre for Astronomy and Astrophysics 
(IUCAA), Pune, India for the Visiting Associateship of the institute.

\appendix
\section{Flux map for some particular energies}\label{appnA}
\begin{figure*}
    \centerline{
    \includegraphics[scale=0.28]{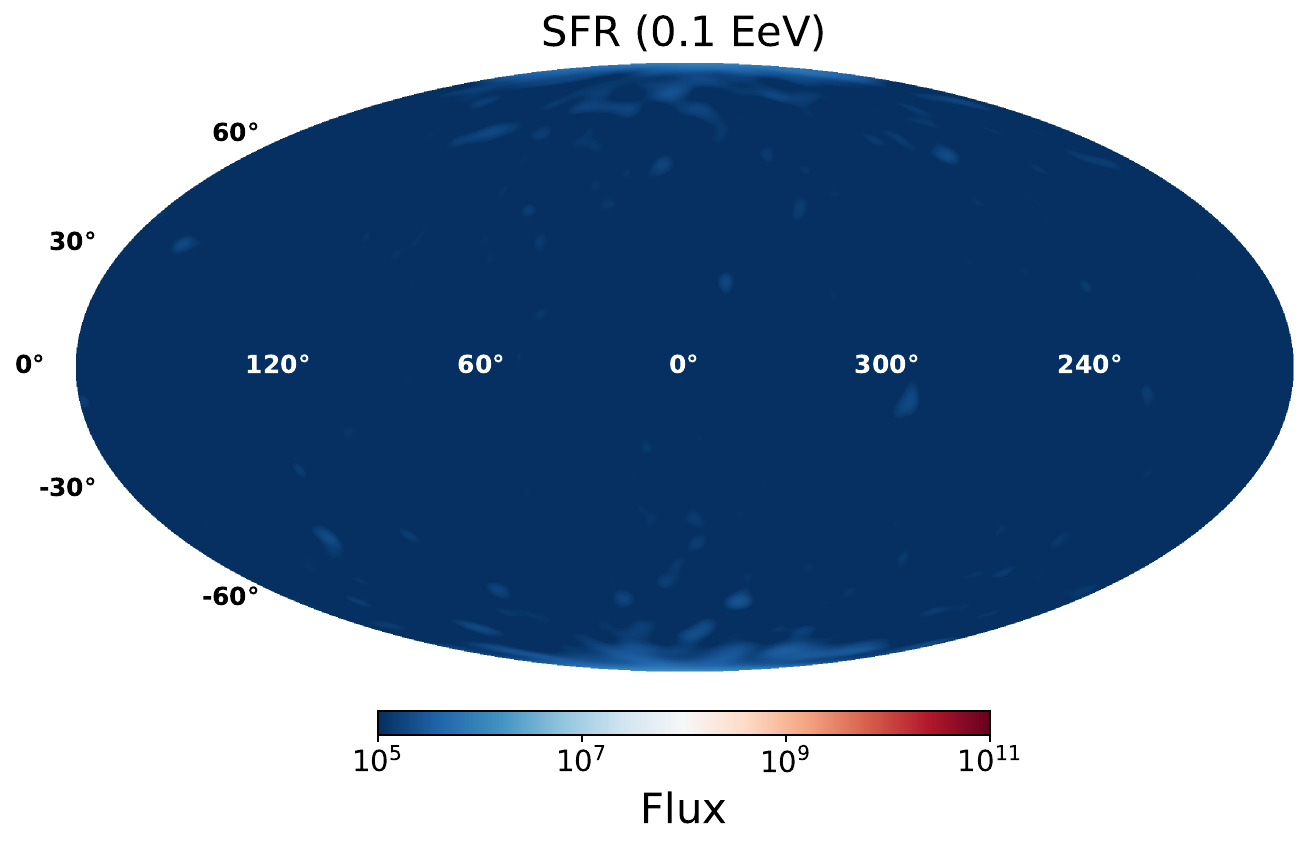}
    \includegraphics[scale=0.28]{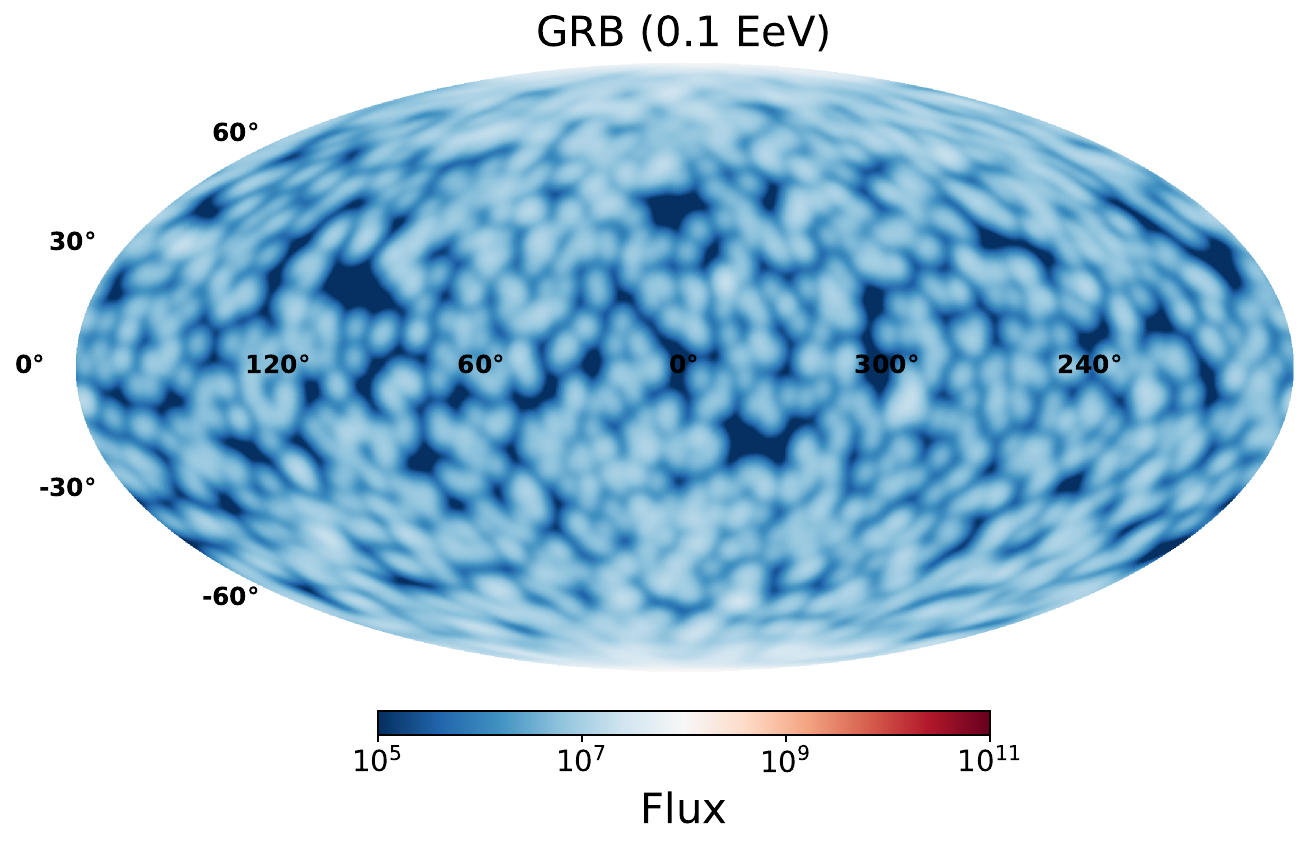}
    \includegraphics[scale=0.28]{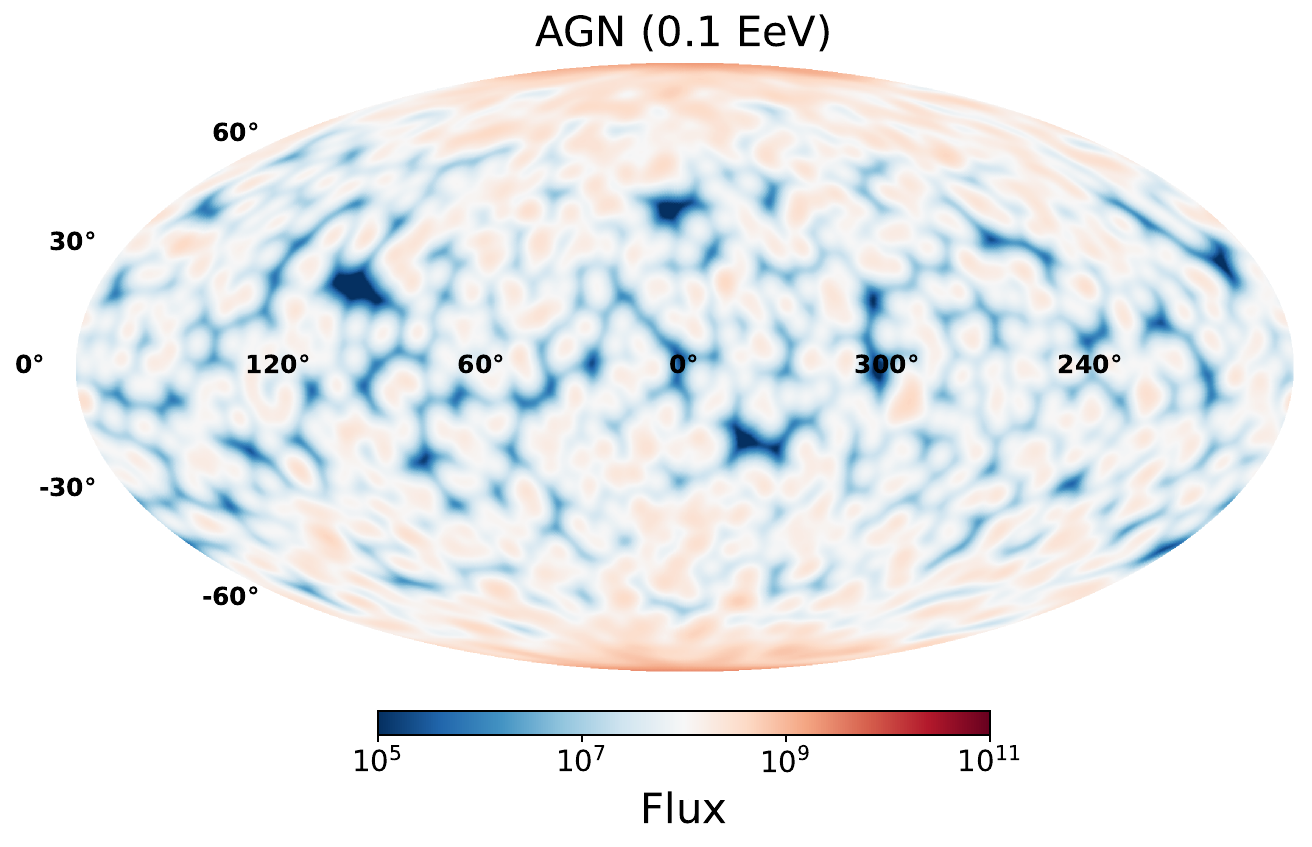}}
    \centerline{
    \includegraphics[scale=0.28]{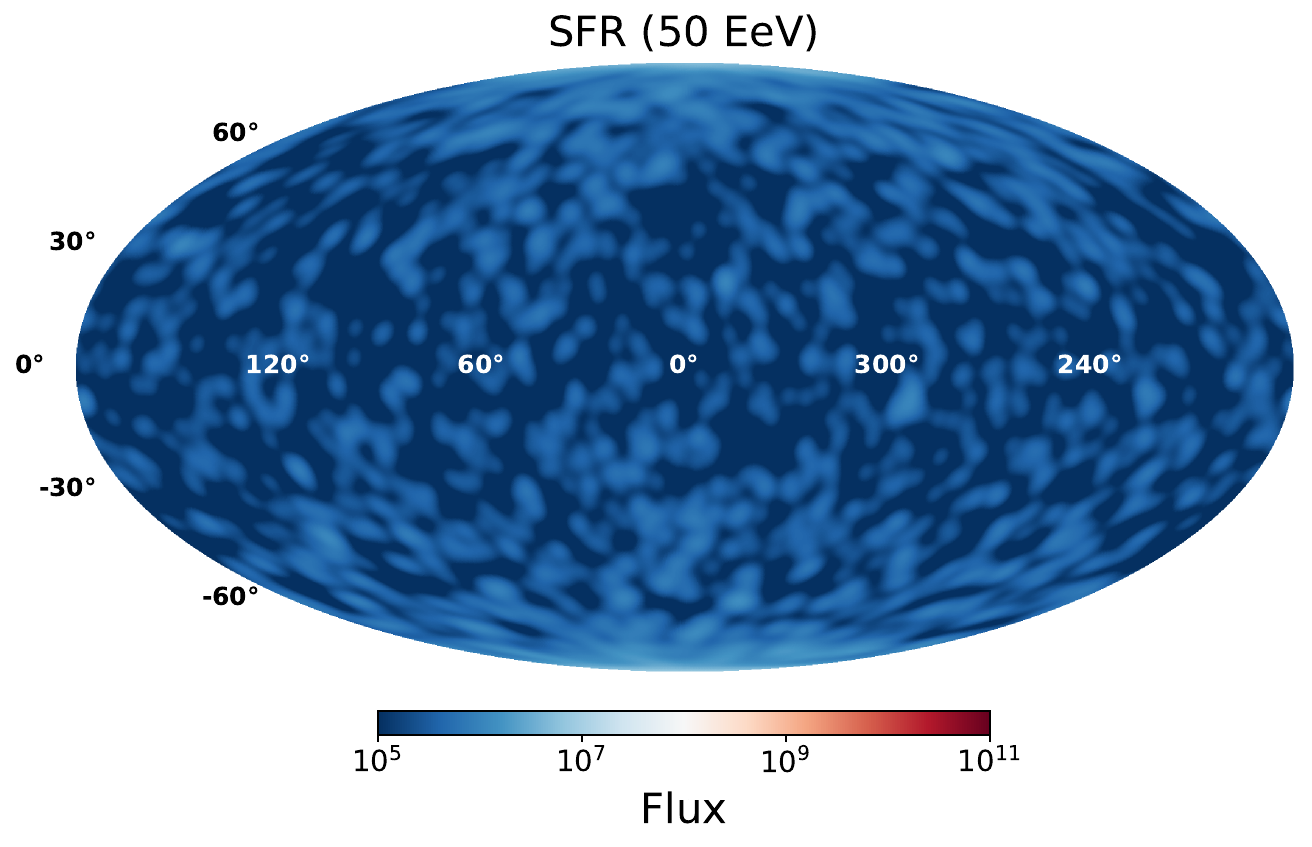}
    \includegraphics[scale=0.28]{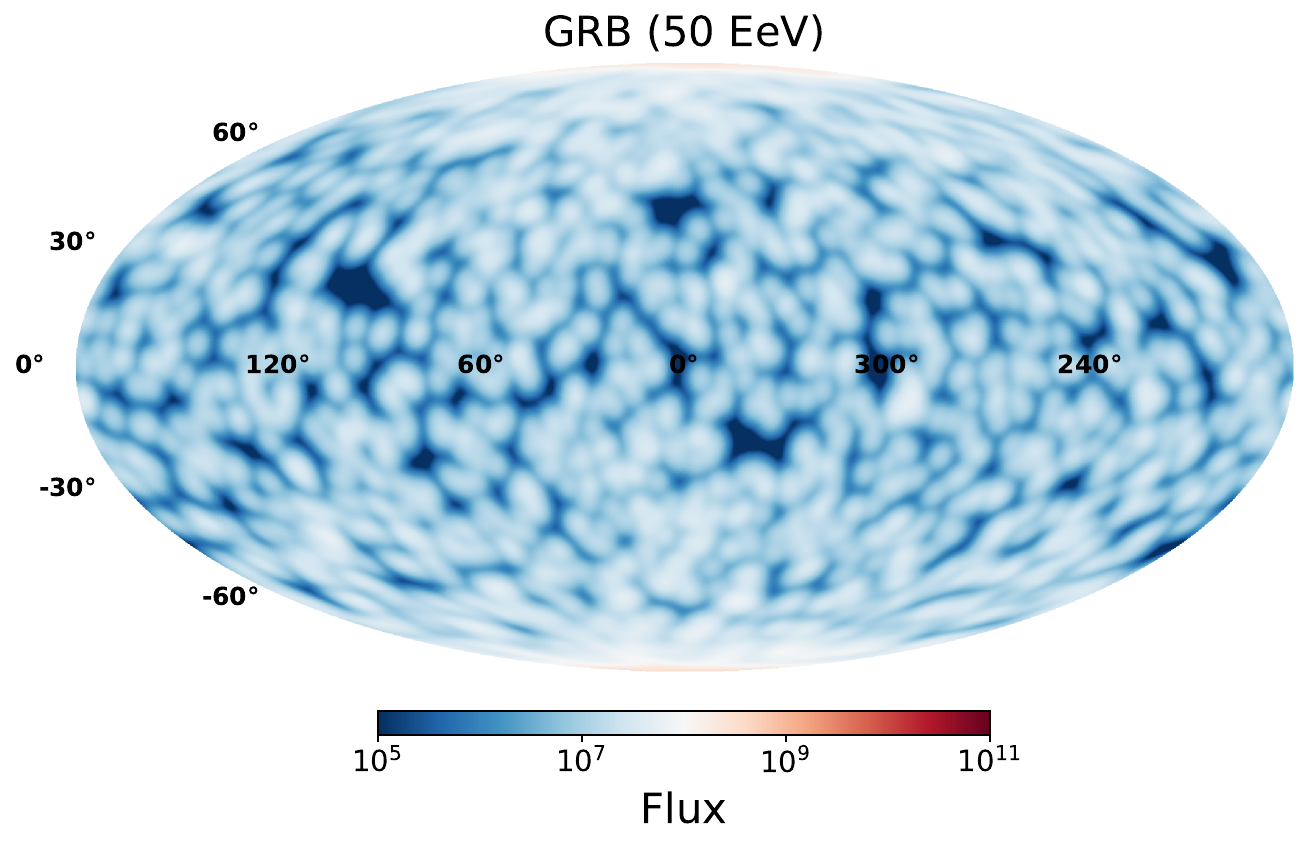}
    \includegraphics[scale=0.28]{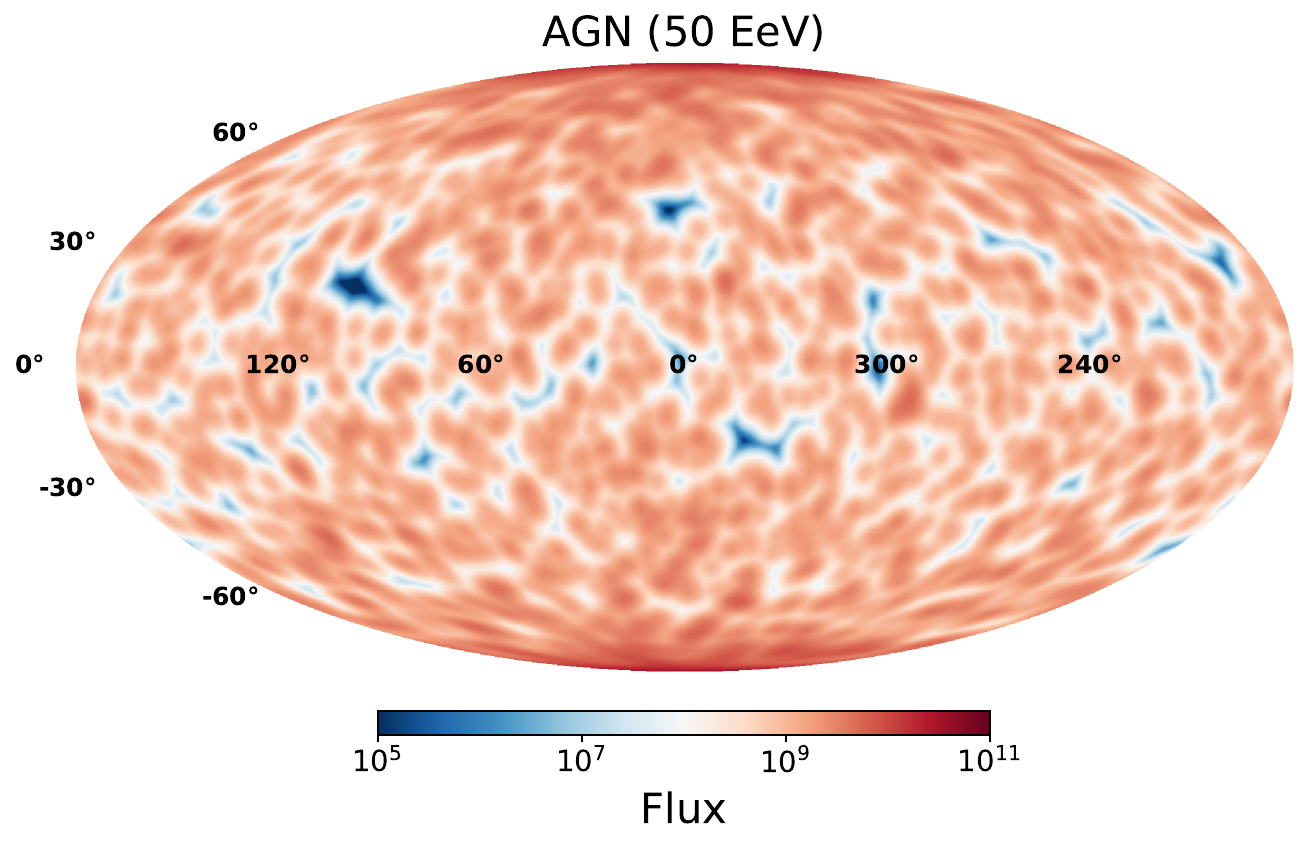}}
   \centerline{
   \includegraphics[scale=0.28]{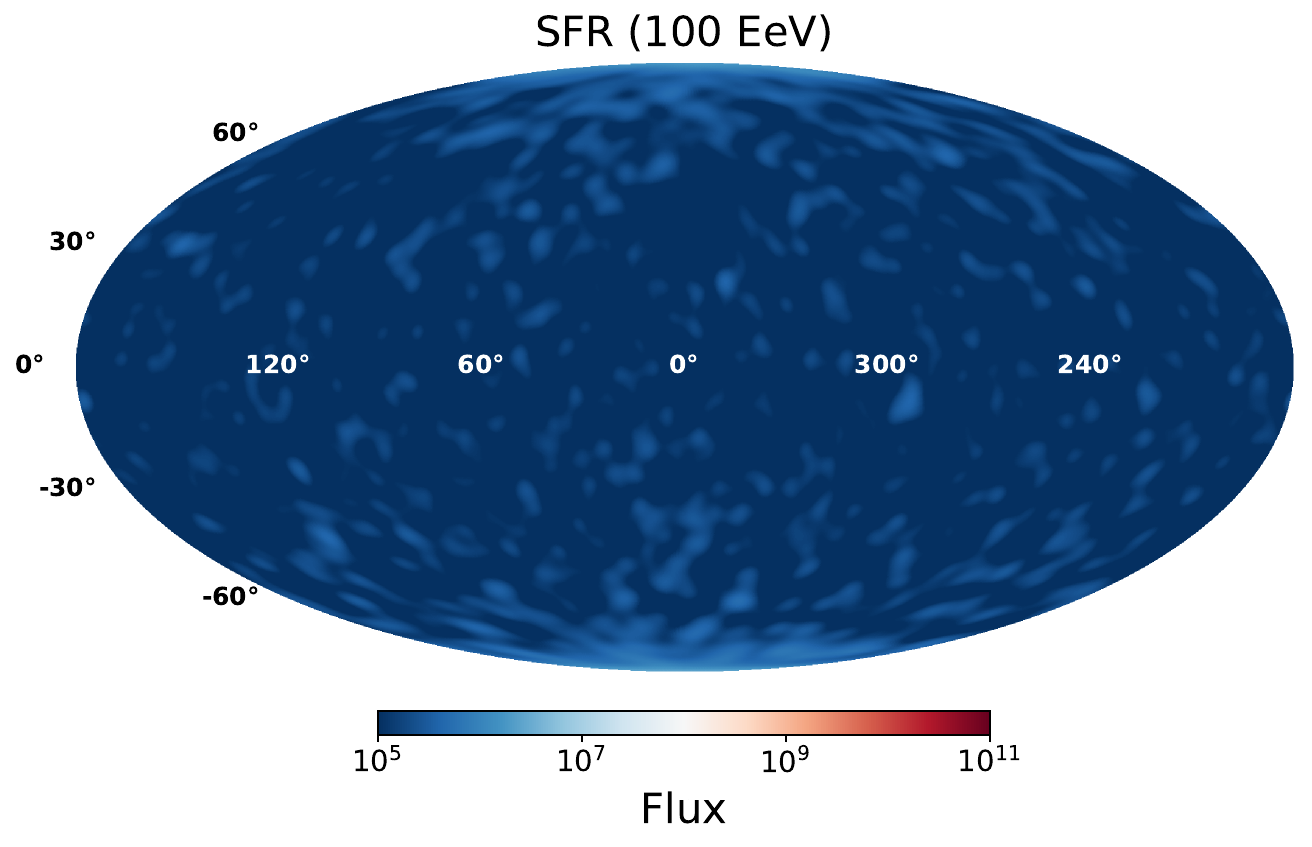}
    \includegraphics[scale=0.28]{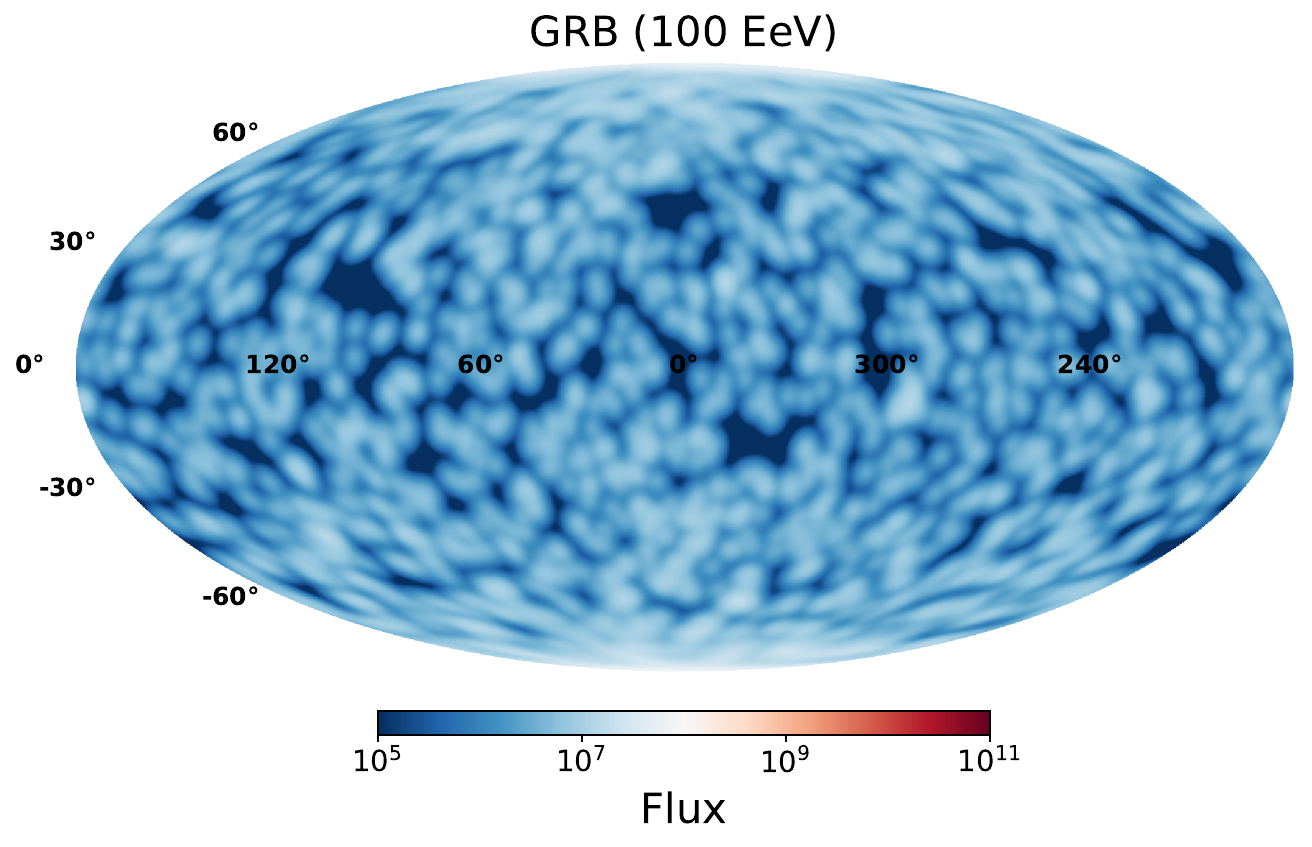}
    \includegraphics[scale=0.28]{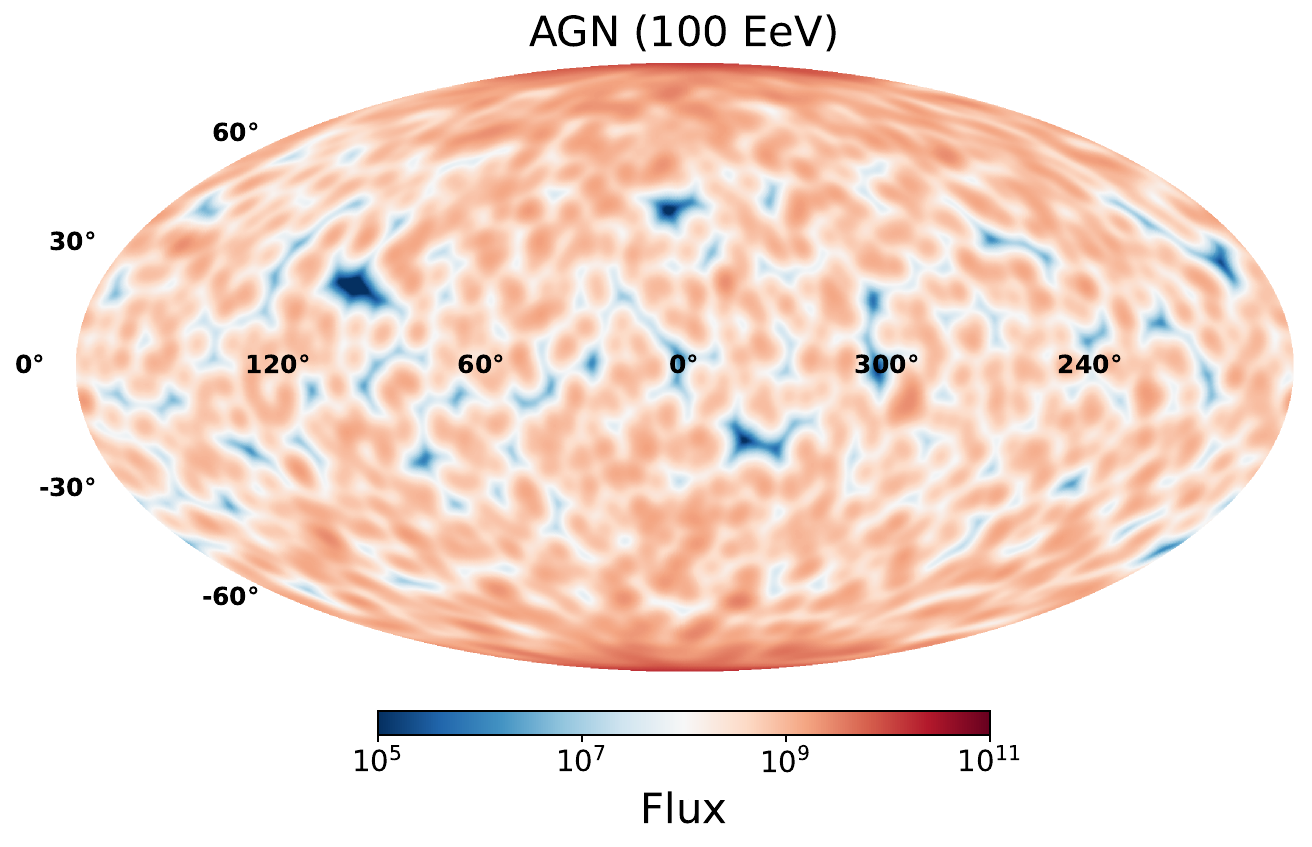}}
\caption{Skymaps of CRs flux distribution for different astrophysical source 
models at varying energies $E = 0.1 $ EeV, $ 50 $ EeV, and $ 100 $ EeV. The
vertical panels correspond to sources following the SFR, GRB, and AGN 
distributions from left to right. The flux is shown on a logarithmic color 
scale, with red indicating higher flux values and blue indicating lower flux 
values.}
\label{fig_app1}
\end{figure*}

Fig.~\ref{fig_app1} presents skymaps of CRs flux distributions for three 
different astrophysical source models at three different energies: 
$ E = 0.1 $ EeV, $ 50 $ EeV, and $100 $ EeV. 
%

\section{Anisotropy map for energy range of $0.1-60$ EeV}\label{appnb}
\begin{figure*}
    \centerline{
 \includegraphics[scale=0.28]{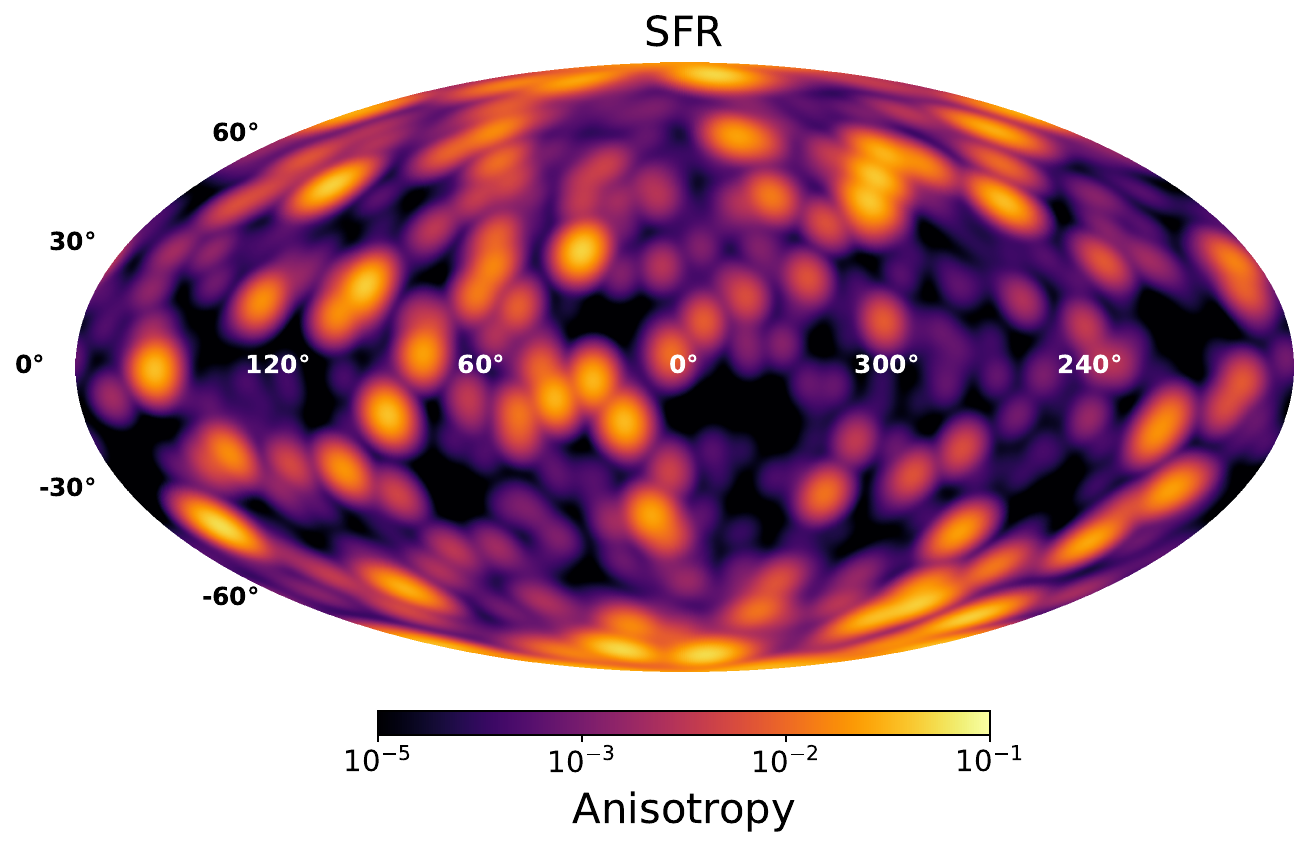}
 \includegraphics[scale=0.28]{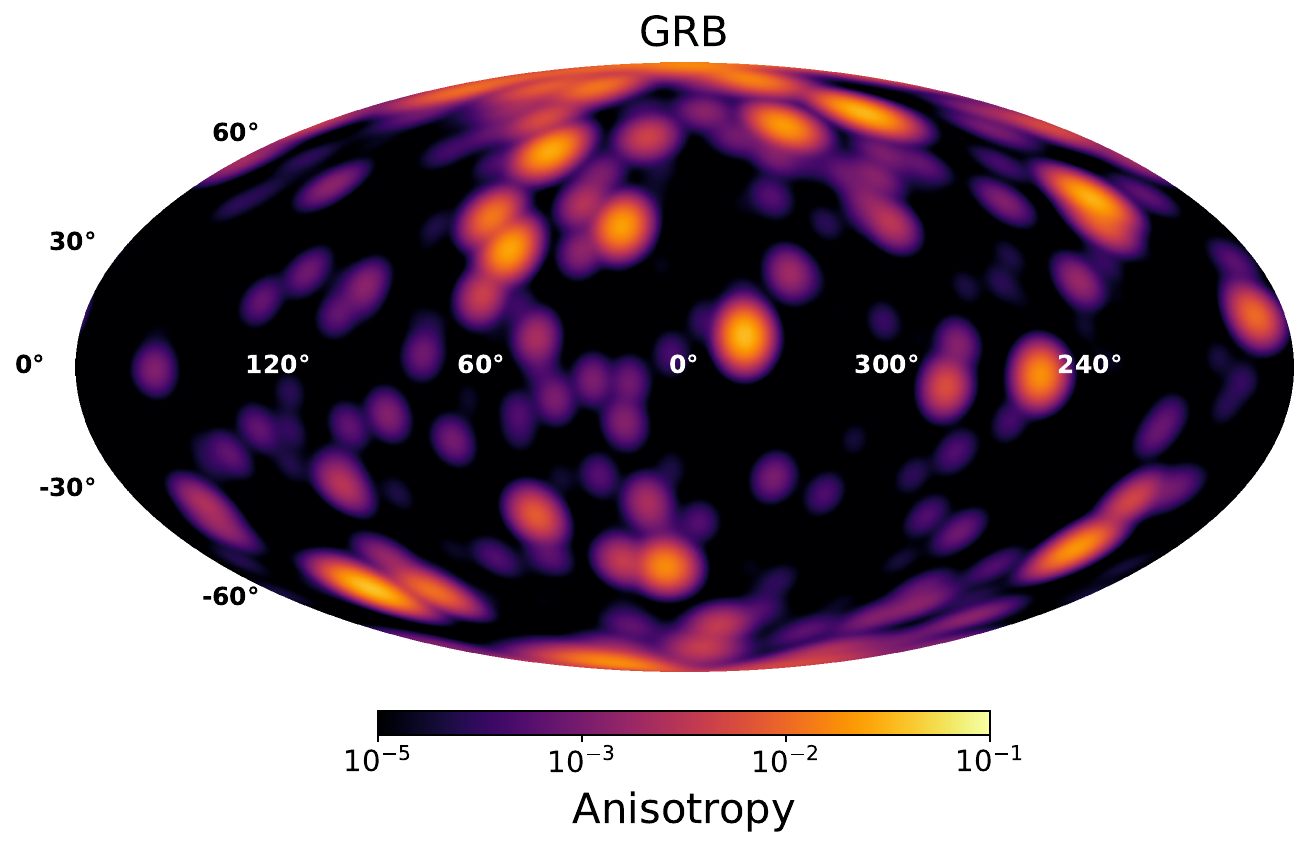}
 \includegraphics[scale=0.28]{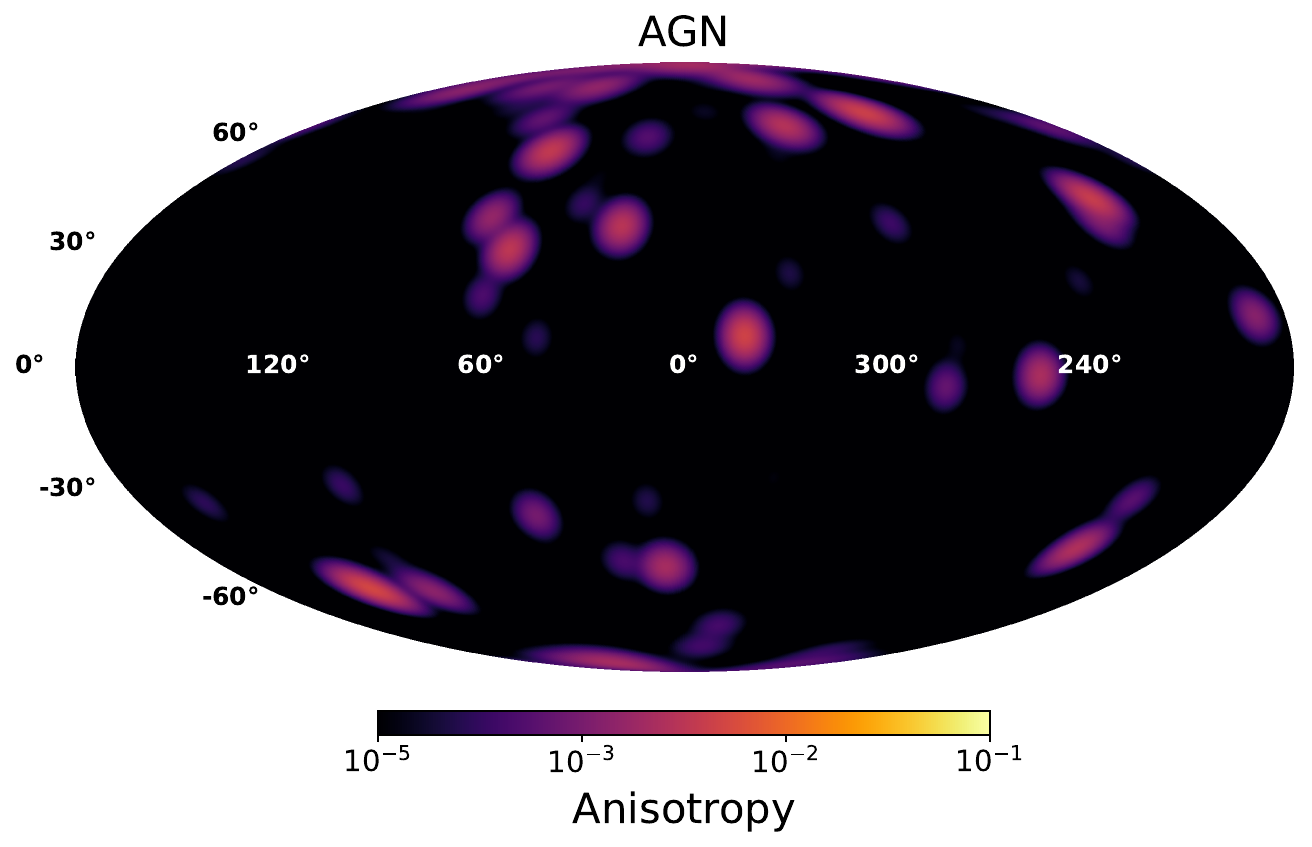}}  
\caption{Skymaps of CRs anisotropy distribution for different astrophysical 
source models at $d_\text{s}=30$ Mpc. Each map of the panel corresponds to 
sources following the SFR, GRB, and AGN distributions from left to right. The 
anisotropy is shown on a logarithmic color scale, with black indicating lower 
anisotropy values and yellow indicating higher anisotropy values.}
\label{fig_app2}
\end{figure*}
The skymaps in Fig.~\ref{fig_app2} illustrate the anisotropy distribution of 
CRs flux in the scale of $0.1-60$ EeV for three different astrophysical source 
models, considering a source separation distance of $ d_s = 30 $ Mpc. 
%

\clearpage


\end{document}